\documentclass[10pt]{article}
\usepackage{epsfig}

\begin{document}
\title{Linear pre-metric electrodynamics and deduction of the light cone}

\author{G. F. Rubilar\footnote{Email: grubilar at udec dot cl.} \\
Departamento de F{\'\i}sica, \\ 
Universidad de Concepci\'on, \\
Casilla 160-C, Concepci\'on, Chile.}

\maketitle
\begin{abstract}
We formulate a general framework for describing the electromagnetic 
properties of spacetime. These properties are encoded in the 
`constitutive tensor of the vacuum', a quantity analogous to that used in the 
description of material media.
We give a {\em generally covariant} derivation of the Fresnel equation 
describing the local properties of the propagation of electromagnetic 
waves for the case of the most general possible linear constitutive tensor. 
We also study the particular case in which a 
light cone structure is induced and the circumstances under which  
such a structure emer\-ges. In particular, we will study the relationship 
between the {\em dual operators} defined by the constitutive tensor under 
certain conditions and the existence of a conformal metric. 
{\em Closure} and {\em symmetry} of the constitutive tensor will be found as 
conditions which ensure the existence of a conformal metric. We will also 
see how the metric components can be explicitly deduced from the constitutive 
tensor if these two conditions are met. 
Finally, we will apply the same method 
to explore the consequences of relaxing the condition of symmetry and how 
this affects the emergence of the light cone.
\end{abstract}

\newpage

\tableofcontents
\newpage

\section{Introduction}

\subsection{Motivation}

In Einstein's theory of gravity, General Relativity (GR), the
fundamental variable describing the gravitational field is the
spacetime metric.  Einstein's equations are partial differential
equations which, together with suitable boundary conditions, determine
the metric field from the energy-momentum distribution of matter.

In GR, the metric tensor fixes all properties of spacetime: its metric
properties (allows to define lengths), its affine structure (allows to
compare object at different points, by parallel transport) and also
its {\em electromagnetic properties} (specifies how electromagnetic
waves propagate and, in particular, defines the light cone at each
event).

In searching for a better understanding of the structures which can be
used to describe spacetime it is interesting and useful to distinguish
each different role that the metric plays in GR. This approach has
been followed before regarding the affine properties of spacetime. As
a result, it is now clear that these properties are in general
described by an affine connection, which is an object logically
independent of the metric. It has been useful to consider the metric
and the affine connection as independent fields describing different
aspects of spacetime.  In this way, it was possible to formulate more
general, so-called metric-affine theories of gravity and spacetime
which have the potential to describe physics beyond GR.  In this more
general framework, Einstein's theory is recovered as a particular case
when certain additional conditions (vanishing of torsion and
non-metricity) are satisfied.  When this happens the affine connection
turns out to be a function of only the metric tensor, as in GR.  On
the other hand, the more general formalism allows us to search
additionally for alternative points of view as, for example, the
possibility to formulate a purely affine theory of gravity, in which
the affine connection is considered to be the fundamental field and
the metric would be a derived quantity.

One of the aims of the present work is to extend this line of thought
also to the description of the electromagnetic properties of
spacetime. For this, it is necessary to recognize as clearly as
possible the role which the metric and affine structures play for the
formulation of classical electrodynamics. It will turn out that the
basic structure of electromagnetic theory, when properly formulated,
is independent of metric and affine concepts.  The electromagnetic
properties of spacetime are encoded in the so-called {\em constitutive
tensor of spacetime}, which is a quantity analogous to that used for
the description of electrodynamics in material media. We consider the
constitutive tensor of spacetime as independent of metric and affine
structures, in the same way as the affine connection was considered to
be independent of the metric, and develop the formalism for the
description of the electromagnetic properties of spacetime. Not
surprisingly, structures richer than in GR are found. In particular,
we study some aspects of wave propagation in the case of an arbitrary
constitutive tensor and derive the Fresnel equation describing the
local properties of the propagation of light. In general, instead of a
light cone structure related to quadratic forms, a general quartic
structure is found. The particular case of Einstein's theory
corresponds to a particular choice of constitutive tensor, completely
determined by the (conformal) metric.

We study the conditions which must be satisfied such that the
electromagnetic properties of spacetime are compatible with a light
cone structure. Finally, it is also possible (similarly to the case of
purely affine theories) to consider the constitutive tensor as a more
fundamental field describing spacetime and the conformal metric as a
derived quantity.

That the conformal metric can indeed be considered as a consequence of
the electromagnetic properties of spacetime should be intuitively
clear.  
The causal properties of spacetime, described by the conformal metric 
in GR, are of an electromagnetic nature.  Light rays are the physical 
realization of null vectors fields, null hypersurfaces are the  
regions through which {\em electromagnetic waves} can propagate. 
{}From an operational point of view it is clear that most of our 
knowledge of spacetime is extracted from the properties of 
electromagnetic fields and how they propagate. Today, we send 
electromagnetic waves towards (reflecting) objects to measure 
distances.  We measure velocities by means of Doppler-shifted 
electromagnetic waves, etc.  The properties of spacetime itself are 
then defined as those independent of the specific configuration of 
the fields and interpreted as describing the underlying `substrate' 
on which we make measurements.

The idea of considering the spacetime {\em metric as a secondary,
derived field}, is not new. Already in 1921 Eddington studied a
`purely affine gravity' model, in which the metric is defined in terms
of a symmetric affine connection, see \cite{Schroedinger60}. Thus, in
Eddingston's theory, the causal properties of spacetime are derived
{}from its affine properties.  Other models have proposed to replace the
metric by a trio of self-dual 2-forms \cite{CJD89,CDJM91}. In
\cite{CDJ91} the metric is obtained from a solution of a theory
formulated only in terms of a $SL(2,C)$ connection, a tetrad, and a
scalar density.  More recently, it has been shown by Barcel\'o et
al. \cite{BLV01a,BLV01b} that an effective metric can be derived
(defined) for {\em almost any} lagrangian theory of scalar fields,
provided one considers perturbations of the fields around some
background configuration. If the theory depends of a {\em single}
scalar field, then a metric can be uniquely defined (so that the
equation for the field perturbation can be written as a Klein-Gordon
equation with respect to that metric) \cite{BLV01a}. If more scalar
fields are involved, multiple metrics can be introduced in general,
which can be described as {\em refringence} (birefringence,
trirefringence, ...) in the sense that different fields of the theory
would `see' different effective metric structures \cite{BLV01b}.  In
these models, the causal properties of space are again a manifestation
of the dynamics of some more fundamental fields.

In a related context, there has been renewed interest in recent years
for so-called `analogue models of gravity', see \cite{VBL01}. These
models are based on the results that some condensed matter systems,
as, for instance, a moving fluid with acoustic perturbations, a moving
dielectric with light, or a moving superfluid with quasiparticles, can
be described in term of some `effective metric'. The metric is also
here a derived, secondary object which turns out to be useful for a
geometrical description (\`a la GR) of the system properties and
depends on the more fundamental degrees of freedom of the system. The
interest in these analogue models lies in the possibility to construct
systems in which {\em kinematical} properties of physics in curved
space could be simulated and tested in the laboratory.  In
\cite{SPS02}, for instance, the possibility of constructing a
`dielectric black hole analogs' is discussed, i.e., dielectric
materials in which the effective metric describes the analog of an
event horizon, see also \cite{LP00,Visser00}.  Dynamics, on the other
hand, is not likely to be simulated, since in general the dynamics of
the effective metric can be completely different to that imposed by
Einstein's equations in GR.

My goal here is not to propose a concrete alternative spacetime
theory, but the more modest one of studying a more general framework
describing classical electrodynamics in which a metric tensor is not
assumed as a basic fundamental field from the very beginning. This
will raise the question on which fundamental structures classical
electrodynamics can be based without using a metric.  The framework
will be called {\em pre-metric} (or sometimes metric-free)
electrodynamics, and it shares many features and analogies with the
theory of electrodynamics in a material medium.  By studying such a
general framework, the conformal properties of spacetime, i.e., the
conformal spacetime metric with its nine independent components, can
eventually emerge as an special case under certain particular
circumstances.  The study of the conditions under which a conformal
metric structure is induced is also one of the subjects of this
work. This framework could then be useful to develop a generalized
theory of spacetime in which the properties of vacuum are treated in
analogy to a material medium and in which GR with its causal
structure could be recovered in suitable limiting cases.

The axiomatic approach to classical electrodynamics presented here has
been developed from the original ideas of Kottler \cite{Kottler22} 
and van Dantzig \cite{vanDantzig34}. They seem
to be the firsts who recognized that the fundamental structure of
Maxwell's equations is independent of the metric and affine structure
of spacetime.  In what concerns the derivation of the spacetime metric
{}from linear electrodynamics, in which we are interested here, Peres
\cite{Peres62}, already in 1962, wrote: `It is therefore suggested to
consider the electromagnetic field as fundamental, and the metric
field only as a subsidiary quantity'. This same idea was also
developed by Toupin \cite{Toupin65} and by Sch\"onberg
\cite{Schoenberg71} who showed that a conformal metric structure is
induced assuming that the `constitutive tensor' defining the
`spacetime relation' between electromagnetic excitations and field
strengths (see section \ref{sectcr}) satisfies the conditions of
symmetry and closure (section \ref{sectcl}). Jadczyk \cite{Jadczyk79}
also showed that a spacetime metric can be introduced under the above
mentioned conditions. However, only very recently and {\em explicit}
derivation of the induced conformal metric has been given by Obukhov
and Hehl \cite{OH99}.

We formulate a general framework for electrodynamics in an arbitrary
linear electromagnetic medium\footnote{We use the word `medium' in a
general sense in order to refer to any `arena' on which
electromagnetic phenomena could take place. In particular, we include
the vacuum as a particular electromagnetic medium.  On the other hand,
`material' or `material medium' will be used to refer to media with a
known (atomic) substructure as, for instance, crystal, liquids,
etc.}. These developments are useful for three different aspects at
least. First, they make the fundamental structures of classical
electrodynamics more transparent. Second, they may provide a basis for
a deeper understanding and for generalized models of spacetime and its
electromagnetic properties.  This includes, for instance, the
application of the formalism to study test theories, see \cite{HL00}
and references therein. Third, the formalism developed in sections
\ref{chax} and \ref{chapwave} can also be interpreted and used in
optics as a {\em general covariant theory of electrodynamics in
inhomogeneous, anisotropic, and in general dissipative material
media}.

Finally, we study the particular case in which a light cone structure
is induced and the circumstances under which such structure
emer\-ges. In particular, we will study the relationship between the
{\em dual operators} defined by the constitutive tensor under certain
conditions and the existence of a conformal metric.  {\em Closure} and
{\em symmetry} of the constitutive tensor will be found to be the
conditions which ensure the existence of a conformal metric.  We will
also see how the metric components can be explicitly derived from the
constitutive tensor if these two conditions are satisfied. We will
also give an alternative, simpler, and more physical derivation of the
metric based on direct use of our general results about the Fresnel
equation describing the local properties of light
propagation. Finally, we will apply the same method to explore the
consequences of relaxing the condition of symmetry and how this
affects the emergence of the light cone.

\subsection{Maxwell-Lorentz equations in 3-vector notation}
We start with electrodynamics within the framework of Special Relativity (SR), 
but in 3D formalism. 

In the usual 3-dimensional vector notation, $\bf V:=(V^x,V^y,V^z)$, the
Maxwell-Lorentz equations in integral form read:
\begin{equation}\label{imev}
\int_{\partial V} {\bf D}\cdot d{\bf S}=\int_V \widetilde{\rho} \, dV , \qquad
\oint_{\partial S}{\bf H}\cdot d{\bf r}=\int_S {\bf J}\cdot d{\bf S} +
\frac{d}{dt}\left(\int_S {\bf D}\cdot d{\bf S}\right) ,
\end{equation}
\begin{equation}\label{hmev}
\int_{\partial V} {\bf B}\cdot d{\bf S}=0 , \qquad
\oint_{\partial S}{\bf E}\cdot d{\bf r}=-
\frac{d}{dt}\left(\int_S {\bf B}\cdot d{\bf S}\right) .
\end{equation}
Here $\bf D$ is the {\em electric excitation} (historically called
`electric displacement'), $\widetilde{\rho}$ the {\em electric charge 
density}, $\bf H$ the {\em magnetic excitation} (historically called
`magnetic field'), $\bf J$ the {\em electric current density}, $\bf B$ the
{\em magnetic field strength} and $\bf E$ the {\em electric field strength}.
The integrals above are defined over arbitrary volumes $V$ with boundary
$\partial V$ and over
arbitrary surfaces $S$ with boundary $\partial S$, respectively.
The corresponding volume elements of the three-, two-, and one-dimensional
regions are denoted by $dV$, $d{\bf S}$, and $d{\bf r}$. Finally, the dot
$\cdot$ denotes the 3-dimensional scalar product of vectors, which is a
metric-dependent operation.

Equation (\ref{imev}a) summarizes the Gauss law. It implies that the field
lines defined by $\bf D$ can be open, the ends of which are located at points
where charge is located.
Equation (\ref{hmev}a) is usually interpreted as expressing the absence of
magnetic monopoles in nature\footnote{For recent (unsuccessful) searches for
magnetic monopoles, see \cite{He97,D098,KMSGSL00} and references therein.}. 
It implies that the magnetic lines defined by
$\bf B$ must be closed. Equation (\ref{hmev}b) summarizes Faraday's induction
law (`a time-variation of an magnetic field induces an electric field').

The field strengths $\bf E$ and $\bf B$ are operationally defined by means of the
{\em Lorentz force law}. On test charges, the force density is given by
\begin{equation}
{\bf F}=\widetilde{\rho}\,{\bf E}+{\bf J}\times{\bf B} .
\end{equation}
Here $\times$ is the 3-dimensional vector product, also a metric-dependent 
operation.

The Maxwell equations are completed by the relations
\begin{equation}\label{crv}
{\bf D}=\varepsilon_0 {\bf E}, \qquad {\bf H}=\frac{1}{\mu_0} {\bf B} .
\end{equation}
The constants $\varepsilon_0$ and $\mu_0$ are called {\em permittivity}
and {\em permeability} of vacuum, respectively. The speed of light,
i.e., the speed with which electromagnetic perturbations propagate,
is given by $c=\frac{1}{\sqrt{\varepsilon_0\mu_0}}$.

As it is well known, the Maxwell equations (\ref{imev}) and (\ref{hmev}) can
be written in differential form as
\begin{equation}\label{imevd}
\nabla\cdot{\bf D}=\widetilde{\rho}, \qquad \nabla\times{\bf H}={\bf J}
+\frac{\partial {\bf D}}{\partial t} ,
\end{equation}
\begin{equation}\label{hmevd}
\nabla\cdot{\bf B}=0, \qquad \nabla\times{\bf E}=-
\frac{\partial {\bf B}}{\partial t} .
\end{equation}

The inhomogeneous Maxwell equations  (\ref{imevd}) are such that the
conservation of electric charge is automatically satisfied, i.e.
\begin{equation}
\frac{\partial \widetilde{\rho}}{\partial t} +\nabla\cdot{\bf J}=0 .
\end{equation}

Actually, Maxwell completed the electromagnetic equations known at his time
by adding the `electric displacement' term
$\frac{\partial {\bf D}}{\partial t}$ {\em such that} the resulting equations
were consistent with charge conservation.

\subsection{Electrodynamics in a material medium}

It is well known that a {\em macroscopic} description of electromagnetic
phenomena inside a material medium (treated as a continuum)
can be achieved using {\em macroscopic
Maxwell equations} which are of the same form as (\ref{imevd}) and
(\ref{hmevd}), but were now
\begin{itemize}
\item $\bf D$ and $\bf H$ denote {\em macroscopic} field strengths related 
to the sources $\rho$ and $\bf J$, which are now {\em external} charge and
current densities, respectively, and
\item the relation between $({\bf D},{\bf H})$ and $({\bf E},{\bf B})$ is now
not given by (\ref{crv}) but by a {\em constitutive relations}
\begin{equation}
{\bf D}={\bf D}\left[{\bf E},{\bf B}\right], \qquad
{\bf H}={\bf H}\left[{\bf E},{\bf B}\right] ,
\end{equation}
which contain the information of the particular electromagnetic properties
of the medium under consideration.
\end{itemize}

Among the many possible particular cases (nonlocal constitutive
laws, ...) we recall here the case of a {\em linear anisotropic medium} for
which the constitutive relations are usually written, in components
(see for instance \cite{LL60}, page 313), as
\begin{equation}\label{c3d}
D_a=\varepsilon_{ab}\,E_b+\alpha_{ab}B_b, \qquad 
H_a=\mu^{-1}_{ab}\,B_b+\beta_{ab}E_b .
\end{equation}
Here $a,b,\ldots=1,2,3$, $\varepsilon_{ab}$ is the permittivity
tensor, and $\mu_{ab}$ the permeability tensor of the medium. The tensors 
$\alpha$ and $\beta$ describe magneto-electrical properties. For examples of 
and further discussions on magneto-electric media, see \cite{ODell70}.

Consider the case in which $\alpha_{ab}$ and $\beta_{ab}$ vanish. 
A {\em non-magnetic} medium corresponds to the
subcase in which $\mu_{ab}=\mu_0 \delta_{ab}$ so that the vacuum relation
(\ref{crv}b) holds.
Interesting properties of non-magnetic anisotropic media regarding
propagation of plane electromagnetic waves include \cite{LL60}:
\begin{itemize}
\item In general, for a triaxial crystal (i.e., when the three eigenvalues of
$\varepsilon_{ab}$ are different), a fourth order Fresnel equation determines
the dispersion relation of plane waves.
\item The wave vector ${\bf k}$ and the ray vector ${\bf s}$ are
in general not parallel: ${\bf k}$ is the vector normal to the wave front,
${\bf s}$ is the direction of energy propagation.
\item Birefringence in uniaxial crystals (two eigenvalues of $\varepsilon_{ab}$
are equal): the Fresnel equation factorizes into two quadratic factors, one
isotropic factor corresponding to `ordinary' waves and an anisotropic one
corresponding to `extraordinary' waves.
\end{itemize}

Effects analogous to the above mentioned will be discussed in section 
\ref{chapwave}.

Other examples of local constitutive laws include:
\begin{itemize}
\item Double refraction induced by an electric field. This so-called Kerr
effect can be induced if an isotropic material is placed in a
constant electric field. The electric field breaks the isotropy of the
medium producing a change in the dielectric constant, leading to effects
similar to those observed in uniaxial crystals. This effect can be
described by the {\em nonlinear} constitutive law corresponding to the following
dielectric tensor:
\begin{equation}
\varepsilon_{ab}=\varepsilon_0\delta_{ab}+\alpha E_aE_b ,
\end{equation}
with some constant $\alpha$.
\item Magneto-optical effects: The dielectric constant depends on the magnetic
field strength $\bf H$.
\end{itemize}

\subsection{Maxwell-Lorentz equations in 3+1 exterior form notation}
\label{secmax31ext}

From the Maxwell-Lorentz equations in their integral form, see (\ref{imev}) and
(\ref{hmev}), one can see that the
different fields appear associated with integrals over regions of different
dimensionality. For the sources, we see that $\widetilde{\rho}$ is integrated over
3-dimensional regions (volumes) while $\bf J$ is integrated over surfaces, 
i.e., 
2-dimensional regions. The excitations $\bf H$ and $\bf D$ are integrated
over 1- and 2-dimensional domains, respectively. Finally, the field strengths
$\bf E$ and $\bf B$ appears in the Maxwell-Lorentz equations under 1- and
2-dimensional
integrals, respectively. From the theory of exterior forms, see 
\cite{Frankel99}, we know that a $p$-form is the natural object to be 
integrated over a $p$-dimensional domain. This means that the Maxwell-Lorentz 
equations can be reformulated in terms of exterior forms according to the 
identifications of table \ref{t1}.
\begin{table}
\begin{center}
\begin{tabular}[h]{|c||c|c|c|c|c|c|}\hline
vector/scalar   & $\widetilde{\rho}$ & $\bf J$ & $\bf H$ & $\bf D$ & $\bf E$ & $\bf B$ \\ \hline
$p$-form & $\rho$ & $j$     & $\cal H$& $\cal D$& $E$     & $B$  \\ \hline
$p$      & 3      & 2       & 1       & 2       & 1       & 2  \\ \hline
\end{tabular}
\caption{Correspondence between vectors and exterior forms.}
\label{t1}
\end{center}
\end{table}
In terms of exterior forms, the Maxwell-Lorentz equations (\ref{imev}) and
(\ref{hmev}) are naturally expressed as
\begin{equation}\label{ihmeext}
\underline{d}\,{\cal D}=\rho, \qquad
\underline{d}\,{\cal H}=\frac{\partial}{\partial t} {\cal D}+j ,
\end{equation}
\begin{equation}\label{hmeext}
\underline{d}\,B=0, \qquad
\underline{d}\,E+\frac{\partial}{\partial t} B=0 .
\end{equation}
Here $\underline{d}=dx^a\wedge\partial_a $ denotes the 3-dimensional exterior
derivative (we will use $d$ for the 4-dimensional exterior derivative). Note the
internal consistency of these equations, since $\underline{d}$ increases by one 
the rank of the exterior form on which it is applied.
Conversely, taking
the equations (\ref{ihmeext}) and (\ref{hmeext}) as starting point, one can
derive the corresponding field equations in terms of field components. We
decompose each form as follows:
\begin{equation} \label{rho3ext}
\rho=\frac{1}{3!}\,\rho_{abc}\, dx^a\wedge dx^ b\wedge dx^c=
\widetilde{\rho}\, dx\wedge dy\wedge dz,
\end{equation}
\begin{equation}
j=\frac{1}{2}\,j_{ab}\, dx^a\wedge dx^b=
\frac{1}{2}\,\hat\epsilon_{abc} j^c\, dx^a\wedge dx^b,
\end{equation}
\begin{equation}
{\cal H}={\cal H}_{a}\, dx^a, \qquad
{\cal D}=\frac{1}{2}\,{\cal D}_{ab}\, dx^a\wedge dx^b=\frac{1}{2}\,
\hat\epsilon_{abc} {\cal D}^c\, dx^a\wedge dx^b,
\end{equation}
\begin{equation} \label{EB3ext}
E=E_{a}\, dx^a, \qquad
B=\frac{1}{2}\,B_{ab}\, dx^a\wedge dx^b=
\frac{1}{2}\,\hat\epsilon_{abc} B^c\, dx^a\wedge dx^b,
\end{equation}
Here $\hat\epsilon_{abc}= \hat\epsilon_{[abc]}$, with $\hat\epsilon_{123}=1$, 
is 
the 3-dimensional Levi-Civita symbol.
Then, substituting (\ref{rho3ext})--(\ref{EB3ext}) into the equations 
(\ref{ihmeext}) and (\ref{hmeext}), one directly finds, see \cite{Schouten89}, 
\begin{equation}\label{ime3c2}
\partial_a {\cal D}^a=\rho, \qquad \epsilon^{abc}\partial_b {\cal H}_c=
\frac{\partial}{\partial t}{\cal D}^a+j^a ,
\end{equation}
\begin{equation}\label{hme3c2}
\partial_a B^a=0, \qquad \epsilon^{abc}\partial_b E_c+
\frac{\partial}{\partial t} B^a=0 ,
\end{equation}
which generalize (\ref{imevd}) and (\ref{hmevd}). Note that ${\cal D}^a$ and
${\cal H}_a$ are 3-dimensional vector {\em densities}.
The formulation in terms
of exterior forms has the advantage that the corresponding $p$-forms are
independent of the 3-dimensional coordinate system used. In other words
(\ref{ihmeext}) and (\ref{hmeext}), and therefore also (\ref{ime3c2}) and
(\ref{hme3c2}), are valid not only in cartesian coordinates but in any
3-dimensional coordinate system. Actually, if one considers
the exterior forms in (\ref{rho3ext})--(\ref{EB3ext}) as basic field variables,
as we do here, then equations (\ref{ihmeext}) and (\ref{hmeext}) are
independent of any metric or affine structure of the 3-dimensional space.

\subsection{Poincar\'e covariant Maxwell equations}

As usually shown in textbooks on SR, the Maxwell equations (\ref{imevd}) and
(\ref{hmevd}) can be written covariantly under Poincar\'e transformations by
defining the field strength $F_{ij}=-F_{ji}$ and the excitation 
$H_{ij}=-H_{ji}$ ($i,j,\ldots=t,x,y,z$) by
\begin{equation}
F_{tx}:=-E^x, \qquad F_{ty}:=-E^y, \qquad F_{tz}:=-E^z,
\end{equation}
\begin{equation}
F_{xy}:=B^z, \qquad F_{yz}:=B^x, \qquad F_{zx}:=B^y,
\end{equation}
\begin{equation}\label{defhv1}
H_{tx}:=H^x, \qquad H_{ty}:=H^y, \qquad H_{tz}:=H^z,
\end{equation}
\begin{equation}\label{defhv2}
H_{xy}:=D^z, \qquad H_{yz}:=D^x, \qquad H_{zx}:=D^y.
\end{equation}
Additionally, the electric current 4-vector density ${\cal J}^i$ is defined as
\begin{equation}
{\cal J}^t:=\rho, \qquad {\cal J}^x:=J^x, \qquad {\cal J}^y:=J^y,
\qquad {\cal J}^z:=J^z .
\end{equation}
Then, we can write (\ref{imevd}) and (\ref{hmevd}) as
\begin{equation}\label{imetd}
\frac{1}{2}\epsilon^{ijkl}\partial_j H_{kl}={\cal J}^i ,\qquad 
\epsilon^{ijkl}\partial_j F_{kl}=0 ,
\end{equation}
where $\epsilon^{ijkl}$ is the Levi-Civita symbol. 
The relation between $F_{ij}$ and $H_{ij}$, namely
equation (\ref{crv}), is translated into
\begin{equation}\label{crt}
H_{ij}=\frac{1}{2}\sqrt{\frac{\varepsilon_0}{\mu_0}}
{\hat\epsilon}_{ijkl}\, \eta^{km}\eta^{ln}F_{mn} ,
\end{equation}
where $\eta^{ij}$ are the components of the Minkowski metric in cartesian
coordinates: $\eta_{ij}={\rm diag}(c^2,-1,-1,-1)$,
$\eta^{ij}={\rm diag}(c^{-2},-1,-1,-1)$.
Inserting (\ref{crt}) into (\ref{imetd}) we obtain (always in cartesian
coordinates) the inhomogeneous Maxwell equations in a form which is often
used in SR, namely
\begin{equation}\label{imetd2}
\sqrt{\frac{\varepsilon_0}{\mu_0}}\,\partial_j F^{ij}={\cal J}^i .
\end{equation}

\subsection{Poincar\'e group invariance and `natural' invariance}
\label{secnat}

It is well known that the Maxwell equations (\ref{imetd}) and
(\ref{crt}) are form invariant under Poincar\'e transformations of the form
\begin{equation}
x^i\rightarrow x^{i^\prime}:=\Lambda^{i^\prime}_{\ i}\, x^i+a^i,
\end{equation}
with $\Lambda^{i^\prime}_{\ i}\in SO(1,3)$, i.e. satisfying
\begin{equation}\label{dlt}
\Lambda^{i^\prime}_{\ i}\Lambda^{j^\prime}_{\ j}\eta^{ij}
=\eta^{i^\prime j^\prime}, \qquad a^i\in {\cal R}.
\end{equation}
This form invariance of the Maxwell equations means that the physics of
the electromagnetic fields remains the same on rotated frames and in frames
moving with constant velocity with respect to each other (i.e. boosted 
frames), as well as at different points in spacetime.

However, a more careful analysis shows that the field equations
(\ref{imetd}) are actually form invariant under {\em any
coordinate transformation} $x^i\rightarrow x^{i^\prime}=x^{i^\prime}(x^i)$,
provided $F_{ij}$, and $H_{ij}$ are considered as components of a second order
(antisymmetric) tensor field, and ${\cal J}^i$ as the components of a vector
density field of weight $+1$. This feature of (\ref{imetd}) is sometimes 
called `natural invariance' of the Maxwell
equations, see \cite{Post97}, chapter 3, for an extended discussion.
The natural invariance shows, on the other hand, that the physical
information about the equivalence of inertial frames under Lorentz 
transformations is contained exclusively in the spacetime relation between 
excitation
and field strength (\ref{crt}). This is a property of the {\em vacuum}.
Furthermore, we note that the Minkowski metric only appears in (\ref{crt}).
Looking at (\ref{crt}) it is then clear that this invariance of the vacuum
is a direct consequence of the form invariance of the Minkowski metric with
respect to Poincar\'e transformations, see (\ref{dlt}). In section 
\ref{secsymlin} we will give a more general definition of symmetries of a 
given electromagnetic medium.

\subsection{Maxwell equations in curved spacetime}

The traditional recipe to construct a theory including the interaction with
the gravitational field is to `replace partial derivatives by covariant
derivatives', also called `minimal coupling' to the gravitational field.
This procedure ensures that the resulting equations are
covariant under an arbitrary change of coordinates, see, for instance,
\cite{MTW73} for more details and examples. The
procedure is not free of ambiguities when one considers the electromagnetic
potential as fundamental variable, since then the Maxwell equations are of
second order and a `normal ordering' problem appears when applying the
recipe above. This is a consequence of the fact that covariant derivatives
do not commute in a curved spacetime.

A not widely recognized fact is, however, that in the case of electrodynamics 
the mentioned recipe is completely unnecessary since the Maxwell equations,
when properly formulated, are `naturally covariant', as we have seen in
section \ref{secnat}.
For the (rather trivial) transition from a Minkowski space to a Riemannian
space with metric $g$ one just needs to replace the spacetime relation by
\begin{equation}\label{crc}
H_{ij}=\sqrt{\frac{\varepsilon_0}{\mu_0}}\hat\epsilon_{ijkl}\,
\sqrt{|g|} g^{km}g^{ln}F_{mn} ,
\end{equation}
with $g:=\det{(g_{ij})}$, so that $F$ and $H$ are tensors, as required by
the natural invariance of (\ref{imetd}). At every event
one can find Riemann normal coordinates which will reduce the metric to its
minkowskian form, i.e. $g_{ij}\stackrel{*}{=}\eta_{ij}$ and then (\ref{crc})
reduces to the form (\ref{crv}), in agreement with the equivalence principle.

Equation (\ref{crc}) can be written in terms of the {\em Hodge dual operator}
${}^*$ of the metric $g$, namely
\begin{equation}\label{crc2}
H=\sqrt{\frac{\varepsilon_0}{\mu_0}}\,{}^*F , \qquad
\left({}^*F\right)_{ij}:=\frac{1}{2}\hat\epsilon_{ijkl}\, \sqrt{|g|}
g^{km}g^{ln}F_{mn}.
\end{equation}

The Maxwell equations in a Riemannian space are therefore, in terms of
$F$ and $g$:
\begin{equation}\label{imec}
\partial_j\left(\sqrt{\frac{\varepsilon_0}{\mu_0}}
\sqrt{|g|} g^{ik}g^{jl}F_{kl}\right)={\cal J}^i ,
\qquad
\epsilon^{ijkl}\partial_j F_{kl}=0 .
\end{equation}

The fact that the Maxwell equations (\ref{imetd}) are
naturally invariant and that the metric structure of space does not enter in
their formulation is not an accident. It is a consequence of the fundamental
property that the basic structure of electrodynamics can be derived from
{\em counting procedures} of charge and magnetic flux, with do not require
a metric (nor an affine) structure of spacetime.
This properties will be further clarified
in the axiomatic approach of section \ref{chax}.

\section{Electrodynamics on an arbitrary 4D-manifold}
\label{chax}

In this section, we would like to present an axiomatic construction of
classical electrodynamics which intends to be as general as possible.
Structures are only introduced when they are indispensable for the
development and not earlier than necessary.
This approach will provide us a very general
framework which can then be applied to many different particular cases.

\begin{figure}[t]
\centering\epsfig{file=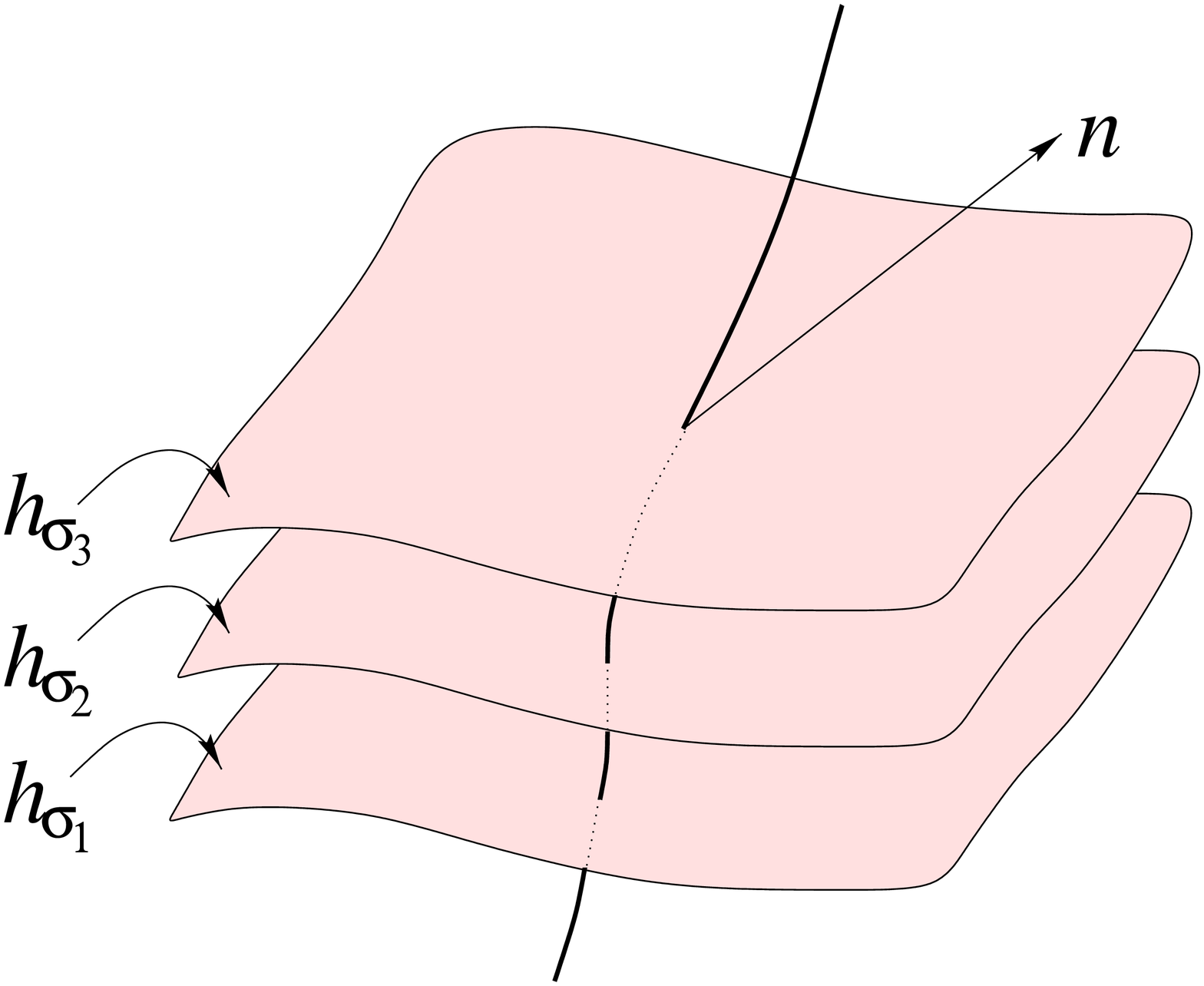, height=7cm, width=10cm}
\caption{Foliation of spacetime: Each hypersurface $h_\sigma$ represents,
   at a `time' $\sigma$, a 3-dimensional submanifold.}
\label{f1}
\end{figure}

We model spacetime as a smooth 4-dimensional manifold $X$ which, at least
in some neighborhood, admits a
foliation into 3-dimensional submanifolds, parameterized by a monotonic
`time' parameter $\sigma$, see figure \ref{f1}.

\subsection{Charge conservation}
Probably {\em the} most important and defining property of electromagnetic
theory is the experimentally well tested fact of charge
conservation\footnote{For experiments
testing charge conservation/violation see, for instance, 
\cite{NB96,SKMW75}}. A basic property of electric
charge is that it is an additive quantity which can be distributed in space,
i.e., it is an {\em extensive quantity}. 
In nature, electric charge is know to be
quantized, its fundamental quanta, the electric charges of the quarks are
$\pm e/3$ and $\pm 2e/3$, and those of the leptons $\pm e$, where $e$ 
denotes the elementary charge. 
In a classical field theory, we describe the distribution of an
extensive quantity in terms of a current density, containing
the information of how many charges are distributed in spacetime, and how
they move.

\begin{figure}[t]
 \centering\epsfig{file=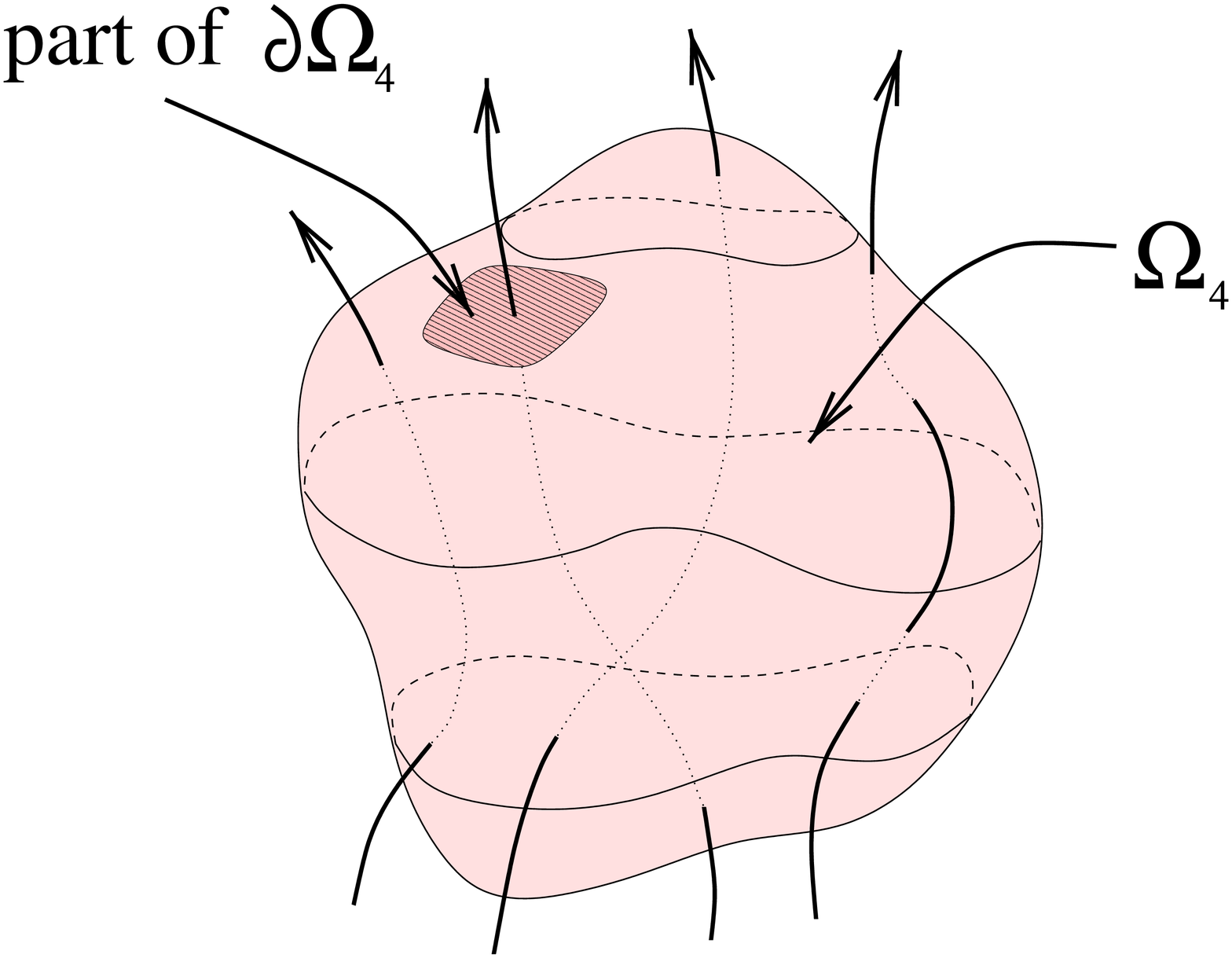, height=7cm, width=10cm}
 \caption{Local conservation of charge: Each worldline of a charged
   particle that enters the finite 4-volume $\Omega_4$ via its
   boundary $\partial \Omega_4$ has also to leave $\Omega_4$.}
 \label{f3}
\end{figure}

Consider the total charge within some 3-dimensional region.
This total charge should be independent of the orientation of any 
coordinate system used. Then, the electric charge distribution is described
by the {\em twisted} electric current 3-form
$J$, see \cite{Frankel99} for definitions.
For a 3-dimensional region $\Sigma_3$ of a hypersurface $h_\sigma$
one can interpret
$\int_{\Sigma_3}J$ as the total charge contained in $\Sigma_3$.
On the other hand, if the 3-dimensional region is of the type
$\Omega_3=\Sigma_2\times[\sigma_1,\sigma_2]$ one can interpret
$\int_{\Omega_3}J$ as the total charge crossing the 2-dimensional surface
$\Sigma_2$ during the `time' interval $[\sigma_1,\sigma_2]$. Consequently,
the 3-form $J$ carries the dimension of charge, i.e. $\left[ J\right] =q$.

Note that no concept of distance or parallel displacement, i.e., no metric or
connection are necessary to define the concepts of charge and
charge current 3-form specifically.
Of course, quantities like `charge per {\em unit volume}' and
`charge per {\em unit area} and {\em unit time}' are useful {\em after} one
provides prescriptions for what `unit volume', `unit area' and `unit time'
are.
However, the latter are concepts not needed to describe, e.g., how many
electrons and therefore how much charge, are contained in a certain region.
Clearly, the total charge is independent of the unit in which volume is measured.

The components $J_{\alpha\beta\gamma}$ of the electric current 3-form with
respect to some coframe basis 
$\vartheta^\alpha$, $\alpha,\beta,\ldots=0,1,2,3$, are determined by
\begin{equation}
J=\frac{1}{3!}\,J_{\alpha\beta\gamma} \vartheta^\alpha\wedge
\vartheta^\beta\wedge
\vartheta^\gamma.
\end{equation}
If one associates a dimension $l$, in the sense of a 
{\em segment} \footnote{i.e., a one dimensional extension on the manifold, 
not in the sense of a {\em unit of length}.}
to the coframe $\vartheta$, i.e. $[\vartheta]=l$, then the components
$J_{\alpha\beta\gamma}$ carry the dimension $[J_{\alpha\beta\gamma}]=ql^{-3}$.

The conservation of electric charge is then expressed as the vanishing
of the integral
\begin{equation}\label{icc}
\oint_{\partial\Omega_4} J=0, \qquad \forall\  \Omega_4 ,
\end{equation}
i.e., for any 3-dimensional boundary of a 4-dimensional region $\Omega_4$.
In particular, for a region $\Omega_4=\Sigma_3\times[\sigma_1,\sigma_2]$,
the integral conservation law (\ref{icc}) requires the balance between the
change of the charge in the region $\Sigma_3$ during the interval
$[\sigma_1,\sigma_2]$ and the flux across its 2-dimensional
boundary $\partial\Sigma_3$.

Since the region $\Omega_4$ in (\ref{icc}) is arbitrary, the Stokes theorem
tells us that the current 3-form must be closed, i.e.
\begin{equation}\label{dj0}
dJ=0 .
\end{equation}

Now, according to the de Rham theorem, see for instance \cite{Frankel99},
the current 3-form is not only closed but also {\em exact}, i.e., it can be 
derived from some 2-form $H$ by exterior derivation, if all its integrals 
vanish over 3-dimensional regions without boundaries. 
Under these assumptions, this means that there must be a 2-form $H$ such that
\begin{equation}\label{me01}
dH = J .
\end{equation}
The twisted 2-form $H$ is called the {\em electromagnetic excitation} and 
carries dimension of charge, $\left[ H\right]=q$. As we
will see, in a 3+1 decomposition its components can be identified with the
metric-free generalization of the electric and magnetic excitations.

However, the conditions above are not enough to uniquely define the 2-form
$H$, since a `gauge' transformation
\begin{equation}
H \rightarrow H':=H+d\Psi
\end{equation}
leaves (\ref{me01}) invariant, for an arbitrary twisted 1-form $\Psi$.

Equation (\ref{me01}) are thus the inhomogeneous Maxwell equations
(4 equations). More than defining the values of $H$, the above arguments show
that the inhomogeneous Maxwell equations {\em must} be of the form
(\ref{me01}), since this is the only kind of field equation which are
compatible with electric charge conservation.

A single electromagnetic excitation will be  picked out by the requirement that
$H=0$ for $F=0$ for the spacetime/constitutive relation, see sections
\ref{secmeasure} and \ref{sectcr}.

\subsection{Lorentz force}
We assume now that the concept of force density is known from mechanics and
use it to define the {\em electromagnetic field strength} 2-form $F$, as
usual, as force per unit charge. We define $F$ by means of
\begin{equation} \label{axiom2}
  f_\alpha=: (e_\alpha\rfloor F) \wedge J\, ,
\end{equation}
where $e_\alpha$ is a frame\footnote{This is again a metric-independent
quantity. A frame is just a basis of the tangent space.}
and $f_\alpha\in\Lambda^4(X)$ are the corresponding components
of the force density covector-valued 4-form in that frame\footnote{Remember,
in classical mechanics force $f_i= \frac{\partial \cal L}{\partial x^i}$ and 
momentum $p_i=\frac{\partial \cal L}{\partial \dot{x}^i}$ are both covectors.}.
From the definition (\ref{axiom2}) $F$ is a
untwisted 2-form, i.e., an intensive quantity. It describes {\em how strongly} 
the electromagnetic field acts on test currents. The dimension of $F$ is
$\left[F\right]=h/q$, with $h$ denoting the dimension of an {\em action}. 
Equivalently $F$ has dimension of magnetic flux. 
The definition above is a very
restrictive one, since it tells us that the force on test charges is
determined by only the six independent components of the 2-form $F$, instead
of the 16 independent quantities that a linear relation between force density
and current density would admit in principle\footnote{Consider for instance
a relation of the form $f_\alpha=\Psi_\alpha\wedge J$. Then $\Psi_\alpha$ is
a covector valued 1-form, and has therefore 16 independent components.}.
This assumption
is part of our axiomatics, and it is suggested by the fact that the
excitation is described by a field with 6 independent components and one 
therefore expects the field strength to have the same number of independent 
components. We
know, of course, that this choice is reasonable since we know that in Maxwell's
theory, the Lorentz force is determined by the six independent
components of the electric and magnetic field strengths.

It is true that in order to have a complete predictive theory, one
still has to specify the relation between velocities and momenta of
test currents. This is required in order to be able to predict the evolution
of test currents in a given electromagnetic field. This
relation includes the metric tensor in the known case of GR.
However, we are interested here in the structure of the general 
electromagnetic theory, and not specifically in the mechanical 
`constitutive relation'.

\subsection{Magnetic flux conservation}

The next step in our axiomatic construction is to find conditions for the
electromagnetic field strength to satisfy.
 The natural operation that can be done with a 2-form is to integrate it on
a given 2-dimensional region.
If we assume, in analogy to conservation of charge, that
\begin{equation}\label{fluxconserv01}
  \oint\limits_{\partial\Omega_3}F=0,
\end{equation}
for an arbitrary 3-dimensional submanifold $\Omega_3$,
then Stokes' theorem provides us with a differential equation for $F$,
namely that the field strength must be closed,
\begin{equation}\label{hme}
dF=0\, ,
\end{equation}
and (at least in some neighborhood of each event) exact, i.e. $F=dA$. The
untwisted 1-form $A$ is then the electromagnetic potential.

We take (\ref{fluxconserv01}), or equivalently (\ref{hme}) as third
axiom. It represents the homogeneous Maxwell equations
(4 equations) and expresses the conservation of magnetic flux.
In general, magnetic
flux is not quantized, as it is the case of electric charge. In type II
superconductors, however, as, e.g., in Niobium, {\em quantized} magnetic flux
lines are possible.

This formulation does {\em not} admit magnetic charges (i.e.
magnetic mono\-po\-les) in a natural way. Since the field strength $F$ is by its
very definition an intensive quantity, a hypothetical magnetic charge density
$\rho_{\rm m}$, such that $dF=\rho_{\rm m}$, would necessarily also be an
intensive quantity (an untwisted 3-form), quite in contrast to the
extensive nature of any charge-like quantity like, e.g., electric charge,
energy-momentum, all twisted 3-forms.

Locally at least, see above, the homogeneous Maxwell equation (\ref{hme})
imply that the field strength $F$ can be derived from a untwisted 1-form $A$,
the electromagnetic {\em potential}, such that
\begin{equation}\label{fda}
F=dA .
\end{equation}
Of course, $A$ is only determined up to a gauge transformation
\begin{equation}
A\rightarrow A^\prime=A+d\Psi,
\end{equation}
for any untwisted 0-form (i.e. scalar) $\Psi$.

The electromagnetic potential carries the same physical dimension as the
electromagnetic field strength, namely $[A]=h/q$.
\begin{figure}[t]
 \centering\epsfig{file=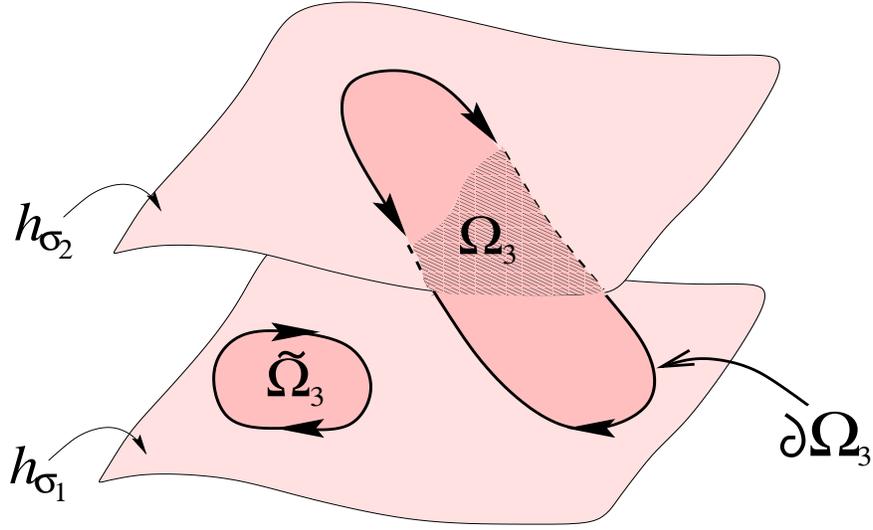, height=7cm, width=11.5cm}
 \caption{Conservation of magnetic flux in spacetime: For an
   arbitrary 3-dimensional integration domain $\Omega_3$,
   the integral $\oint\limits_{\partial\Omega_3}F$ vanishes.}
 \label{f6}
\end{figure}

\subsection{`Space-time' Decomposition}
\label{3mas1}
Consider a foliation characterized by the monotonic `time-like' parameter $\sigma$.
Consider also a vector field $n$ not lying on a $\sigma=$\ const. surface, i.e.
$n\rfloor d\sigma\neq 0$, so that it can be used to `evolve' the folia, see
figure \ref{f1}.
In particular, one can rescale the vector field to make it fulfill the 
normalization condition $n\rfloor d\sigma=1$, which turns out to be useful.

We will call any $p$-form $\omega$ `3-dimensional with respect to $n$' if $n\rfloor\omega=0$. 
The `time-like' components of a `3-dimensional form with respect to 
$n$' in a basis adapted to the folia ($e_0\sim n$) vanish. 
Then, given any $p$-form $\Psi$ one can define the $p$-form $\underline{\Psi}$
and the $(p-1)$-form $\Psi_{\perp}$, both of which are 3-dimensional with respect to $n$ in the above sense, as
\begin{equation}\label{defbar}
\underline{\Psi}:=n\rfloor\left( d\sigma\wedge\Psi\right)  ,
\end{equation}
and
\begin{equation}
\Psi_{\perp}:=n\rfloor\Psi      .
\end{equation}
For $p=4$ we have $\underline{\Psi}=0$. Similarly, for $p=0$, $\Psi_{\perp}=0$.

These two `3-dimensional' quantities $\underline{\Psi}$ and $\Psi_{\perp}$ contain the
complete information of the original form $\Psi$. The latter can be written as
\begin{equation}
\Psi= d\sigma\wedge\Psi_{\perp}+\underline{\Psi}.
\end{equation}

One may also call $\underline{\Psi}$ and $d\sigma\wedge\Psi_{\perp}$ the `transverse' 
and `longitudinal' parts of $\Psi$ with respect to $n$, respectively, see \cite{HO02}.

From their definition, $\underline{\Psi}$ and $\Psi_{\perp}$ satisfy the 
following properties
\begin{itemize}
\item $_{\perp}$ of a product,
\begin{equation}
\left( \Psi\wedge\Phi\right)_{\perp}=\Psi_{\perp}\wedge\underline{\Phi}
+(-1)^p\, \underline{\Psi}\wedge\Phi_{\perp},
\end{equation}
\item $\underline{\ }$ of a product,
\begin{equation}
\underline{\Psi\wedge\Phi}=\underline{\Psi}\wedge\underline{\Phi},
\end{equation}
\item Lie derivative ${\cal L}_n$ and $_{\perp}$ commute,
\begin{equation}
\left( {\cal L}_n\Psi\right)_{\perp} ={\cal L}_n \Psi_{\perp},
\end{equation}
\item Lie derivative ${\cal L}_n$ and  $\underline{\ }$ commute,
\begin{equation}
\underline{{\cal L}_n\Psi}= {\cal L}_n\underline{\Psi}.
\end{equation}
\end{itemize}
Additionally, it is useful to define a 3-dimensional part of the exterior
derivative, $\underline{d}$, such that for any $p$-form $\Psi$
\begin{equation}
\underline{d}\,\Psi:=n\rfloor\left( d\sigma\wedge\Psi\right) .
\end{equation}
It has the following properties:
\begin{itemize}
\item $\underline{d}$  is 3-dimensional with respect to $n$,
\begin{equation}
\underline{\underline{d}\,\Psi}=n\rfloor\left(\underline{d}\,\Psi\right)\equiv 0 ,
\end{equation}
\item $_{\perp}$ of a derivative,
\begin{equation}
\left(d \Psi\right)_{\perp}={\cal L}_n\underline{\Psi} -
\underline{d}\,\Psi_{\perp},
\end{equation}
\item $\underline{\ }$ of a derivative,
\begin{equation}
\underline{d\Psi}=\underline{d}\,\underline{\Psi}. 
\end{equation}

\end{itemize}

By means of this general decomposition procedure, we decompose the Maxwell 
equations (\ref{me01}) and (\ref{hme}) in terms of `3-dimensional' quantities. 
First decompose the fields
$J$, $H$, $F$ and $A$ and introduce the notation
\begin{equation}
j:=-J_{\perp}, \qquad  \rho:=\underline{J} , \qquad
{\cal H}:=H_{\perp}, \qquad  {\cal D}:=\underline{H} ,
\end{equation}
\begin{equation}
E:=-F_{\perp}, \qquad  B:=\underline{F} , \qquad 
\varphi:=-A_{\perp}, \qquad  {\cal A}:=\underline{A} .
\end{equation}
The 4-dimensional quantities can be reconstructed according to
\begin{equation}
J=-j\wedge d\sigma + \rho  , \qquad
H=-{\cal H}\wedge d\sigma + {\cal D}   ,\label{hhd}
\end{equation}
\begin{equation}
F=E\wedge d\sigma + B    ,\qquad 
A=-\varphi\wedge d\sigma + {\cal A} .
\end{equation}
Now one can take the inhomogeneous Maxwell equation (\ref{me01}) and find
\begin{equation}
\left( dH-J\right)_{\perp}=\left( dH\right)_{\perp}-J_{\perp} 
={\cal L}_n\underline{H} -\underline{d}\,H_{\perp}-J_{\perp} 
={\cal L}_n {\cal D} -\underline{d}\,{\cal H}+j ,
\end{equation}
and
\begin{equation}
\underline{dH-J}= \underline{dH}-\underline{J}
= \underline{d}\,\underline{H}-\underline{J}
= \underline{d}\,{\cal D}-\rho .
\end{equation}
Analogously, from the homogeneous Maxwell equations (\ref{hme}) we obtain
\begin{equation}
\left( dF\right)_{\perp}
= {\cal L}_n\underline{F} -\underline{d}\,F_{\perp} 
={\cal L}_n B +\underline{d}\,E ,
\end{equation}
and 
\begin{equation}
\underline{dF}=\underline{d}\,\underline{F} = \underline{d}\,B .
\end{equation}
Thus, we find the Maxwell equations in a 3+1 decomposed form to be
\begin{equation}\label{ihme31}
\underline{d}\,{\cal D}=\rho, \qquad
\underline{d}\,{\cal H}={\cal L}_n {\cal D}+j , \qquad
\underline{d}\,B=0, \qquad
\underline{d}\,E+{\cal L}_n B=0 .
\end{equation}

\begin{figure}[t]
\centering\epsfig{file=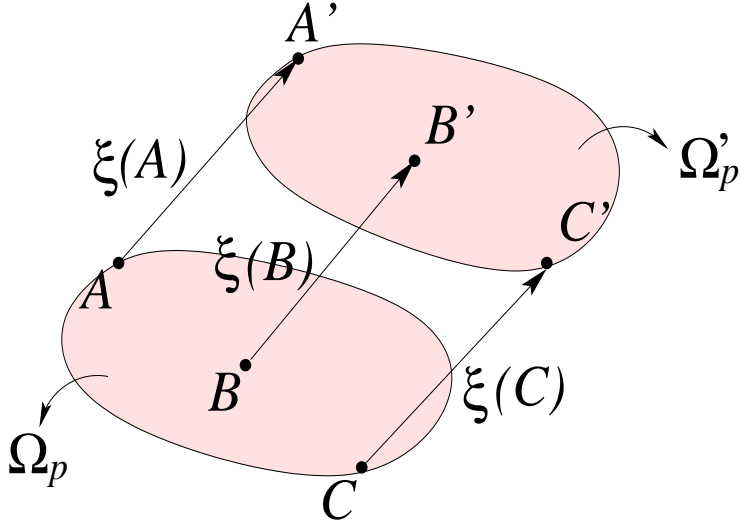, height=7cm, width=10cm}
\caption{Under the displacement $x^i\rightarrow
x^{\prime i}=x^i+a\, n^i(x)$ ($a$ being an infinitesimal constant
parameter) the integral $\int_{\Omega^\prime_p} \Psi$ of a $p$-form $\Psi$
over the corresponding $p$-dimensional mapped region ${\Omega^\prime_p}$
is given by $\int_{\Omega^\prime_p} \Psi= \int_{\Omega_p} \Psi
+a\int_{\Omega_p} {\cal L}_n \Psi$.}
\label{figlie}
\end{figure}

We can also decompose the law of charge conservation (\ref{dj0}).
It is important
to note that no information is obtained from $\underline{dJ}$ since this
quantity vanishes for {\em any} 3-form $J$, as can be seen from the definition
(\ref{defbar}). We then compute
\begin{equation}
\left( dJ\right)_{\perp}
= {\cal L}_n\underline{J} -\underline{d}\,J_{\perp} 
={\cal L}_n \rho +\underline{d}\,j ,
\end{equation}
so that charge conservation means
\begin{equation}
{\cal L}_n \rho +\underline{d}\,j=0.
\end{equation}
Finally, we decompose equation (\ref{fda}). We find
\begin{equation}
\left( dA\right)_{\perp}
= {\cal L}_n\underline{A} -\underline{d}\,A_{\perp} 
={\cal L}_n {\cal A} +\underline{d}\,\varphi ,
\end{equation}
and 
\begin{equation}
\underline{dA}=\underline{d}\,\underline{A} = \underline{d}\,{\cal A} ,
\end{equation}
so that
\begin{equation}
E=-\underline{d}\,\varphi - {\cal L}_n {\cal A}, 
\qquad B=\underline{d}\,{\cal A} .
\end{equation}
The decomposed Maxwell equations (\ref{ihme31}) naturally
generalize equations (\ref{ihmeext}) and (\ref{hmeext}). 
The Lie derivative ${\cal L}_n$ is the natural
generalization of the time derivative since it measures the change of the
(integral of the) corresponding field between different folia, according to 
the displacement induced by the vector field $n$, see figure \ref{figlie}.

\subsection{Measuring $H$}
\label{secmeasure}
In this section, we discuss a general procedure for measuring the excitation 
$H$. This can be done by using an idealized object, namely an {\em ideal 
conductor}. This special material is assumed to have the following two 
properties: 
\begin{enumerate}
\item In an ideal conductor, all charges are located {\em on its surface}.
In other words, the ideal conductor is such that inside it no free
charges can be found. If $\Omega_3$ is the 3-dimensional region describing the
conductor, then the free charges are all located on $\partial\Omega_3$.
\item In the `rest frame' of the conductor the electric excitation $\cal D$
vanishes. If at some event inside the conductor a volume element is spanned
by the vectors $e_a$, $a,b,\ldots=1,2,3$ and $e_0=n$ is a vector pointing in
the zeroth
independent `time' direction, then we assume 
$H_{ab}:=e_b\rfloor e_a\rfloor H=0$, or
equivalently $H=d\sigma\wedge{\cal H}$, with $n\rfloor d\sigma=1$ and 
$e_a\rfloor d\sigma=0$, see (\ref{hhd}).
\end{enumerate}
\begin{figure}[t]
 \centering\epsfig{file=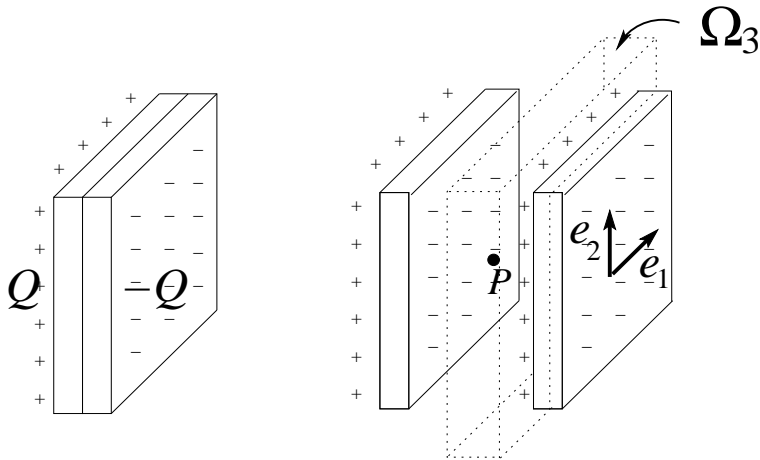, height=7cm, width=10cm}
 \caption{Maxwellian plates. Here $e_1$ and $e_2$ span the surface element
 parallel to the plates.}
 \label{plates}
\end{figure}

With these assumptions one can measure the excitation by use of 
`Max\-we\-llian double plates'. Consider two (uncharged) parallel plates 
made of an ideal conductor
and locate them at the point $P$ where the excitation should be measured.
The field strength (whatever value it may have) will
induce, through the Lorentz force, surface charges in the conductor. 
Separate now the plates and measure the
charge $Q$ induced in one of its surfaces. One can then integrate the
inhomogeneous Maxwell equation over a volume with one side in one conducting
plate and the other between the plates, see figure \ref{plates}. Due to 
property one, in the limit of vanishing `thickness' of $\Omega_3$, 
see fig. \ref{plates}, one finds
\begin{equation}
\int_{\Omega_3}dH=\int_{\partial\Omega_3}H=
\left( e_2\rfloor e_1\rfloor H\right)_{\rm P} -
\left( e_2\rfloor e_1\rfloor H\right)_{\rm cond},
\end{equation}
and $\int_{\Omega_3}J=Q$, so that
\begin{equation}\label{e1e2hq}
\left( H_{12}\right) _{\rm P} -
\left( H_{12}\right)_{\rm cond}=Q.
\end{equation}
The second term on the left hand side of (\ref{e1e2hq}) vanishes because of 
property 2 of ideal conductors. Therefore, the induced charge $Q$ determines
the component $H_{12}$ of the excitation.
Similarly, by orienting the plates differently one can measure, e.g., 
$H_{13}$. Furthermore, by changing the state of
motion of the conductor (i.e., different 4-velocities) one can measure 
components of $H$ which are, in the notation we are using, of the form
$\left( e_1\rfloor e_0\rfloor H\right)=H_{01}$.

\subsection{Constitutive relations}
\label{sectcr}
As it is clear from the last sections, the Maxwell equations in their form
(\ref{me01}) and (\ref{hme}) are valid for {\em any} electromagnetic medium 
(i.e. including the vacuum). They describe
the general features of electrodynamics which follow from charge and magnetic
flux conservation. In this form, Maxwell equations are valid, for instance, in
vacuum in Special Relativity, but also if gravitational effects are included in
the context of General Relativity or alternative theories, as for instance in
those formulated in a general metric-affine spacetime. They are also applicable
to any material medium, if one interprets $J$ and $H$ as macroscopic quantities. 
In other words, (\ref{me01}) and (\ref{hme}) are general
structures of electromagnetic theory. The additional structure that really
defines the particular physical properties of the system under consideration, 
is the subject of this section.
\begin{figure}[t]
\centering\epsfig{file=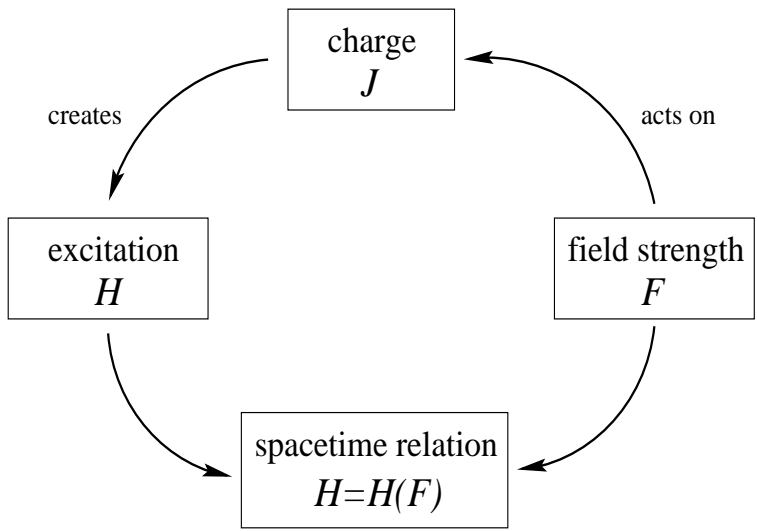, height=7cm, width=10cm}
\caption{The spacetime/constitutive relation connecting  excitation and
field strength.}
\label{const}
\end{figure}

 We saw that the Maxwell equations (\ref{me01}) and (\ref{hme}) amount to only
8 independent equations for the 12 independent fields, $H$ and $F$.
A further relation between the excitation and field strength, i.e., a
connection between the field {\rm generated} by the charges and the field
{\em acting}
on test currents is therefore necessary in order to make the theory complete.
Such a relation is referred to as
the {\em constitutive relation} when corresponding to some material medium, and
as the {\em spacetime relation} when describing the vacuum properties. The
formalism presented here can thus be used in both situations.
In the first one, the
current 3-form describes the so-called `external' (`free') currents flowing
through a material medium. In vacuum, $J$ represents the total current
density. In any case, the relation between $H$ and $F$ describes the
properties of the arena in which the electromagnetic phenomena of interest
take place. Clearly, however, the spacetime relation is of more fundamental 
importance, since it describes a universal property of spacetime. 
Its knowledge 
allows, in principle, to {\em deduce} the constitutive relation of a material 
medium from the spacetime relation and the (charge) structure of the material, 
see the appendix \ref{appmat}. 
Finally, it is also possible, when interested in material media, to abandon 
the 
deductive point of view and adopt a phenomenological description of the 
constitutive relation between the macroscopic field excitation related to the 
macroscopic (free) charges and the field strength without trying to deduce it 
from the underlying structure of the system. From this point of view, the only 
physical property which differentiates what we call vacuum and what we refer 
to as 
a material medium, is that the former does not (by definition) have any charge 
substructure. It is a truly physical question whether the charges described 
by the charge density $J$ are really the total charges within a certain 
region, or 
whether they are macroscopically averaged charges. It is important to realize 
that in any case, as soon as the considered charge is conserved, a 
corresponding 
electromagnetic excitation $H$ can be introduced by mean of the inhomogeneous 
Maxwell equations. As a result, $H$ will be just as `microscopic' as the 
corresponding charge current $J$ is.

In general, spacetime/constitutive relations $H=H(F)$ can have any functional 
dependence,
i.e, they can be non-linear and non-local. Non-local constitutive relations are
common for material media because in general the macroscopic electromagnetic
properties at some point of the material are influenced by the field
and charge configuration at other points of it. Additionally, the
finite propagation speed of the microscopical electromagnetic interaction 
between
the different parts of the material leads to time-like non-locality. In many
cases (typically, for slowly varying fields), however, this non-local effects
are negligible and the medium can be treated as if the field excitation $H$ 
were depending only on the value of the field strength at the same point, i.e. 
with a local constitutive law. 
In what follows, we restrict ourselves only to this later case.

\subsubsection{Linear constitutive relations}

We now concentrate on the particular case in which the field strength and
excitations are proportional. A great number of material media
are known in which this is a good description for a variety of conditions, 
see for instance \cite{LL60}. We also expect the
spacetime relation (i.e. the vacuum relation) to be simple and, in
particular, linear. Linearity is taken for granted in all traditional
approaches to vacuum electrodynamics.

 Given a local frame $\vartheta^\alpha$, with
$\alpha=\hat 0,\hat 1,\hat 2,\hat 3$, we can decompose the
exterior forms $H$ and $F$ as
\begin{equation}\label{geo1}
H = {\frac 1 2}\,H_{\alpha\beta}\,\vartheta^\alpha\wedge\vartheta^\beta,
\qquad
F = {\frac 1 2}\,F_{\alpha\beta}\,\vartheta^\alpha\wedge\vartheta^\beta ,
\end{equation}
and write a general linear spacetime relation (constitutive law) as
\begin{equation}\label{cl}
H_{\alpha\beta}=\frac{1}{2}\,\kappa_{\alpha\beta}^{\ \ \ \gamma\delta} 
F_{\gamma\delta}=\frac{1}{4}\,{\hat\epsilon}_{\alpha\beta\gamma\delta}\,
\chi^{\gamma\delta\epsilon\theta}\,F_{\epsilon\theta} ,
\end{equation}
where ${\hat\epsilon}_{\alpha\beta\gamma\delta}$ is the Levi-Civita symbol with
${\hat\epsilon}_{\hat 0\hat 1\hat 2\hat 3}:=1$ and
$\chi^{\alpha\beta\gamma\delta}$ is called the {\em constitutive
tensor density}, which is an untwisted tensor density of weight $+1$, carrying
dimension $\left[\chi\right]=\left[H\right]/\left[F\right]=q^2/h$ or,  
equivalently, dimension of conductance or 1/(resistance).

 From its definition,
the constitutive tensor satisfies the following symmetry properties
\begin{equation} \label{symm01}
{\chi}^{\alpha\beta\gamma\delta}= -{\chi}^{\beta\alpha\gamma\delta}
=-{\chi}^{\alpha\beta\delta\gamma} ,
\end{equation}
which means that it has 36 independent components. Due to these symmetry
properties, the constitutive tensor can also be represented by $6\times 6$
matrix, say $\chi^{IJ}$, where each index $I,J,\ldots$ corresponds to a
pair of antisymmetrized 4-dimensional indices, i.e., $I\rightarrow [i_1 i_2]$,
etc. In this notation, e.g., $\epsilon^{IJ}\rightarrow \epsilon^{i_1i_2j_1j_2}$.
One can enumerate the
6-dimensional indices according to $I,J,\ldots=01,02,03,23,31,12$.

We want to identify some irreducible components of the constitutive tensor.
Usually, irreducible pieces of a covariant object have different physical 
properties. 
Remember that so far no metric structure has been introduced. However, we
always have the Levi-Civita tensor density
$\epsilon^{\alpha\beta\gamma\delta}$ at
our disposal, and thus we can decompose the constitutive tensor according to, 
see \cite{ROH02},
\begin{equation}\label{decomp}
{\chi}^{\alpha\beta\gamma\delta}={}^{(1)}\chi^{\alpha\beta\gamma\delta}+
{}^{(2)}\chi^{\alpha\beta\gamma\delta} + {}^{(3)}\chi^{\alpha\beta\gamma\delta} ,
\end{equation}
where the different irreducible pieces ${}^{(1)}\chi$, ${}^{(2)}\chi$, and
${}^{(3)}\chi$ are determined by the symmetry properties
\begin{equation}
{}^{(1)}\chi^{\alpha\beta\gamma\delta}={}^{(1)}\chi^{\gamma\delta\alpha\beta},
\qquad {}^{(1)}\chi^{[\alpha\beta\gamma\delta]}= 0 ,
\end{equation}
\begin{equation}
{}^{(2)}\chi^{\alpha\beta\gamma\delta}=-
{}^{(2)}\chi^{\gamma\delta\alpha\beta} ,\qquad
{}^{(2)}\chi^{[\alpha\beta\gamma\delta]}=0 ,
\end{equation}
\begin{equation}
{}^{(3)}\chi^{\alpha\beta\gamma\delta}=
{}^{(3)}\chi^{[\alpha\beta\gamma\delta]}, \qquad
{}^{(3)}\chi^{\alpha\beta\gamma\delta}=
{}^{(3)}\chi^{\gamma\delta\alpha\beta},
\end{equation}
or, explicitly,
\begin{equation}
{}^{(3)}\chi^{\alpha\beta\gamma\delta}:= \chi^{[\alpha\beta\gamma\delta]} ,
\end{equation}
\begin{equation}
{}^{(2)}\chi^{\alpha\beta\gamma\delta}:= \frac{1}{2}\,\left[
\chi^{\gamma\delta\alpha\beta}-\chi^{\alpha\beta\gamma\delta}\right] ,
\end{equation}
\begin{equation}
{}^{(1)}\chi^{\alpha\beta\gamma\delta}:=
\chi^{\alpha\beta\gamma\delta}-{}^{(2)}\chi^{\alpha\beta\gamma\delta}-
{}^{(3)}\chi^{[\alpha\beta\gamma\delta]}.
\end{equation}
One can introduce an alternative, but equivalent, parametrization of the
15 independent components of the asymmetric piece ${}^{(2)}\chi$
in terms of a {\em traceless} second rank tensor $S_\alpha^{\ \beta}$ (thus, also
with 15 independent components) \cite{HOR02} as
\begin{equation} \label{2chis}
{}^{(2)}\chi^{\alpha\beta\gamma\delta}=\epsilon^{\alpha\beta\epsilon[\gamma}
S_\epsilon^{\ \delta]}-\epsilon^{\gamma\delta\epsilon[\alpha}S_\epsilon^{\ \beta]},
\qquad S_\alpha^{\ \alpha}=0 .
\end{equation}
Now, contracting (\ref{2chis}) with the Levi-Civita symbol, one finds that
\begin{equation} \label{stij}
S_\alpha^{\ \beta} =\frac{1}{4}\,\hat\epsilon_{\alpha\gamma\delta\epsilon}
\,{}^{(2)}\chi^{\gamma\delta\epsilon\beta},
\end{equation}
which shows that the traceless tensor $S_\alpha^{\ \beta}$
is uniquely determined by (\ref{2chis}).

The third piece ${}^{(3)}\chi$ can always be written as
\begin{equation}
{}^{(3)}\chi^{\alpha\beta\gamma\delta}=
\alpha(x)\,\epsilon^{\alpha\beta\gamma\delta},
\end{equation}
where $\alpha(x)$ is uniquely determined to be
\begin{equation}
\alpha:=\frac{1}{4!}\,
\hat\epsilon_{\alpha\beta\gamma\delta}{\chi}^{\alpha\beta\gamma\delta} .
\end{equation}
It is called an abelian axion field.

Denote the contribution of each piece of the constitutive
tensor to the excitation as ${}^{(1)}H$, ${}^{(2)}H$ and ${}^{(3)}H$, such that
\begin{equation}
{}^{(1)}H_{\alpha\beta}=\frac{1}{4}\,{\hat\epsilon}_{\alpha\beta\gamma\delta}\,
{}^{(1)}\chi^{\gamma\delta\epsilon\theta}\,F_{\epsilon\theta} ,
\end{equation}
and analogously for the other pieces. Then
\begin{equation}
H={}^{(1)}H+{}^{(2)}H+{}^{(3)}H,
\end{equation}
and
\begin{equation}\label{3halpha}
{}^{(3)}H=\alpha\,F .
\end{equation}
Using (\ref{me01}) and (\ref{hme}), we find
\begin{eqnarray}
dH&=&d\left({}^{(1)}H+{}^{(2)}H+{}^{(3)}H\right) \nonumber \\
&=&d\left({}^{(1)}H+{}^{(2)}H\right)+d\left(\alpha\wedge F\right) \nonumber \\
&=& d\left({}^{(1)}H+{}^{(2)}H\right)+d\alpha\wedge F ,
\end{eqnarray}
showing that the axion piece contributes to the Maxwell equations only if
$d\alpha\neq 0$.

Equation (\ref{decomp}) represents the irreducible decomposition of the
corresponding $6\times 6$ matrix with respect to the linear group into a
symmetric traceless piece, an antisymmetric piece, and a trace piece. The
Levi-Civita symbol serves as a kind of `metric' in the 6-dimensional
formulation ($\epsilon^{ijkl}\rightarrow \epsilon^{IJ}$) which is used to
construct the traces of $\chi^{IJ}$. The constitutive tensor is therefore
reduced as ${\chi}={}^{(1)}\chi+{}^{(2)}\chi + {}^{(3)}\chi$ in $36=20+15+1$
independent components, respectively. No further decomposition of the
constitutive tensor is possible at this point, since no additional geometric
objects are available.

For later application, see section \ref{secsingu}, we abbreviate the action of
the constitutive tensor by defining the operator
$^\#:\Lambda^2\rightarrow \Lambda^2$ such that
\begin{equation}\label{defdual}
^\#\Omega:=\frac{1}{4}\,\hat\epsilon_{\alpha\beta\gamma\delta}
\,\chi^{\gamma\delta\epsilon\theta}\,
\Omega_{\epsilon\theta}\, \vartheta^\alpha\wedge \vartheta^\beta ,
\end{equation}
for any 2-form $\Omega$ with frame components $\Omega_{\alpha\beta}$.
Then we can rewrite our spacetime/constitutive relation (\ref{cl}) as
\begin{equation}\label{cl03}
H=\,^\#F.
\end{equation}

Each irreducible piece is expected to
describe different aspects of the medium. Additional information about the
different properties of each piece will be obtained from the study of the
electromagnetic energy-momentum current and of wave propagation.
Notice for example that, if ${}^{(2)}\chi\neq 0$ the Maxwell equations
cannot follow as Euler-Lagrange equations from a Lagrangian of the usual form
$V:=H\wedge F$, since ${}^{(2)}\chi\neq 0$ drops out from $V$ due to its
symmetry properties.
However, our intention is to try to develop our electromagnetic theory as
generally as possible. Therefore we want to also include systems for which no
Lagrange density can be found. Typical examples of such kind of physical systems
are those including
dissipative effects. One can therefore expect the `extra' irreducible piece
${}^{(2)}\chi$ to be related to some kind of intrinsic dissipative property of the medium.
As we will see in section \ref{secsymem}, this is indeed the case. Furthermore,
constitutive
laws (for matter) with ${}^{(2)}\chi\neq 0$
(non-vanishing ``skewon fields'') have been discussed by Nieves and Pal
\cite{NP89,NP94}. They yield $T$- and $P$-violating terms in the field equations. 

Non-abelian axions were
postulated for the first time by Peccei and Quinn \cite{PQ77}, see also
\cite{Weinberg78,Wilczek78}. Abelian axions coupling to electromagnetism
were first considered by Ni \cite{Ni77} and correspond to a non-vanishing 
piece ${}^{(3)}\chi$. 
There have been intensive experimental searches for axions, see
\cite{Ni99,Stedman97,vBK01} and references therein.
To date, no evidence of such a field has
been found. Constraints on the axion mass and coupling to the
electromagnetic field have been obtained both from astrophysical
observations as from laboratory experiments, see \cite{Ni99,Stedman97,vBK01}
for details. However, as we will see later, the axion-like term
does not enter into important quantities as the energy-momentum current and
the Fresnel equation. The axion remains a serious candidate
for a particle search in experimental high energy physics and is a candidate
for cold dark matter.
The discussion about the possible existence of such
field for some {\em material medium} has been rather controversial, see
\cite{WL98,WL98b,T99} and references therein. In this context the vanishing 
on the axion piece is referred to as the `Post constraint' (PC), after the
work of Post \cite{Post97}. Sihvola and collaborators have correctly
recognized that the axion piece is allowed by the basic structure of
electrodynamics. However, if the medium is homogeneous,
$\alpha$ is constant\footnote{by the very definition of an homogeneous medium.}
and then the corresponding term drops out completely from the Maxwell field
equations. Therefore, a possible axion field can only be detected if it is
inhomogeneous ($d\alpha\neq 0$), or by its effects on the
boundary separating two homogeneous media with different axion fields each.
Some theoretical work on the reflection
and transmission properties of this kind of medium can be found in \cite{T99} 
and references therein.
No clear example of a material medium with a nontrivial 
axion-like term in its constitutive law has been reported in the literature.

Within a $\chi$-$g$ formalism, i.e., when a (symmetric) constitutive tensor $\chi$ is assumed 
for the description of the electromagnetic properties of spacetime {\em in addition to a 
metric} $g$, astrophysical constraints on `nonmetric' theories, i.e., in which the 
constitutive tensor differs from the one on a riemannian space, were considered in 
\cite{Ni84}. The particular case of deviations from the constitutive tensor of Minkowski 
vacuum is studied in \cite{HK95,EFMMN00,KM01}. 

\subsubsection{Three dimensional decomposition of the constitutive tensor}

In some particular applications in which a $3+1$ decomposition is used, as for
instance in nonrelativistic and/or noncovariant formulations of 
electrodynamics, see section \ref{secnievespal} for a particular example, it is convenient to express the constitutive tensor (36 components) in terms of four $3\times 3$ matrices $\cal A, B, C$, and $\cal D$ (each with 9 independent components). We define\footnote{The conventions in these definitions are taken such that they are consistent with those of \cite{HO02}.} them as follows:
\begin{equation}\label{cd01}
{\cal A}^{ba}:=\chi^{0a0b},
\qquad {\cal B}_{ba}:=\frac{1}{4}\,\hat\epsilon_{acd} 
\chi^{cdef}\hat\epsilon_{efb},
\end{equation}
\begin{equation}  \label{cd02}
{\cal C}^b_{\ a}:=\frac{1}{2} \,\hat\epsilon_{acd}\chi^{cd0b},
\qquad {\cal D}_b^{\ a}:=\frac{1}{2} \, \chi^{0acd} \hat\epsilon_{cdb} .
\end{equation}
Here $a,b,c,\ldots=1,2,3$. The inverse relations are
\begin{equation}
\chi^{0a0b}={\cal A}^{ba},
\qquad {\chi}^{0abc}={\cal D}_d^{\ a}\epsilon^{dbc},
\end{equation}
\begin{equation}
\chi^{ab0c}=\epsilon^{abd}{\cal C}^c_{\ d},
\qquad
\chi^{abcd}=\epsilon^{abe}{\cal B}_{fe}\epsilon^{fcd}.
\end{equation}

Additionally, one can $(3+1)$-decompose the coordinate components of 
excitation and field strength as
\begin{equation}\label{3vect2a}
{\cal D}^a := \left(H_{23},H_{31},H_{12}\right), \quad
{\cal H}_a := \left(H_{01}, H_{02},H_{03}\right),
\end{equation}
\begin{equation}\label{3vect2b}
B^a := \left(F_{23},F_{31},F_{12}\right), \quad
E_a := \left(F_{10},F_{20},F_{30}\right).
\end{equation}
With these definitions, the spacetime/constitutive relation can be written as
\begin{equation}\label{31cr}
{\cal D}^a := -{\cal A}^{ba}E_b+{\cal D}_b^{\ a}B^b, \quad
{\cal H}_a := -{\cal C}^b_{\ a}E_b+{\cal B}_{ba}B^b .
\end{equation}
From these relations, one sees that the components of ${\cal A}$ correspond 
to the components of the usual dielectric tensor $\varepsilon$, whereas the components 
of ${\cal B}$ correspond to those of the inverse permeability tensor $\mu$, see (\ref{c3d}). 
Furthermore, using the 6-dimensional notation, we have
\begin{equation}
H_I=\left({\cal H}_a, {\cal D}^a\right), \qquad
F_I=\left( -E_a, B^a \right) ,
\end{equation}
and the spacetime relation is written as
\begin{equation}
H_I=\epsilon_{IJ}\chi^{JK}F_K ,
\end{equation}
with
\begin{equation}\label{chiepsIJ}
\chi^{IJ}= \left( \begin{array}{cc} 
                {\cal B}_{ab}& {\cal D}_a{}^b \\ 
                {\cal C}^a{}_b & {\cal A}^{ab} 
                \end{array}\right),
\qquad
\epsilon^{IJ}=\epsilon_{IJ}=\left(\begin{array}{cc}
                0_3 & 1_3 \\
                1_3 & 0_3
                \end{array} \right)  .
\end{equation}
Then the irreducible decomposition of $\chi$ can be found to be
\begin{equation}\label{chi1epsIJ}
{}^{(1)}\chi^{IJ}=\left(\begin{array}{cc}
{\cal B}_{(ab)} &
\frac{1}{2}\left(\not\!\!{\cal D}_a^{\ b}+\not\!{\cal C}^b_{\ a}\right)  \\
\frac{1}{2}\left(\not\!{\cal C}^a_{\ b}+\not\!\!{\cal D}_b^{\ a}\right)
& {\cal A}^{(ab)}
                \end{array} \right) ,
\end{equation}
\begin{equation}\label{chi2epsIJ}
{}^{(2)}\chi^{IJ}=\left(\begin{array}{cc}
{\cal B}_{[ab]} &  \frac{1}{2}\left({\cal C}^b_{\ a}-{\cal D}_a^{\ b}\right) \\
\frac{1}{2}\left({\cal D}_b^{\ a}-{\cal C}^a_{\ b}\right) & {\cal A}^{[ab]}
                \end{array} \right)  ,
\end{equation}
\begin{equation}\label{chi3epsIJ}
{}^{(3)}\chi^{IJ}=
\frac{1}{6}\left( {\cal C}^c_{\ c}+{\cal D}_c^{\ c}\right)\,\epsilon^{IJ} .
\end{equation}
We introduced the 3-dimensional traceless quantities
\begin{equation}
\not\!{\cal C}^a_{\ b}
:={\cal C}^a_{\ b}- \frac{1}{3}\,{\cal C}^c_{\ c}\delta^a_b, \qquad
\not\!\!{\cal D}_a^{\ b}
:={\cal D}_a^{\ b}- \frac{1}{3}\,{\cal D}_c^{\ c}\delta_a^b ,
\end{equation}
so that
\begin{equation}
\not\!{\cal C}^a_{\ a}=0, \qquad \not\!\!{\cal D}_a^{\ a}=0.
\end{equation}

In terms of $S$, see (\ref{stij}), the second irreducible piece of the 
3-dimensional matrices (\ref{cd01})-(\ref{cd02}) are found to be given by
\begin{equation}
{}^{(2)}{\cal A}^{ba}=\epsilon^{abc} S_c^{\ 0} ,\qquad
{}^{(2)}{\cal B}_{ba}=-\hat\epsilon_{abc}\, S_0^{\ c} ,
\end{equation}
\begin{equation}
{}^{(2)}{\cal C}^b_{\ a} =-S_a^{\ b}+\delta_a^b\,S_c^{\ c} , \qquad
{}^{(2)}{\cal D}_b^{\ a}=S_b^{\ a}-\delta_b^a\,S_c^{\ c} .
\end{equation}
Therefore
\begin{equation}
  ^{(2)}\chi^{IJ}= \left( \begin{array}{cc} 
                   {}^{(2)}{\cal B}_{ab} & {}^{(2)} {\cal D}_a^{\ b} \\ 
                   {}^{(2)}{\cal C}^a_{\ b} & {}^{(2)}{\cal A}^{ab}
    \end{array}\right)
    =\left(\begin{array}{cc}
    \hat\epsilon_{abc}\,S_0{}^c & +S_a^{\ b}-\delta^b_a\,S_c^{\ c} \\
    -S_b^{\ a}+\delta_b^a\,S_c^{\ c} & -\epsilon^{abc} S_c{}^0
    \end{array}
\right)
\end{equation}
and
\begin{eqnarray}\label{Dfield}
  ^{(2)}{\cal D}^a&=&
    \epsilon^{abc}S_b{}^0E_c+\left(S_b{}^a-\delta^a_b S_c{}^c\right) B^b  ,\\
    {}^{(2)}{\cal H}_a&=& \hat\epsilon_{abc}\,S_0{}^bB^c
    +\left( S_a{}^b-\delta_a^bS_c{}^c\right) E_b
     .\label{Hfield}
\end{eqnarray}

\subsubsection{The asymmetric constitutive tensor of Nieves and Pal}
\label{secnievespal}

Consider the particular case in which $S_i^{\ j}$ is the traceless part of the
product of a covector $\omega$ and a vector $v$, i.e.,
\begin{equation}\label{omegav}
S_i^{\ j}=\omega_iv^j-\frac{1}{4}\,\left(\omega_kv^k\right)\,\delta_i^j .
\end{equation}
Then we have
\begin{equation}
{}^{(2)}{\cal A}^{ba}=v^0\epsilon^{abc} \omega_c ,\qquad
{}^{(2)}{\cal B}_{ba}=-\omega_0\,\hat\epsilon_{abc}\, v^c ,
\end{equation}
\begin{equation}
{}^{(2)}{\cal C}^b_{\ a} =-\left[\omega_av^b+\frac{1}{2}\delta_a^b
\left(\omega_0v^0-\omega_cv^c\right)\right] ,
\end{equation}
\begin{equation}
{}^{(2)}{\cal D}_b^{\ a}=\omega_bv^a+\frac{1}{2}\delta_b^a
\left(\omega_0v^0-\omega_cv^c\right) .
\end{equation}
As we mentioned, Nieves and Pal \cite{NP89,NP94} discussed
P- and T-violating generalizations of the Maxwell-Lorentz equations for
isotropic material media. The constitutive tensor they study has a nonvanishing
skew-symmetric piece ${}^{(2)}\chi$. A 4-dimensional formulation of the
constitutive tensor (Nieves and Pal used the 3+1 form of Maxwell's equations in
cartesian coordinates) can be given by considering, in addition to the Minkowski
metric $\eta$, a time-like vector $v$, which explicitly breaks Lorentz invariance.

Consider, in addition to the usual `metric' piece
\begin{equation}
{}^{(1)}\chi^{ijkl}=\sqrt{\frac{\varepsilon_0}{\mu_0}}
\left(\eta^{ik}\eta^{jl}-\eta^{jk}\eta^{il}\right)
=2\sqrt{\frac{\varepsilon_0}{\mu_0}}\,\eta^{i[k}\eta^{l]j},
\end{equation}
an antisymmetric piece ${}^{(2)}\chi$ defined by (\ref{2chis}) and 
(\ref{omegav})
for the particular case in which $\omega$ and $v$ are, in cartesian coordinates
$\omega_a\stackrel{*}{=}(\omega,0,0,0)$ and $v^a\stackrel{*}{=}(v,0,0,0)$. This
choice implies in particular that the material will look spatially
isotropic in this frame.
The corresponding 3-dimensional constitutive
matrices ${\cal A}^{ab}$, ${\cal B}_{ab}$, ${\cal C}^a_{\ b}$, and
${\cal D}_a^{\ b}$, according to their
definitions (\ref{cd01}) and (\ref{cd02}), are then found to be
\begin{equation}
{\cal A}^{ab}\stackrel{*}{=}-\varepsilon_0\delta^{ab},
\qquad {\cal B}_{ab}\stackrel{*}{=}\mu_0^{-1}\delta_{ab},
\end{equation}
\begin{equation}
{\cal C}^b_{\ a}\stackrel{*}{=}-\frac{\omega vc^2}{2}\,\delta^b_a, \qquad
{\cal D}_b^{\ a}\stackrel{*}{=}\frac{\omega vc^2}{2}\,\delta^a_b.
\end{equation}
Here we have also used  $\eta_{00}\stackrel{*}{=}c^2$,
$\eta_{ab}\stackrel{*}{=}-\delta_{ab}$, $\eta^{00}\stackrel{*}{=}c^{-2}$ and
$\eta^{ab}\stackrel{*}{=}-\delta^{ab}$.

With these constitutive matrices, the 3+1 decomposition of the constitutive law
takes the form
\begin{equation}
{\cal D}^a\stackrel{*}{=}\varepsilon_0\delta^{ab}E_b+\frac{\omega vc^2}{2}
\, B^a,
\qquad
{\cal H}_a\stackrel{*}{=}\frac{\omega vc^2}{2}\, E_a +\mu_0^{-1}\delta_{ab}B^b ,
\end{equation}
or, in the vector notation used by Nieves and Pal \cite{NP89,NP94},
\begin{equation}
{\bf D}\stackrel{*}{=}\epsilon_0{\bf E}+\gamma\, {\bf B}, \qquad
{\bf H}\stackrel{*}{=}\gamma\, {\bf E}+\mu_0^{-1}{\bf B}.
\end{equation}
Therefore, the constitutive relation of Nieves and Pal can be recovered 
provided $\gamma=\omega vc^2/2$.

Since we have introduced an additional timelike vector, it is natural that
the material will violate $T$ invariance, since $v$ defines a preferred `time
direction'.

\subsubsection{Nonlinear constitutive relations}

We are mainly interested in {\em linear electrodynamics}. However, the formalism 
developed here can also be applied to nonlinear models, in the case that one concentrates 
on the properties of small perturbations of the electromagnetic configuration. This is, 
for instance, the case when one wishes to study propagation of electromagnetic waves in a 
nonlinear model. Examples of such models are the effective Heisenberg-Euler nonlinear 
theory \cite{HE36,IZ85,PLH97} and Born-Infeld electrodynamics \cite{BI34}. 
In this cases, our linear formalism can be applied by defining an {\em effective constitutive 
tensor}. 

Consider an arbitrary {\em local} relation $H=H(F)$.
The corresponding Maxwell equation in absence of charges is written, in components, as
\begin{equation}\label{mehc}
\partial_j\left(\epsilon^{ijkl}H_{kl}\right)=0 .
\end{equation}
Consider now the properties of a small
perturbation $\Delta F$ of the electromagnetic field around some background
configuration $\bar F$. We write the total electromagnetic field strength
$F$ as $F={\bar F}+\Delta F$.
Then the field excitation can be written, to first order in the perturbation
$\Delta F$, as
\begin{equation}\label{expH}
H_{ij}(F)=H_{ij}({\bar F})+\frac{1}{2}
\left.\frac{\partial H_{ij}}{\partial F_{kl}}\right|_{\bar F} \Delta F_{kl} .
\end{equation}
Inserting (\ref{expH}) into (\ref{mehc}) and assuming that the background
field $\bar F$ is a solution of (\ref{mehc}), i.e.
$\partial_j\left[\epsilon^{ijkl}H_{kl}({\bar F})\right]=0$, we obtain an
equation for the perturbation:
\begin{equation}
\partial_j\left(\frac{1}{2}\epsilon^{ijkl}
\left.\frac{\partial H_{kl}}{\partial F_{mn}}\right|_{\bar F}
\Delta F_{mn}\right)=0 .
\end{equation}
We can write this equation in the same form as in the linear case, i.e. as
\begin{equation}
\partial_j\left(\chi^{ijkl}_{\rm eff} \Delta F_{kl}\right)=0 ,
\end{equation}
with the `effective constitutive tensor'
\begin{equation}\label{ect}
\chi^{ijkl}_{\rm eff}:=\frac{1}{2}\epsilon^{ijmn}
\left.\frac{\partial H_{mn}}{\partial F_{kl}}\right|_{\bar F}  .
\end{equation}
The tensor $\chi^{ijkl}_{\rm eff}$ will, in general, depend on the
local constitutive law and on the background field $\bar F$. This result
shows that most of the results obtained for linear constitutive/spacetime
relations can also be applied to every local electromagnetic theory provided
one considers perturbations on some background solution.
This is the case, for instance, of the propagation of waves in nonlinear media.

\subsection{Symmetries and Energy-momentum}
\label{secsymem}
\subsubsection{Symmetry of a linear medium}
\label{secsymlin}
Here we define the concept of a {\em symmetry of an electromagnetic medium}. This
definition applies to the linear case, when a constitutive tensor $ \chi $ is
available.

To motivate our definition, consider the particular case in which the field
configuration is such that the Lie-derivative of the electromagnetic field strength
along some vector field $\xi$ vanishes, i.e.
\begin{equation}\label{lief0}
{\cal L}_\xi F=0 .
\end{equation}
The geometric interpretation of the condition  (\ref{lief0}) is clear.
It means that the electromagnetic field strength $F$ is `constant' along the
direction $\xi$. More precisely, the {\em integral} $\int_{\Omega_2} F$
on some 2-dimensional surface $\Omega_2$ (which is the natural physical quantity
associated to the 2-form $F$) is invariant under the displacement of the 
integration
region $\Omega_2$ induced by $\xi$. This property follows from the very definition
of the Lie derivative, see, for instance, the discussion in \cite{Post97} and
figure \ref{figlie} for $\Psi=F$ and $n=\xi$.

Now, in linear electrodynamics the excitation is determined by the
field strength $F$ and the constitutive tensor $\chi$. It is clear that even if 
(the integral of) the field strength $F$ is constant under the displacements 
defined by $\xi$, i.e., if (\ref{lief0}) holds, the excitation will not be constant
unless the {\em medium itself}, i.e. the constitutive tensor,
satisfies some condition. We will call this condition a {\em symmetry} condition 
for the medium. 
Clearly, the condition we are referring to is the vanishing of the Lie
derivative of the constitutive tensor, since in this case (\ref{lief0}) implies
that $H$ is also constant, i.e. ${\cal L}_\xi H=0$. This lead us to the following
definition:

\underline{Definition:}
A linear electromagnetic medium is said to have a symmetry under the displacement
induced by a vector field $\xi$ if the Lie derivative along $\xi$ of its
constitutive tensor vanishes, i.e. if
\begin{equation}\label{liechi}
{\cal L}_\xi\,\chi^{ijkl}=0,
\end{equation}
for some vector field $\xi$.

Notice that this condition implies the independent vanishing of the Lie
derivative of each irreducible piece of the constitutive tensor, see the
general decomposition formula (\ref{decomp}).
From this definition and the property about the noncommutativity of the Lie 
derivative, it follows that if $\xi_1$ and $\xi_2$ are two vector fields
describing symmetries of the medium, then the new vector
$\left[\xi_1,\xi_2\right]$ is also a symmetry of the medium.

This immediately raises questions analogous to those one encounters in the study
of isometries in GR, namely a) What is the maximum number $n_{\rm max}$
of symmetries that a constitutive tensor allows?
And then, assuming that $n_{\rm max}$ is
finite, b) What is the form of a `maximally symmetric' constitutive tensor?,
i.e., of a constitutive tensor allowing the maximum number of symmetries?

It can be speculated that the maximally symmetric constitutive tensor 
can correspond to $\chi=\chi_{\{\eta\}}$, i.e., to the (conformal) Minkowski 
vacuum, 
since we know that it is a highly symmetric case, see also the particular case 
below. This result would be
interesting since it would provide a further way to `derive' the vacuum spacetime
relation by postulating that it has to be maximally symmetric.
These questions will be investigated in the future.

\subsubsection{Riemannian case}
\label{secconsriem}

Here I show that in the special case for which $\chi=\chi_{\{g\}}$,
i.e. for the `vacuum' on a riemannian space \footnote{or a material medium in
Minkowski space with an `effective optical metric' $g$.}, 
the equation (\ref{liechi}) defining the
symmetries of the medium is equivalent to the {\rm conformal Killing equation} 
for the metric $g$, i.e. to
\begin{equation}\label{confkilling}
{\cal L}_\xi\left(\left|g\right|^{1/4}g^{ij}\right)=0,
\end{equation}
or, explicitly
\begin{equation}\label{confkill01}
g^{mj}\partial_m\xi^i +g^{im}\partial_m\xi^j+\xi^m\partial_m g^{ij}
= \frac{1}{4}g^{ij}\left[2\partial_m\xi^m+\xi^mg_{kl}\partial_m
g^{kl}\right] .
\end{equation}
One proves this as follows. From (\ref{chig2}) one finds
\begin{eqnarray}\label{liechig0}
{\cal L}_\xi\,\chi^{ijkl}_{\{g\}}&=&2\left(\left|g\right|^{1/4}g^{k[i}\right)
{\cal L}_\xi\left(\left|g\right|^{1/4}g^{j]l}\right) \nonumber \\
&&+2 \left(\left|g\right|^{1/4}g^{l[j}\right)
{\cal L}_\xi\left(\left|g\right|^{1/4}g^{i]k}\right).
\end{eqnarray}
Contracting (\ref{liechig0}) with $\left(\left|g\right|^{-1/4}g_{jl}\right)$,
one obtains
\begin{equation}\label{liechig}
\left(\left|g\right|^{-1/4}g_{jl}\right){\cal L}_\xi\,\chi^{ijkl}_{\{g\}}=
g^{ik}g_{jl}{\cal L}_\xi\left(\left|g\right|^{1/4}g^{jl}\right)+
2{\cal L}_\xi\left(\left|g\right|^{1/4}g^{ik}\right) ,
\end{equation}
so that the condition (\ref{liechi}) implies
\begin{equation}\label{liechig2}
g^{ik}g_{jl}{\cal L}_\xi\left(\left|g\right|^{1/4}g^{jl}\right)+
2{\cal L}_\xi\left(\left|g\right|^{1/4}g^{ik}\right)=0.
\end{equation}
Contracting this equation with $g_{ik}$ one finds 
$g_{jl}{\cal L}_\xi\left(\left|g\right|^{1/4}g^{jl}\right)=0$ which, when
substituted back into (\ref{liechig2}), results in the {\rm conformal Killing
equation} for the metric $g$, namely (\ref{confkilling}). This means that
the conformal Killing equation is a necessary condition for $\xi$ to be a symmetry
of the medium. On the other
hand, from (\ref{liechig0}) one sees that if $\xi$ is a conformal Killing vector
then (\ref{confkilling}) is automatically satisfied. Therefore, (\ref{confkilling})
is also a sufficient condition.

In other words, {\em in the riemannian
case, all symmetries of the medium are conformal symmetries of the metric}.

\subsubsection{Conservation of energy-momentum}
\label{seccons}

In physics conservation laws play a central role. Among other properties, they
allow the definition of conserved quantities in terms of which the description 
of the system and its evolution becomes simpler. In particular, energy and 
momentum are quantities associated to any field. In GR and similar theories 
they are the source of the gravitational field.

In field theory, the energy-momentum distribution and its flow are described by
a covector valued 3-form which, in the electromagnetic case we will denote as
$ \Sigma_\alpha $. Usually, one also needs an additional vector field $\xi$ 
for constructing a frame-independent 3-form 
$Q:=\Sigma_\alpha\,\xi^\alpha$ following the
general scheme `conserved quantity' $\sim$ `momentum' $\times$ `vector field'.
The vector field is usually related to some symmetry of the system. 
As an example 
of this general scheme recall the case of GR. There, a conserved quantity
along the trajectory of a free test particle (i.e., a particle moving along
a geodesic) can be defined provided the spacetime admits a symmetry. This
symmetry is described by a Killing vector field $\xi_{\rm K}$, i.e. an isometry
of the spacetime metric, so that ${\cal L}_{\xi_{\rm K}}\, g_{ij}=0$.
The corresponding conserved quantity is then given by the projection
(contraction) of the {\em momentum} \footnote{Momentum is always a 1-form, 
i.e., a covector, see, for instance, figure 2 in \cite{Post97}.} of the test 
particle of mass 
$m$, $p_i:=mg_{ij}{dx^j}/{d\tau}$, along 
the direction of the Killing vector i.e. $q:=p_i\,\xi^i_{\rm K}$.

Back to our original problem, the main property that $Q$ has to fulfill in order
to be interpreted as energy-momentum is that
it has to be related to the corresponding force law through a derivative of the
form $\xi^\alpha f_\alpha\sim dQ$. If, for instance, $\xi$ happens to be a 
4-velocity field,
$\xi^\alpha f_\alpha$ can be interpreted as the rate of energy transfer to the
particles via the Lorentz force. A second condition on $Q$ is that it
must be conserved under some conditions, i.e. $dQ=0$. One expects again the
condition for energy-momentum conservation to be related to the symmetries of the
system, at least when a Lagrangian is available,
as we know from the Noether theorem. In linear electrodynamics this latter can
happen when the irreducible piece ${}^{(2)}\chi$ of the constitutive tensor
vanishes.

Therefore, to find an adequate energy-momentum $\Sigma_\alpha$ for the
electromagnetic field, we try to express $\xi^\alpha f_\alpha$ as a total
derivative of some 3-form $dQ$, with $Q:=\Sigma_\alpha \xi^\alpha$.

First, we use the expression for the Lorentz force law (\ref{axiom2}),
replace the current $J$ from the inhomogeneous Maxwell equation (\ref{me01}) and
`partially integrate'. One finds
\begin{eqnarray}
\xi^\alpha f_\alpha
&=& \xi^\alpha \left( e_\alpha\rfloor F\right)\wedge J     \nonumber \\
&=& \left( \xi\rfloor F\right)\wedge J     \nonumber \\
&=& \left( \xi\rfloor F\right)\wedge dH    \nonumber \\
&=& -d\left[ \left( \xi\rfloor F\right) \wedge H \right]
+H\wedge d\left(\xi\rfloor F\right)     \nonumber \\
&=& -d\left[ \left( \xi\rfloor F\right) \wedge H \right]
+H\wedge \left({\cal L}_\xi F\right) \label{casi}.
\end{eqnarray}
In the last step we have used the definition of the Lie derivative of $F$ and the
homogeneous Maxwell equation, $dF=0$.

The choice of $Q$ that equation (\ref{casi}) suggests, namely
$Q=-\left( \xi\rfloor F\right) \wedge H $, is not what we are looking for
since it would lead to conservation of $Q$ only if the field strength is constant
in the direction $\xi$, i.e. if ${\cal L}_\xi F=0$, so that the last term in 
(\ref{casi}) vanish. But this is clearly a too restrictive condition.

To improve the situation we use now the fact that (\ref{casi}) can be rewritten as
\begin{equation}\label{casi2}
\xi^\alpha f_\alpha= d\left[ Z-\left( \xi\rfloor F\right) \wedge H \right]
+H\wedge \left({\cal L}_\xi F\right) -dZ   ,
\end{equation}
where $Z$ is an arbitrary 3-form.

To be sure that one is covering all the available possibilities, one should
consider the most general 3-form $Z$ that can be constructed using the
available objects. In our case, they are only $F$, $H$, and $\xi$ \footnote{In principle,
one could also use the potential $A$ as a further object. However,
it seems that no 3-form {\em linear in} $\xi$, as required by the left hand side of
(\ref{casi2}), can be constructed using this 1-form together with $F$ and $H$.}.
Under these conditions, $Z$ can only be
a linear combination of the terms $\left( \xi\rfloor F\right)\wedge H$ and
$\left( \xi\rfloor H\right)\wedge F$. Therefore, we make the ansatz
\begin{equation}
Z=a_1  \left( \xi\rfloor F\right)\wedge H +
a_2 \left( \xi\rfloor H\right)\wedge F ,
\end{equation}
with arbitrary constants $a_1$ and $a_2$. This implies
\begin{eqnarray}
dZ&=&a_1\, d\left( \xi\rfloor F\right)\wedge H
-a_1 \left( \xi\rfloor F\right) \wedge dH
+ a_2\, d\left( \xi\rfloor H\right)\wedge F 
-a_2 \left( \xi\rfloor H\right)\wedge dF     \nonumber \\
 &=&a_1\, d\left( \xi\rfloor F\right)\wedge H
-a_1 \left( \xi\rfloor F\right)\wedge J
+ a_2\, d\left( \xi\rfloor H\right)\wedge F    \nonumber \\
 &=&a_1 \left( {\cal L}_\xi F\right) \wedge H  -a_1\xi^\alpha f_\alpha
+ a_2 \left[ {\cal L}_\xi H-\left( \xi\rfloor dH\right)\right] \wedge F
\nonumber \\
 &=&a_1 \left( {\cal L}_\xi F\right) \wedge H  -a_1\xi^\alpha f_\alpha
+ a_2  \left( {\cal L}_\xi H\right)\wedge F
-a_2 \left( \xi\rfloor J\right) \wedge F \nonumber \\
 &=&a_1 \left( {\cal L}_\xi F\right) \wedge H
 + a_2 \left( {\cal L}_\xi H\right)  \wedge F
+\left(a_2-a_1\right) \xi^\alpha f_\alpha   .
\end{eqnarray}
Thus, we rewrite (\ref{casi}) as
\begin{eqnarray}
\xi^\alpha f_\alpha&=& d\left[ \left( a_1-1\right)
\left( \xi\rfloor F\right)\wedge H +
a_2 \left( \xi\rfloor H\right)\wedge F\right]
+\left( 1-a_1\right)H\wedge \left({\cal L}_\xi F\right)\nonumber \\
&&  -a_2 \left( {\cal L}_\xi H\right)  \wedge F
-\left(a_2-a_1\right) \xi^\alpha f_\alpha .
\end{eqnarray}
This expression, which is an identity as soon as $F$ and $H$ satisfy the Maxwell
equations and $f_\alpha$ is given by (\ref{axiom2}), leads us to define
\begin{equation}\label{q}
Q:= \frac{1}{1-a_1+a_2}\,\left[ \left( a_1-1\right)
\left( \xi\rfloor F\right)\wedge H +
a_2 \left( \xi\rfloor H\right)\wedge F\right] ,
\end{equation}
\begin{equation}\label{x}
X:= \frac{1}{1-a_1+a_2}\,
\left[\left(1- a_1\right)H\wedge \left( {\cal L}_\xi F \right)
-a_2 \left( {\cal L}_\xi H\right)  \wedge F  \right]  ,
\end{equation}
so that
\begin{equation} \label{fqx}
\xi^\alpha f_\alpha= dQ+X .
\end{equation}

The identity (\ref{fqx}) summarizes all the possibilities to write
$\xi^\alpha f_\alpha$ as a total derivative of the available fields plus some 
rest. The problem now is to find suitable values for $a_1$ and $a_2$ so that
$Q$ and $X$ satisfy our requirements. In particular, we need $X$ to vanish under 
some reasonable circumstances.

Up to here our results in this section are valid for any electromagnetic medium, 
since the spacetime/constitutive relation has not been used.
To make up our minds about a reasonable choice of the
constants it is useful to study the particular case of linear electrodynamics and
compute $X$ explicitly in terms of the field strength
and the irreducible pieces of the constitutive tensor.

We use ${\cal L}_\xi F=\frac{1}{2}\,{\cal L}_\xi F_{ij}\, dx^i\wedge dx^j$ and
similarly for $H$,
${\cal L}_\xi H=\frac{1}{2}\,{\cal L}_\xi H_{ij}\, dx^i\wedge dx^j$. 
Furthermore, we introduce the abbreviation
$a:={1}/[{4\left(1-a_1+a_2\right)}]$ and compute
\begin{eqnarray}
X&=&a
\left[\left(1- a_1\right)H_{ij}\left( {\cal L}_\xi F_{kl} \right)
-a_2 \left( {\cal L}_\xi H_{ij}\right)  F_{kl}  \right]
dx^i\wedge dx^j\wedge dx^k\wedge dx^l    \nonumber \\
&=&a
\left[\left(1- a_1\right)H_{ij} \left( {\cal L}_\xi F_{kl} \right)
-a_2 \left( {\cal L}_\xi H_{ij}\right)  F_{kl}  \right]
\epsilon^{ijkl} \hat\epsilon  \nonumber \\
&=&2a
\left[\left(1- a_1\right){\cal H}^{kl} \left( {\cal L}_\xi F_{kl} \right)
-a_2 \left( {\cal L}_\xi {\cal H}^{kl}\right)   F_{kl}  \right]
\hat\epsilon  \nonumber \\
&=&a
\left[\left(1- a_1\right)\chi^{klij}F_{ij}\left( {\cal L}_\xi F_{kl}
\right)
-a_2 \left( {\cal L}_\xi \chi^{klij}F_{ij}\right)   F_{kl}  \right]
\hat\epsilon  \nonumber \\
&=&a
\left[\left(1- a_1\right)\chi^{klij}F_{ij}\left( {\cal L}_\xi F_{kl}
\right)
-a_2 \left( {\cal L}_\xi \chi^{klij}\right) F_{ij}F_{kl} 
 -a_2 \chi^{klij}\left( {\cal L}_\xi F_{ij}\right) F_{kl} \right]
\hat\epsilon  \nonumber \\
&=&a
\left[\left[ \left(1- a_1\right)\chi^{klij}-a_2 \chi^{ijkl}\right]
F_{ij}\left( {\cal L}_\xi F_{kl}\right)
-a_2 \left( {\cal L}_\xi \chi^{klij}\right) F_{ij}F_{kl}\right]
\hat\epsilon  \nonumber \\
&=&a
\left[\left(1- a_1-a_2\right)
\left({}^{(1)}\chi^{klij}+{}^{(3)}\chi^{klij}\right)
F_{ij}\left( {\cal L}_\xi F_{kl}\right)   \right.  \nonumber \\
&& \quad  +\left(1- a_1+a_2\right){}^{(2)}\chi^{klij}
F_{ij}\left( {\cal L}_\xi F_{kl}\right) \nonumber \\
&& \quad \left.-a_2 {\cal L}_\xi\left({}^{(1)}\chi^{klij}+{}^{(3)}
\chi^{klij} \right) F_{ij}F_{kl}\right]\hat\epsilon .
\end{eqnarray}
Therefore, we find
\begin{eqnarray}
X&=& \frac{1}{4}\, F_{ij}\,\left[
\frac{\left(1- a_1-a_2\right)}{\left(1-a_1+a_2\right)}
\left({}^{(1)}\chi^{klij}+{}^{(3)}\chi^{klij}\right)
\left( {\cal L}_\xi F_{kl}\right)+ {}^{(2)}\chi^{klij}\right.
\left( {\cal L}_\xi F_{kl}\right) \nonumber \\
&&\qquad\quad \left. -\frac{a_2}{\left(1-a_1+a_2\right)}
 {\cal L}_\xi\left({}^{(1)}\chi^{klij}+{}^{(3)}
\chi^{klij} \right) F_{kl}\right]\,\hat\epsilon . \label{efdec}
\end{eqnarray}
From this result, one recognizes first that no choice of
$a_1$ and $a_2$ makes $X$ to vanish in general. Thus, no conservation
law is possible, unless additional conditions are fulfilled. This is, however,
what one expects. One also sees that the irreducible pieces enter 
differently in (\ref{efdec}). 
In particular, notice that the factor $\left(1- a_1+a_2\right)$
in front of the irreducible piece ${}^{(2)}\chi$ cancels completely.
This means that the term
${}^{(2)}\chi^{klij} F_{ij}\left( {\cal L}_\xi F_{kl}\right)$ {\em will always 
be present} in
the decomposition of $X$, no matter which values of $a_1$ and $a_2$
one chooses. In other words, the irreducible piece ${}^{(2)}\chi$ of the 
constitutive tensor will always prevent $Q$ from being a conserved quantity
(in charge-free regions), unless of course a much more
restrictive condition, like ${\cal L}_\xi F=0$, is satisfied by the field
configuration.
This result is consistent with our interpretation of ${}^{(2)}\chi$ as related
to intrinsic dissipative properties of the medium.

The `best' one can do is to choose 
\begin{equation}
1- a_1-a_2=0,
\end{equation}
so that the terms proportional to the symmetric irreducible pieces
${}^{(1)}\chi$ and ${}^{(3)}\chi$ in the first term on the right hand side of
(\ref{efdec}) vanish.
In this case, after substituting $a_2=1-a_1$ into (\ref{q}) and
(\ref{x}), one finds that $Q$ and $X$ do not depend on $a_1$, so that one
obtains a unique result, namely
\begin{equation}\label{Q}
Q= \frac{1}{2}\,\left[\left( \xi\rfloor H\right)\wedge F
-\left( \xi\rfloor F\right)\wedge H \right] ,
\end{equation}
\begin{equation}\label{X}
X= \frac{1}{2}\,\left[H\wedge \left( {\cal L}_\xi F \right)
-\left( {\cal L}_\xi H\right)  \wedge F  \right] .
\end{equation}

Furthermore, for linear electromagnetic media we obtain,  from (\ref{efdec}),
\begin{equation}\label{Xdecomp}
X=-\frac{1}{8}\,F_{ij}\left[2\,{}^{(2)}\chi^{ijkl}\left( {\cal L}_\xi F_{kl}
\right)+{\cal L}_\xi \left( {}^{(1)}\chi^{ijkl}+{}^{(3)}\chi^{ijkl}\right)
F_{kl}\right]\hat\epsilon .
\end{equation}

That the choice (\ref{Q}) and (\ref{X}) is reasonable can be seen as follows. 
Consider the case in which ${}^{(2)}\chi=0$ so that a Lagrangian is available
for the description of the system. In a charge-free region, (\ref{Xdecomp})
shows that $Q$ is conserved provided $\xi$ is a symmetry of the medium, i.e., 
${\cal L}_\xi \chi^{klij}=0$. Furthermore, our choice agrees with the results
valid in a Riemannian space.

Notice also that $Q$ does not depend on the
axion-like piece ${}^{(3)}\chi$  entering the linear constitutive law. This can
directly be seen using (\ref{Q}) and (\ref{3halpha}), since
\begin{eqnarray}
\left( \xi\rfloor {}^{(3)}H\right)\wedge F
-\left( \xi\rfloor F\right)\wedge {}^{(3)}H&=&
\left( \xi\rfloor \left(\alpha F\right)\right)\wedge F
-\left( \xi\rfloor F\right)\wedge \left(\alpha F\right) \nonumber \\
&=& \alpha \left( \xi\rfloor F\right)\wedge F
-\alpha \left( \xi\rfloor F\right)\wedge F \nonumber \\
&=& 0.  \label{3sigma0}
\end{eqnarray}
However, the axion-like piece {\em does} appear in (\ref{Xdecomp}) and therefore
can contribute to the non-conservation of $Q$ provided 
${\cal L}_\xi\,\alpha\neq0$.

\subsubsection{Axiom 4: Energy-momentum tensor}
\label{sectem}

The results in section \ref{seccons} show how to construct a
3-form which for a linear electromagnetic medium leads to conservation laws, 
under reasonable
conditions. We take these results and generalize them to our fourth axiom, by
means of which the {\em kinematic energy-momentum 3-form}
${}^{\rm k}\Sigma_\alpha$ of {\em any electromagnetic medium} is defined by
\begin{equation}\label{defsigma}
{}^{\rm k}\Sigma_\alpha:=
\frac{1}{2}\left[F\wedge\left(e_\alpha\rfloor H\right)-
H\wedge\left(e_\alpha\rfloor F\right)\right] .
\end{equation}
Then the 3-form $Q$ of section \ref{seccons} is given by
$Q={}^{\rm k}\Sigma_\alpha \xi^\alpha $.

The adjective `kinematic' is used here to emphasize that our definition of
${}^{\rm k}\Sigma_\alpha$ was not based on dynamical properties of the system, but
rather on kinematic arguments.

One can also now define the components ${}^{\rm k}{\cal T}_\alpha^{\ \beta}$ of
the kinetic energy-momentum tensor density by
\begin{equation}\label{sigmacomp}
\vartheta^\beta\wedge{}^{\rm k}\Sigma_\alpha=:
{}^{\rm k}{\cal T}_\alpha^{\ \beta}\,\hat\epsilon,
\end{equation}
where $\hat\epsilon:=\vartheta^{\hat 0}\wedge\vartheta^{\hat 1}
\wedge\vartheta^{\hat 2}\wedge\vartheta^{\hat 3}$. In particular, in a
coordinate basis, with ${\cal H}^{ij}:=\epsilon^{ijkl}H_{kl}/2$,
\begin{equation}
{}^{\rm k}{\cal T}_i^{\ j}=\frac{1}{4}\delta^j_i F_{kl}{\cal H}^{kl}-
F_{ik}{\cal H}^{jk}.
\end{equation}

For a general electromagnetic medium, we summarize our results as
\begin{equation} \label{fqx2}
\xi^\alpha f_\alpha= dQ+X   ,
\end{equation}
with
\begin{equation}\label{Q2}
Q:= {}^{\rm k}\Sigma_\alpha \xi^\alpha ,
\end{equation}
\begin{equation}\label{X2}
X= \frac{1}{2}\,\left[H\wedge \left( {\cal L}_\xi F \right)
-\left( {\cal L}_\xi H\right)  \wedge F  \right] .
\end{equation}

For a linear electromagnetic medium, we already showed that the axion-like 
piece ${}^{(3)}\chi$ does
not contributes to the kinematic energy-momentum tensor, i.e.,
\begin{equation}\label{axenergy}
  ^{\rm k}\Sigma_\alpha(\chi) = \,^{\rm k}\Sigma_\alpha
  \left(^{(1)}\chi +{}^{(2)}\chi +{}^{(3)}\chi \right)=\,^{\rm k}\Sigma_\alpha
  \left(^{(1)}\chi +{}^{(2)}\chi \right) .
\end{equation}

Additionally, we have shown, see (\ref{efdec}) and (\ref{fqx2}), that the
corresponding conservation law takes the form
\begin{equation}\label{diqi}
dQ=f_\alpha\xi^\alpha +\frac{1}{8}\,F_{ij}
\left[{\cal L}_\xi \left( {}^{(1)}\chi^{ijkl}+{}^{(3)}\chi^{ijkl}\right)
F_{kl}+2{}^{(2)}\chi^{ijkl}
\left( {\cal L}_\xi F_{kl}\right)\right]\hat\epsilon,
\end{equation}
for any vector field $\xi$.
In this equation the contributions of the three irreducible pieces
${}^{(1)}\chi$, ${}^{(2)}\chi$, and ${}^{(3)}\chi$ are isolated from each other.

From these results, a consistent interpretation of (\ref{diqi}) can be given as a
`balance equation' for the energy-momentum content of the electromagnetic field.

Recall first that even if $\xi$ is a symmetry of the medium, in which case the
second and third terms on the right hand side of (\ref{diqi}) vanish, there is
still a
contribution proportional to the irreducible piece ${}^{(2)}\chi$ of the
constitutive tensor. This is interpreted as an intrinsic dissipative property
of the medium generated by the irreducible piece ${}^{(2)}\chi$.

In the case that $\xi$ is a 4-velocity field the result above can be 
interpreted
as an `energy balance' equation for the change of the total electromagnetic
energy. The integration of (\ref{diqi}) over a
4-dimensional region of the form $\Omega_4=\Sigma\times [\sigma_0,\sigma]$
will produce \footnote{We assume that the fields vanish at 
$\partial\Sigma_\sigma$ (`spatial infinity').} 
at the left hand side the `change of the total energy of the electromagnetic
field' between the `times' $\sigma_0$ and $\sigma$, i.e.,
$\int_{\Sigma_\sigma}Q-\int_{\Sigma_{\sigma_0}}Q$.
The first term on the right hand side, $f_\alpha\xi^\alpha$, will lead
to the energy transfered from the electromagnetic field to the test particles
via the Lorentz force, i.e., $\int_{\Omega_4}f_\alpha\xi^\alpha$.
The term proportional to ${}^{(2)}\chi$ can be then interpreted as
the rate at which the energy of the electromagnetic field is dissipated,
$\frac{1}{4}\int_{\Omega_4}F_{ij}{}^{(2)}\chi^{ijkl}\left( {\cal L}_\xi F_{kl}
\right) \hat\epsilon$.
The other terms will be proportional to the `time' derivative of the
constitutive tensor, $\frac{1}{8}\int_{\Omega_4}F_{ij}
{\cal L}_\xi \left( {}^{(1)}\chi^{ijkl}+{}^{(3)}\chi^{ijkl}\right)
F_{kl}\,\hat\epsilon$. When the medium is `time' independent this term
vanishes. If the time derivative does not vanishes, it means that the
properties of the medium are changing in `time'.
Therefore, one could try to interpret this term as the energy per unit
time needed to change the material properties \footnote{Maybe with a change of
some suitably defined `internal energy' of the medium.}.

\section{Wave propagation}
\label{chapwave}

We turn our attention to the wave propagation properties in our general
framework of linear pre-metric electrodynamics.

Consider a region in spacetime without charges, i.e., $J=0$. Maxwell's
equations take then the form
\begin{equation}\label{vme}
dF=0\, \qquad dH=0,
\end{equation}
completed by the linear spacetime/constitutive relations (\ref{cl}).
These equations
will allow for solutions propagating in spacetime, the behavior of which is
determined by the electromagnetic properties of the spacetime/medium.

Since the constitutive tensor can have a very complicated spacetime
dependence, many specific features of the propagation of waves over finite
distances cannot be studied in general terms here. However, an important
{\em local} property of the propagation of waves,
namely the dispersion relation that the covectors tangent to a
wave front must satisfy, can be derived in general. This is done by deriving
the so-called {\em Fresnel equation} for the wave covectors.

\subsection{Propagation of singularities}
\label{secsingu}

In the theory of partial differential equations, the propagation of
waves is described by Hadamard discontinuities of solutions across a
characteristic (wave front) hypersurface $S$ \cite{Hadamard03}.
One can locally
define $S$ by the equation $\Phi(x^i) = const$, for some function $\Phi(x)$. 
The Hadamard discontinuity
of any function ${\cal F}(x)$ across the hypersurface $S$ is defined
as the difference $\left[{\cal F}\right]_S(x) := {\cal F}(x_+) -
{\cal F}(x_-)$, where $x_{\pm}:=\lim\limits_{\varepsilon\rightarrow 0}
\,(x\pm\varepsilon)$ are points on the opposite sides of $S\ni x$.
An electromagnetic wave is described as a solution of the vacuum Maxwell
equations
(\ref{vme}) for which the {\em derivatives} of $H$ and $F$ have
discontinuities across the wave front hypersurface $S$.
\begin{figure}[t]
\centering\epsfig{file=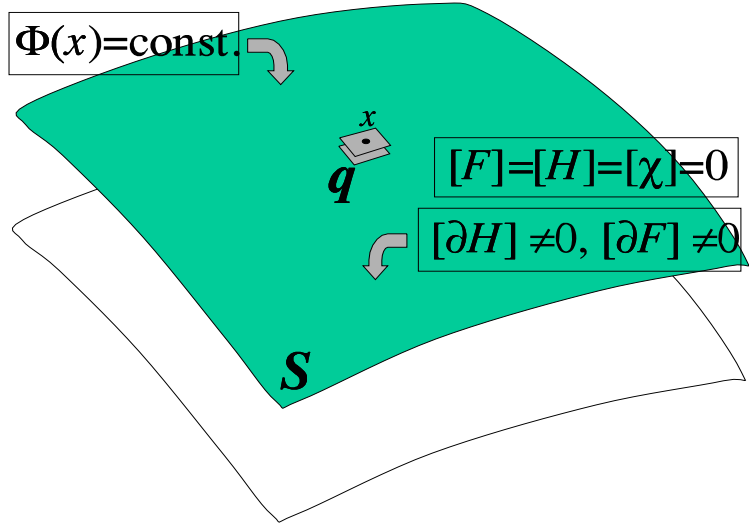, height=7cm, width=10cm}
\caption{Characteristic surface for propagation of electromagnetic
disturbances. The 1-form $q:=d\Phi$ corresponds geometrically to two small
parallel planes tangent to $S$ at each point $x$.}
\label{f2}
\end{figure}

Thus, in terms of field components, we have on the characteristic
hypersurface $S$,
\begin{eqnarray}
&& [F_{ij}]_S = 0,\qquad [\partial_i F_{jk}]_S = q_i\, f_{jk}, \label{had1}\\
&& [H_{ij}]_S = 0,\qquad [\partial_i H_{jk}]_S = q_i\,  h_{jk},\label{had2}
\end{eqnarray}
where $f_{ij}$ and $h_{ij}$ are the components of the 2-forms $f$ and $h$
describing the corresponding jumps
of the derivatives of field strength and excitation across $S$, respectively,
and the covector tangent to the wave front is given by
\begin{equation}
q:=d\Phi=q_i\,dx^i .
\end{equation}
Notice that (\ref{had1}) and (\ref{had2}), taken together, are covariant
conditions. In particular, although
(\ref{had1}b) and (\ref{had2}b) may not look covariant at first sight, they
are, provided (\ref{had1}a) and (\ref{had2}a) are also satisfied. This can be
seen by considering the transformation law of 
$\partial_i F_{jk}$ under coordinate transformations. Consider the
coordinate transformation $x^i\rightarrow x^{i^\prime}$. Then we have
\begin{equation}
\partial_{i^\prime}F_{j^\prime k^\prime}=\frac{\partial}{\partial x^{i^\prime}}
\left(\frac{\partial x^j}{\partial x^{j^\prime}}
\frac{\partial x^k}{\partial x^{k^\prime}}\right)\, F_{jk}+
\frac{\partial x^i}{\partial x^{i^\prime}}
\frac{\partial x^j}{\partial x^{j^\prime}}
\frac{\partial x^k}{\partial x^{k^\prime}}\, \partial_i F_{jk} .
\end{equation}
Assuming that the coordinate transformation is smooth across $S$, i.e.
$[\frac{\partial x^i}{\partial x^{i^\prime}}]_S=0$,
$[\frac{\partial^2 x^i}{\partial x^{i^\prime}\partial x^{j^\prime}}]_S=0$, we
find that the jump of $\partial_i F_{jk}$ across $S$ is given by
\begin{equation}
\left[\partial_{i^\prime}F_{j^\prime k^\prime}\right]_S=
\frac{\partial}{\partial x^{i^\prime}}
\left(\frac{\partial x^j}{\partial x^{j^\prime}}
\frac{\partial x^k}{\partial x^{k^\prime}}\right)\, \left[F_{jk}\right]_S+
\frac{\partial x^i}{\partial x^{i^\prime}}
\frac{\partial x^j}{\partial x^{j^\prime}}
\frac{\partial x^k}{\partial x^{k^\prime}}\,
\left[\partial_i F_{jk}\right]_S ,
\end{equation}
which reduces to the transformation law of a tensor field, provided
(\ref{had1}a) is satisfied, i.e.
\begin{equation}
\left[\partial_{i^\prime}F_{j^\prime k^\prime}\right]_S=
\frac{\partial x^i}{\partial x^{i^\prime}}
\frac{\partial x^j}{\partial x^{j^\prime}}
\frac{\partial x^k}{\partial x^{k^\prime}}\,
\left[\partial_i F_{jk}\right]_S .
\end{equation}
We use (\ref{had1}), (\ref{had2}) and the Maxwell equations (\ref{vme}) to 
find the compatibility conditions that the field discontinuities most 
satisfy, and find 
\begin{equation}\label{ce01}
q\wedge h=0, \qquad q\wedge f=0 \, .
\end{equation}
 This ensures that the Maxwell equations (\ref{vme})
are satisfied. The Hadamard method used here is equivalent to the usual
{\em geometric optics} limit made by expanding a solution for the
electromagnetic potential of the form $A=a e^{i\Phi}$. The covector
$q:=d\Phi$ corresponds thus to the {\em wave covector}.

Assuming now that the constitutive tensor is regular on $S$
(in a geometric optics limit this means that we assume the scale of
variations of the constitutive tensor to be much bigger that the scale of
variation of the wave field) we obtain from (\ref{cl}) the corresponding
relation between $h$ and $f$, namely
\begin{equation}\label{cld}
h={}^\# f.
\end{equation}
Substituting (\ref{cld}) into (\ref{ce01}a) and using (\ref{ce01}b),
we see that the equations (\ref{ce01}) reduce to
\begin{equation}\label{ce02}
q\wedge {}^\#f=0, \qquad q\wedge f=0 .
\end{equation}

Now, the general solution of (\ref{ce02}b) is of the form
\begin{equation}\label{fqa}
f=q\wedge a\, ,
\end{equation}
where the 1-form $a$ is defined up to `gauge transformations'
\begin{equation}\label{agt}
a \rightarrow a+\varphi q ,
\end{equation}
for any scalar function $\varphi$. This is an algebraic consequence of
the gauge freedom in the definition of the electromagnetic potential.
Then the compatibility condition (\ref{ce02}a) reduces to
\begin{equation}\label{e2}
q\wedge {}^\#( q\wedge a)=0 .
\end{equation}

Furthermore, in order to allow for solutions, the fields $f$ and $h$ must satisfy 
some `integrability' conditions. Since the form of the general solutions of
(\ref{ce01}) is  $f=q\wedge a$ and $h=q\wedge b$, we find 
\begin{equation} \label{integrability00}
f\wedge f=0, \qquad h\wedge h=0, \qquad f\wedge h=0 ,
\end{equation}
or, in components,
\begin{equation}\label{integrability}
{\epsilon}^{\,ijkl}\, f_{ij}\, f_{kl}=0, \qquad
{\epsilon}^{\,ijkl}\, h_{ij}\, h_{kl}=0, \qquad
{\epsilon}^{\,ijkl}\, f_{ij}\, h_{kl}=0 .
\end{equation}

\subsection{Fresnel equation}
\label{secfresnel}

As has been noticed before, see \cite{OFR00}, not all the equations in
(\ref{e2}) are independent. This fact makes the derivation of the Fresnel
equation more involved.

Let us isolate the trivial parts of (\ref{e2}). We use a 
covector basis $\vartheta^\alpha$ in order to write the covector $a$
in terms of its frame components $a_\alpha$ as $a=a_\alpha\vartheta^\alpha$
and {\em choose the covector $\vartheta^{\hat 0}$} as the
covector $q$, i.e. $\vartheta^{\hat 0}=q$.
This can, of course, always be done at each point of spacetime, since we assume 
$q\neq 0$.
With this choice, a gauge transformation (\ref{agt}) can then be completely
accounted for by a transformation
$a_{\hat 0}\rightarrow a_{\hat 0} + \varphi$. This means that the other frame
components $a_A$, $A,B,\ldots=\hat 1, \hat 2, \hat 3$,
are gauge invariant. Since the term coming from the basis covector 
$\vartheta^{\hat 0}$ identically vanishes, i.e. 
$q\wedge {}^\#( q\wedge \vartheta^{\hat 0})\equiv 0$, (\ref{e2}) can be written as
\begin{equation}\label{e3}
q\wedge {}^\#( q\wedge \vartheta^{B})\, a_{B}=0 .
\end{equation}
Furthermore, only 3 of the 4 components in the 3-form (\ref{e3}) are
non-trivial. Multiplying (\ref{e3}) by $\vartheta^{\hat 0}$, one finds
\begin{equation}\label{e4}
\vartheta^{\hat 0}\wedge q\wedge {}^\# (q\wedge \vartheta^{B})\,
a_{B}\equiv 0 \, .
\end{equation}
The remaining 3 components are
\begin{equation}\label{e5}
\vartheta^{A}\wedge q\wedge {}^\# (q\wedge \vartheta^{B})\,
a_{B}= 0 \, .
\end{equation}
This equation can be written as
\begin{equation}\label{e6}
\hat\epsilon\, W^{AB}\, a_{B}= 0 \, ,
\end{equation}
with a volume element $\hat\epsilon$ that we choose to be
$\hat\epsilon:=\vartheta^{\hat 0}\wedge\vartheta^{\hat 1}\wedge\vartheta^{\hat 2}
\wedge\vartheta^{\hat 3}$ (it doesn't matter very much which 4-form one uses,
since the equation is homogeneous) and a $3\times 3$ matrix
$W^{AB}$.
Using (\ref{defdual}) and (\ref{e5}) we compute
\begin{eqnarray}
\vartheta^{A}\wedge q\wedge {}^\#( q\wedge \vartheta^{B})
&=&\frac{1}{4}\vartheta^{A}\wedge \vartheta^{\hat 0} \wedge
\left(\hat\epsilon_{\alpha\beta\gamma\delta}
\chi^{\gamma\delta\epsilon\theta}\,
\delta^{\hat 0}_{\epsilon}\delta^{B}_{\theta}\,
\vartheta^\alpha\wedge \vartheta^\beta\right) \nonumber \\
&=&\frac{1}{4}\, \epsilon^{A\hat0\alpha\beta}
\hat\epsilon_{\alpha\beta\gamma\delta}
\chi^{\gamma\delta\hat 0B}\,
\vartheta^{\hat 0}\wedge \vartheta^{\hat 1}\wedge
\vartheta^{\hat 2}\wedge \vartheta^{\hat 3} \nonumber \\
&=&\delta^{A}_{[\gamma}\delta^{\hat 0}_{\delta ]}
\chi^{\gamma\delta\hat 0B}\, \hat\epsilon \nonumber \\
&=&\chi^{A \hat 0 \hat 0B}\, \hat\epsilon \nonumber \\
&=&-\chi^{\hat 0 A \hat 0B}\, \hat\epsilon\, .
\end{eqnarray}
Thus, since the negative sign is unimportant, we can define 
\begin{equation}
W^{AB}:=\chi^{\hat 0 A \hat 0B} .
\end{equation}
The corresponding Fresnel equation is then obtained from the vanishing of
the determinant of $W$, as the necessary and sufficient condition for
existence of non-vanishing
solutions $a_{B}$. Now, since $W$ is a $3\times 3$ matrix, its
determinant is given by
\begin{eqnarray}\label{det01}
{\cal W}&:=&\det(W) \nonumber \\
&=&\frac{1}{3!}\hat\epsilon_{ABC}\,
\hat\epsilon_{DEF}\,W^{AD}W^{BE}W^{CF}
\nonumber \\
&=&\frac{1}{3!}\hat\epsilon_{ABC}\,
\hat\epsilon_{DEF}\,
\chi^{\hat 0 A \hat 0D}
\chi^{\hat 0 B \hat 0E}
\chi^{\hat 0 C \hat 0F} .
\end{eqnarray}
We apply now the following procedure to rewrite  $\cal W$ as a fully
4-dimensional covariant expression. We `complete' the 3-dimensional
Levi-Civita symbols $\hat\epsilon_{ABC}$ to obtain the
4-dimensional one, by first using $\hat\epsilon_{ABC}\equiv
\hat\epsilon_{\hat 0ABC}$ and then taking one of the
$\hat 0$-components of the constitutive tensors as fourth summation
index. This leads us to consider the following identity:
\begin{equation}\label{id01}
\hat\epsilon_{\hat 0ABC}
\chi^{\hat 0 A \hat 0D}
\chi^{\hat 0 B \hat 0E}
\chi^{\hat 0 C \hat 0F}
\equiv \frac{1}{2}\, \hat\epsilon_{\alpha\beta\gamma\delta}
\chi^{\alpha\beta \hat 0D}
\chi^{\hat 0 \gamma \hat 0E}
\chi^{\hat 0 \delta \hat 0F} .
\end{equation}
This identity holds because on the right hand side of (\ref{id01}), due to the
properties of the Levi-Civita symbol, one of the indices $\alpha,\beta,\gamma$
or $\delta$ must be zero, but on the other hand only $\alpha$ and $\beta$
will contribute, due to the (anti)symmetry properties of the
constitutive tensor, see (\ref{symm01}), which would otherwise make
one of the two last $\chi$-factors vanish.
Finally, the two remaining
contributions are equal, canceling the factor $1/2$ and proving the identity.
This allows us to rewrite (\ref{det01}) as
\begin{equation}\label{det02}
{\cal W}=\frac{1}{3!}\frac{1}{2}\hat\epsilon_{\alpha\beta\gamma\delta}\,
\hat\epsilon_{DEF}\,
\chi^{\alpha\beta \hat 0D}
\chi^{\hat 0 \gamma \hat 0E}
\chi^{\hat 0 \delta \hat 0F} .
\end{equation}
We repeat the same procedure to complete the remaining 3-dimensional
Levi-Civita symbol. Now, we can use the following identity: 
\begin{equation}\label{id02}
\hat\epsilon_{\hat 0DEF}
\chi^{\alpha\beta \hat 0D}
\chi^{\hat 0 \gamma \hat 0E}
\chi^{\hat 0 \delta \hat 0F}
\equiv \frac{1}{2}\, \hat\epsilon_{\lambda\rho\sigma\tau}
\chi^{\alpha\beta \hat 0\rho}
\chi^{\hat 0 \gamma \hat 0\sigma}
\chi^{\hat 0 \delta \lambda\tau} .
\end{equation}
Using (\ref{id02}) in (\ref{det02}), we finally obtain
\begin{equation}\label{det03}
{\cal W}=\frac{1}{4!}\hat\epsilon_{\alpha\beta\gamma\delta}\,
\hat\epsilon_{\lambda\rho\sigma\tau}
\chi^{\alpha\beta \hat 0\rho}
\chi^{\hat 0 \gamma \hat 0\sigma}
\chi^{\hat 0 \delta \lambda\tau}  .
\end{equation}
Since $\vartheta^{\hat 0}=q=q_i\,dx^i$, the above result can
be written in coordinate components as
\begin{equation}\label{wgen}
{\cal W}=\frac{\theta^2}{4!}\,\,\hat{\epsilon}_{mnpq}\,
 \hat{\epsilon}_{rstu}\, \chi^{\,mnri}\,
 \chi^{\,jpsk}\, \chi^{\,lqtu }\, q_iq_jq_kq_l ,
\end{equation}
with $\theta:=\det(e^\alpha_i)$.
We define the fourth order tensor density of weight $+1$, the `Fresnel tensor'
${\cal G}^{ijkl}$, as
\begin{equation}\label{G4}
 {\cal G}^{ijkl}:=\frac{1}{4!}\,\hat{\epsilon}_{mnpq}\,
 \hat{\epsilon}_{rstu}\,\chi^{\,mnr(i}\, \chi^{\,j|ps|k}\, \chi^{\,l)qtu }, 
\qquad {\cal G}^{ijkl}={\cal G}^{(ijkl)},
\end{equation}
which has 35 independent components. Then the Fresnel equation can finally be 
written as
\begin{equation} \label{Fresnel}
{\cal G}^{ijkl}q_i q_j q_k q_l = 0 .
\end{equation}

At each point, the Fresnel equation (\ref{Fresnel}) defines in the space of
wave covectors the {\em wave (co-)vector surface}, see for instance 
\cite{LL60}. I would like to emphasize the generality of the above result. 
The Fresnel equation (\ref{Fresnel}) is valid for {\em any linear 
electromagnetic medium}. This means for media that in general are inhomogeneous, 
anisotropic, and dissipative. As we have already discussed, this result can even be applied 
to study the properties of propagation of perturbation of any local medium, 
by using an effective constitutive tensor, see \cite{OR02}. 
Furthermore, the above  result is, as the whole formalism, {\em generally covariant}.

The general Fresnel equation (\ref{Fresnel}) is in general a {\em quartic}
equation in $q_i$ despite the fact that it was derived from a
determinant of a $3\times 3$ matrix quadratic in the wave covectors. This is 
because the remaining quartic term is multiplied by the trivial factor 
$\theta^2$, see (\ref{wgen}). 
This corrects Denisov \& Denisov \cite{DD99} who claim that a
particular case of the general linear constitutive law may yield a
sixth order Fresnel equation.

\subsubsection{Coordinate $3+1$ decomposition of the Fresnel equation}

We rewrite now our general result (\ref{wgen}) by performing a $3+1$ coordinate
decomposition, and obtain
\begin{equation}
{\cal W} = \theta^2\left( q_0^4 M + q_0^3q_a\,M^a + q_0^2q_a q_b\,M^{ab} +
q_0q_a q_b q_c\,M^{abc}  + q_a q_b q_c q_d\,M^{abcd}\right),\label{fresnel}
\end{equation}
with
\begin{equation}\label{compare01}
M:={\cal G}^{0000},\quad M^a:=4{\cal G}^{000a},
\quad M^{ab}:=6{\cal G}^{00ab},\quad
\end{equation}
\begin{equation}\label{compare02}
M^{abc}:=4{\cal G}^{0abc}, \quad  M^{abcd}:={\cal G}^{abcd} ,
\end{equation}
or explicitly in terms of the $3\times 3$ matrices defined in
(\ref{cd01}) and (\ref{cd02}):
\begin{eqnarray}  
M&=&\det{\cal A} \,, \\  
M^a&=& -\hat{\epsilon}_{bcd}\left( {\cal A}^{ba}\,{\cal A}^{ce}\,  
{\cal C}^d_{\ e} + {\cal A}^{ab}\,{\cal A}^{ec}\,{\cal D}_e^{\ d}  
\right)\,,\label{ma1}\\  
M^{ab}&=& \frac{1}{2}\,{\cal A}^{(ab)}\left[({\cal C}^d{}_d)^2 +   
({\cal D}_c{}^c)^2 - ({\cal C}^c{}_d + {\cal D}_d{}^c)({\cal C}^d{}_c +   
{\cal D}_c{}^d)\right]\nonumber\\   
&&+({\cal C}^d{}_c + {\cal D}_c{}^d)({\cal A}^{c(a}{\cal C}^{b)}{}_d  +   
{\cal D}_d{}^{(a}{\cal A}^{b)c}) - {\cal C}^d{}_d  
{\cal A}^{c(a}{\cal C}^{b)}{}_c \nonumber\\   
&& - {\cal D}_c{}^{(a}{\cal A}^{b)c}{\cal D}_d{}^d    
- {\cal A}^{dc}{\cal C}^{(a}{}_c {\cal D}_d{}^{b)} 
 +  \left({\cal A}^{(ab)}{\cal A}^{dc}-   
{\cal A}^{d(a}{\cal A}^{b)c}\right){\cal B}_{dc}\,,\label{ma2}\\   
M^{abc} &=& \epsilon^{de(c|}\left[{\cal B}_{df}(  
{\cal A}^{ab)}\,{\cal D}_e^{\ f} - {\cal D}_e^{\ a}{\cal A}^{b)f}\,)  
+ {\cal B}_{fd}({\cal A}^{ab)}\,{\cal C}_{\ e}^f \right. \nonumber \\
&& \left. - {\cal A}^{f|a}{\cal C}_{\ e}^{b)}) 
+{\cal C}^{a}_{\ f}\,{\cal D}_e^{\ b)}\,{\cal D}_d^{\ f}  
+ {\cal D}_f^{\ a}\,{\cal C}^{b)}_{\ e}\,{\cal C}^{f}_{\ d} \right]   
\, ,\label{ma3}\\  
M^{abcd} &=& \epsilon^{ef(c}\epsilon^{|gh|d}\,{\cal B}_{hf}  
\left[\frac{1}{2} \,{\cal A}^{ab)}\,{\cal B}_{ge}  
- {\cal C}^{a}_{\ e}\,{\cal D}_g^{\ b)}\right] \,.\label{ma4}  
\end{eqnarray}  
Each 3-dimensional tensor $M$ is totally symmetric, i.e. $M^{ab}=M^{(ab)}$, 
$M^{abc}=M^{(abc)}$, $M^{abcd}=M^{(abcd)}$. 
These results have been verified by using the Maple computer algebra
system, together with its tensor package GrTensor \footnote{See
http://grtensor.org.}.

\subsubsection{Properties of the Fresnel tensor density}
\label{secpropfresnel}

First, one notices that the Fresnel equation is
independent of the axion-like piece ${}^{(3)}\chi$ of the constitutive tensor, 
\begin{equation}\label{propg1}
 {\cal G}^{ijkl}(\chi)= {\cal G}^{ijkl}({}^{(1)}\chi+{}^{(2)}\chi) ,
\end{equation}
which, in particular, implies 
\begin{equation} \label{propg2}
  {\cal G}^{ijkl}(^{(3)}\chi)=0 .
\end{equation}
Furthermore, due to the antisymmetry property of  ${}^{(2)}\chi$, one verifies
that also
\begin{equation}\label{propg3}
  {\cal G}^{ijkl}({}^{(2)}\chi)= 0 .
\end{equation}
Actually, properties (\ref{propg2}) and (\ref{propg3}) generalize to 
\begin{equation} \label{propg4}
  {\cal G}^{ijkl}({}^{(2)}\chi+{}^{(3)}\chi)=0 ,
\end{equation}
which can be verified, for instance, by using computer algebra. Notice that 
this identity is {\em not} trivial since $\cal G$ depends cubicly on the 
constitutive tensor $\chi$. The identity (\ref{propg4}) shows
that the symmetric piece $^{(1)}\chi$ is indispensable for obtaining  
well-behaved wave properties: If $^{(1)}\chi=0$, the Fresnel equation is 
trivially satisfied and thus no light cone structure could be induced.

Furthermore, since
\begin{equation} \label{propg5}
 {\cal G}^{ijkl}({}^{(1)}\chi+{}^{(2)}\chi)\neq {\cal G}^{ijkl}({}^{(1)}\chi), 
\end{equation}
the `skewon' field does influences the Fresnel equation, and therefore,
eventually, the light cone structure. An example of this general result can be
found again in the asymmetric constitutive tensor studied by Nieves and Pal, see
section \ref{secnievespal} and references \cite{NP89,NP94}.

Actually, after some algebra, one finds
\begin{eqnarray} \label{propg6}
 {\cal G}^{ijkl}({}^{(1)}\chi+{}^{(2)}\chi)&=&{\cal G}^{ijkl}({}^{(1)}\chi)
 + \frac{2}{4!}\,\hat{\epsilon}_{mnpq}\,
 \hat{\epsilon}_{rstu}\,{}^{(1)}\chi^{\,mnr(i}\, {}^{(2)}\chi^{\,j|ps|k}\,
 {}^{(2)}\chi^{\,l)qtu }
 \nonumber \\
 &&+ \frac{1}{4!}\,\hat{\epsilon}_{mnpq}\,
 \hat{\epsilon}_{rstu}\,{}^{(2)}\chi^{\,mnr(i}\, {}^{(1)}\chi^{\,j|ps|k}\,
 {}^{(2)}\chi^{\,l)qtu } ,
\end{eqnarray}
or, in a (more of less) obvious notation (see the definition (\ref{G4})), 
\begin{eqnarray} \label{propg7}
 {\cal G}^{ijkl}(\chi,\chi,\chi)&=&
 {\cal G}^{ijkl}({}^{(1)}\chi,{}^{(1)}\chi,{}^{(1)}\chi)
 + 2\, {\cal G}^{ijkl}({}^{(1)}\chi,{}^{(2)}\chi,{}^{(2)}\chi)   \nonumber \\
 && + {\cal G}^{ijkl}({}^{(2)}\chi,{}^{(1)}\chi,{}^{(2)}\chi) .
\end{eqnarray}
The other terms vanish due to the symmetry properties of each irreducible 
piece.

Take now (\ref{propg6}) and substitute the parametrization of ${}^{(2)}\chi$ in
terms of $S_i^{\ j}$, see (\ref{2chis}).
After some lengthy but straightforward algebra, one finds that the two last
contributions to the right hand side of (\ref{propg6}) are actually equal, namely
\begin{equation}
 {\cal G}^{ijkl}({}^{(1)}\chi,{}^{(2)}\chi,{}^{(2)}\chi)=
 {\cal G}^{ijkl}({}^{(2)}\chi,{}^{(1)}\chi,{}^{(2)}\chi) 
 = \frac{1}{3}\,{}^{(1)} \chi^{\,m(i|n|j}S_m^{\ k} S_n^{\ l)}.
\end{equation}
Therefore, the final result reads 
\begin{equation} \label{propg8}
 {\cal G}^{ijkl}({}^{(1)}\chi+{}^{(2)}\chi)={\cal G}^{ijkl}({}^{(1)}\chi)
 + {}^{(1)}\chi^{\,m(i|n|j}S_m^{\ k} S_n^{\ l)}, 
\end{equation}
a very simple expression, indeed. This equation summarizes how the skewon 
piece `perturbates' the Fresnel equation that one would obtain only from the 
principal piece ${}^{(1)}\chi$. The skewon piece adds a second term which is 
linear in  ${}^{(1)}\chi$ and quadratic in ${}^{(2)}\chi$.

For the particular case of Nieves and Pal, see (\ref{2chis}) and (\ref{omegav}),
one finds
\begin{equation}
{\cal G}^{ijkl}q_iq_jq_kq_l
=-\sqrt{\frac{\varepsilon_0}{\mu_0}}
\left[\frac{\varepsilon_0}{\mu_0}
(q\cdot q)^2 -(q\cdot q)(v\cdot v)(v\cdot q)^2+(v\cdot q)^4\right] .
\end{equation}
Compare this result with equation (5.7) in \cite{NP94}.
Use $q_i\stackrel{*}{=}(\omega,-\vec{k})$,
$v^i\stackrel{*}{=}(v,0,0,0)$,
$g_{ij}=\eta_{ij}\stackrel{*}{=}(c^2,-1,-1,-1)$,
$c^2={1}/{\varepsilon_0\mu_0}$ and obtain
\begin{equation}
{\cal G}^{ijkl}q_iq_jq_kq_l\stackrel{*}{=}-\varepsilon_0^3
\left[\left(\omega^2-k^2c^2\right)^2
+\frac{c^4}{\varepsilon_0^2}k^2\omega^2v^4\right] .
\end{equation}
This results agrees\footnote{Nieves and Pal use a different system of units, 
compare, for instance, their 
vacuum equations (2.1) and (2.2) with our Maxwell equations (\ref{imev}) and
(\ref{hmev}). Their equations can be obtained from our ones by substituting
${\cal D}={\cal D}_{\rm N}$, ${\cal H}={\cal H}_{\rm N}/c$,
$E=E_{\rm N}/\varepsilon_0$,
$B=B_{\rm N}/(c\varepsilon_0)$, $\rho=\rho_{\rm N}$, $j=j_{\rm N}$, and
$t=t_{\rm N}$, where the subindex {\rm N} refers to  the fields in the notation
of Nieves and Pal.} with that in \cite{NP94} (in their $\varepsilon=\mu=1$ 
case) for $\zeta={cv^2}/{\varepsilon_0}$.

Notice that in general, for $\zeta\neq 0$, the Fresnel equation will have
complex solutions. This is again a manifestation of the dispersive properties
described by the antisymmetric piece ${}^{(2)}\chi$ of the constitutive tensor.

\subsection{Light rays}

 In a riemannian space, light rays are defined as the integral lines of the
vector field defined by $V_{\rm riem}:=g^{ij}q_j\partial_i$, which is then a
null vector.
It satisfies $V_{\rm riem}\rfloor q=0$, so that the vector $V_{\rm riem}$ is
orthogonal to the wave front $S$. This definition makes direct use of a
metric, it is a nontrivial question whether it is possible to define light
rays in our more general, pre-metric framework.

In analogy to the riemannian case, one can try to define light rays as the
integral lines of a vector field $V$ which satisfies
\begin{equation}\label{vinq}
V\rfloor q=0.
\end{equation}
Equation (\ref{vinq}) represents one condition for the 4 independent components
of $V$. Therefore, this information is not enough to define light rays
uniquely. We know, however, that in the riemannian case the light rays defined
by $V_{\rm riem}$ are also orthogonal to the polarization vector $a$ (such
that $f=q\wedge a$), i.e. $V_{\rm riem}\rfloor a=0$. The same is true for $b$
with $h=q\wedge b$, i.e. $V_{\rm riem}\rfloor b=0$. These known
properties suggest to define the vector $V$ for a given
solution of the Maxwell equations as those satisfying
\begin{equation}\label{vinqab}
V\rfloor q=0, \qquad V\rfloor a=0, \qquad  V\rfloor b=0.
\end{equation}
The three conditions (\ref{vinqab}) determine then the vector $V$ up to a
scalar factor. 
This ambiguity in the definition of $V$ is however
irrelevant, since the integral lines of $V$ are independent of that scalar
factor \footnote{If one computes the integral line by ${dx^i}/{dp}=V^i$,
then the freedom of choosing the scalar factor corresponds to the freedom
of choosing the parameter $p$ along the curve. In other words, an arbitrary
scalar factor can always be absorbed by a reparametrization
$p^\prime=p^\prime(p)$.}.

One can construct immediately a solution of (\ref{vinqab}), since they imply
that the coordinate components $V^i$ of $V$ have to be proportional to 
$\epsilon^{ijkl}q_ja_kb_l$. We therefore define
\begin{equation}\label{defv}
{\cal V}^i:=\epsilon^{ijkl}q_ja_kb_l .
\end{equation}
This quantity is a vector density of weight $+1$. Again, the fact that 
(\ref{defv}) does not define a true vector field represents no
problem for the definition of light rays as its integral lines.
[Alternatively, one could try to `normalize' $\cal V$ in order to construct
a true vector, for example by dividing $\cal V$ by the density
$|a|^4:={\cal G}^{ijkl}a_ia_ja_ka_l$, with $|a|$ being a kind of `norm' of
$a$. Obviously, this procedure only makes sense if $|a|\neq 0$].
One can write (\ref{defv}) in a coordinate-free way as
\begin{equation}\label{defvv}
{\cal V}={}^\diamond\left( q\wedge a\wedge b \right),
\end{equation}
where ${}^\diamond$ is the (metric-free) operator mapping 3-form into vector 
densities of weight $+1$ according to ${}^\diamond\left(  \vartheta^\alpha\wedge\vartheta^\beta\wedge\vartheta^\gamma\right)
=\epsilon^{\delta\alpha\beta\gamma}\, e_\delta$, so that 
${\cal V}={\cal V}^\alpha \, e_\alpha$.

Now, using the identities $q\wedge b=h={}^\# f={}^\#(q\wedge a)$ we can write
(\ref{defvv}) in terms of only $q$, $a$ and the operator ${}^\#$ as
\begin{equation}
{\cal V}={}^\diamond\left[ a\wedge{}^\#\left( a\wedge q\right) \right],
\end{equation}
or, in terms of the components of the constitutive tensor, 
\begin{equation}\label{vchiaq}
{\cal V}^\alpha=\chi^{\alpha\beta\gamma\delta}a_\beta a_\gamma q_\delta .
\end{equation}

As a consequence of our definitions, $\cal V$ depends explicitly
not only on the wave 1-form $q$ but also on the polarization 1-form $a$. This
property agrees with the result that, for a general
constitutive tensor, different polarization states will propagate in different
directions. Birefringence is again a particular example of the above mentioned 
fact. One can also verify that the axion-like piece of the constitutive 
tensor drops out from ${\cal V}$, so it depends only on ${}^{(1)}\chi$ and 
${}^{(2)}\chi$.

To gain more physical insight on our definition of light rays, we evaluate
the kinematic energy-momentum 3-form ${}^{\rm k}\Sigma_\alpha$ for our
particular solutions. From its definition (\ref{defsigma}) and using
$F\rightarrow f=q\wedge a$ and $H\rightarrow h=q\wedge b$, we find
\begin{eqnarray}
{}^{\rm k}\Sigma_\alpha&=&\frac{1}{2}\left[q\wedge a\wedge\left(q_\alpha\, b-
q\, b_\alpha \right)-q\wedge b\wedge\left(q_\alpha\, a- q\, a_\alpha
\right)\right] \nonumber \\
&=&\frac{1}{2}\left[q_\alpha\, q\wedge a\wedge b- q_\alpha\, q\wedge b\wedge a
\right] \nonumber \\
&=&q_\alpha\, q\wedge a\wedge b .
\end{eqnarray}
Using (\ref{defvv}), we thus arrive at  
\begin{equation}
{}^\diamond\left( {}^{\rm k}\Sigma_\alpha\right) =q_\alpha\,{\cal V}.
\end{equation} 
In terms of the components ${}^{\rm k}{\cal T}_\alpha^{\ \beta}$ of the kinetic energy-momentum density, see (\ref{sigmacomp}), one finds
\begin{equation}
{}^{\rm k}{\cal T}_\alpha^{\ \beta}=q_\alpha {\cal V}^\beta .
\end{equation}
This result shows that $\cal V$ determines the direction in spacetime in which the energy-momentum of the wave is transported. In other words, $\cal V$ 
represents the {\em 4-velocity of the energy transport}.

\subsection{Fresnel equation for $V$}
Now, since $f=q\wedge a$ and  $h=q\wedge b$, we see that the conditions
(\ref{vinqab}) imply
\begin{equation}\label{defv2}
V\rfloor h=0, \qquad V\rfloor f=0 .
\end{equation}

Kiehn \cite{Kiehn01} uses (\ref{defv2}) to define $V$ as an `extremal
vector field'. Furthermore, it can be shown that (\ref{defv2}) are equivalent to
(\ref{vinqab}). Take for instance (\ref{defv2}b), then we have
\begin{equation}\label{vqa}
V\rfloor f=V\rfloor \left(q\wedge a\right)=
\left(V\rfloor q\right)a-q\left(V\rfloor a\right)=0.
\end{equation}
Suppose now that the scalar $V\rfloor q$ does not vanish, then (\ref{vqa})
implies that $a$ is proportional to $q$. But this would imply that
$f=q\wedge a\equiv 0$. Therefore, nontrivial solutions ($f\neq 0$) require
(\ref{vinqab}a) to hold. Substituting this into (\ref{vqa}) shows that also
(\ref{vinqab}b) must be satisfied. Similarly, starting from (\ref{defv2}a),
the same argument shows that (\ref{vinqab}c) is required.

The equations (\ref{defv2}) for the vector $V$ are similar to the Hadamard
compatibility conditions (\ref{ce01}) for the wave 1-form $q$. This suggests
to apply a  method similar to that used in section \ref{secfresnel} for 
deriving the Fresnel equation for $q$.

First, we solve (\ref{defv2}b) for $f$. A general solution can be written as
\begin{equation}
f=V\rfloor c ,
\end{equation}
where $c$ is an arbitrary 3-form. We write the 4 independent components of $c$
as $c_{\alpha\beta\gamma}=:\hat\epsilon_{\alpha\beta\gamma\delta}\, c^\delta$.
Using this in (\ref{defv2}a) and (\ref{cl03}) we find
\begin{equation}\label{veq}
V\rfloor{}^\#\left(V\rfloor c\right)=0 .
\end{equation}
This equation is analogous to (\ref{e2}). As in the case of the wave
1-form $q$, not all the equations in (\ref{veq})
are independent, since interior product of $V$ with (\ref{veq}) vanishes
identically. In analogy to the method used in section \ref{secfresnel}, it is
convenient to use a vector frame basis $e_\alpha$ and its dual
$\vartheta^\alpha$, with the special `gauge' $e_{\hat 0}=V$. Then, after
some algebra, one finds that (\ref{veq}) is equivalent to
\begin{equation}\label{veq2}
M_{A B}\, c^{B}=0 ,
\end{equation}
with the $3\times 3$ matrix $M_{A B}$ defined as
\begin{equation}\label{mdd}
M_{A B}:=\hat\epsilon_{\hat 0ACD}\,
\chi^{CDEF}\,\hat\epsilon_{EFB\hat 0}.
\end{equation}
Again, nontrivial solutions ($c^{B}\neq 0$) exist only if
$\det{\left(M_{AB}\right)}=0$, i.e. if and only if 
\begin{equation}\label{detmdd}
\frac{1}{3!}\epsilon^{ABC}\epsilon^{DEF}
M_{AD}M_{BE}M_{CF}=0.
\end{equation}
Using (\ref{mdd}) and (\ref{detmdd}), and applying the same method as in section
\ref{secfresnel} to complete two Levi-Civita symbols, we finally find also a
quartic equation that now the components of $V$ must satisfy, namely
\begin{equation}\label{fresnelv}
{\cal M}_{ijkl}\, V^iV^jV^kV^l=0,
\end{equation}
where
\begin{equation}
{\cal M}_{ijkl}:=\frac{1}{4!}\,
\chi^{mtef}\hat\epsilon_{efh(i}\hat\epsilon_{j|mnp}
\chi^{npqr}\hat\epsilon_{qrs|k}\hat\epsilon_{l)tuv}
\chi^{uvhs},
\end{equation}
is a totally symmetric tensor density of weight $-1$.

Equation (\ref{fresnelv}) defines in the space of ray vectors the {\em ray
surface}, see \cite{LL60}.

One can verify after some algebra that in the riemannian case 
the tensor density ${\cal M}_{ijkl}$ reduces to
\begin{equation}
{\cal M}_{ijkl}:=4{|g|}^{-1/2}\,g_{(ij} g_{kl)}.
\end{equation}
Additionally, one verifies from (\ref{vchiaq}) that the vector 
density $\cal V$ reduces to
\begin{equation}
{\cal V}^i=4{|g|}^{1/2}\left[\left(g^{ik}a_k\right)
\left(g^{jl}a_jq_l\right)-\left(g^{il}q_l\right)\left(g^{jk}a_ja_k\right)
\right] .
\end{equation}
Since in a riemannian space we can additionally choose a `Lorenz gauge' for 
$a$, namely $g^{ij}a_i q_j=0$, the first term vanishes and the above result 
shows that ${\cal V}^i\sim q^i:=g^{ij}q_j$, as expected.

\section{Light cone structure}
\label{sectcl}

In this section, we want to study the particular case of our general theory in 
which a {\em light cone} structure is induced. In terms of the Fresnel tensor 
$\cal G$, this means that the spacetime/medium is such that
\begin{equation}
{\cal G}^{ijkl}={\cal G}^{(ij} {\cal G}^{kl)}
=\frac{1}{3}\left({\cal G}^{ij} {\cal G}^{kl}+{\cal G}^{kj} {\cal G}^{il}
+{\cal G}^{lj} {\cal G}^{ik}\right),
\end{equation}
for {\em some} symmetric second order tensor density ${\cal G}^{ij}$ of weight 
$+1/2$. Then the Fresnel equation,
which in general is quartic in the wave-covector, reduces to a second order
equation, namely to a light cone condition:
\begin{equation}\label{lc}
{\cal G}^{ij} \,q_iq_j=0 .
\end{equation}

Obviously, we are interested in this subcase because in GR (and its special 
case of SR) the local properties of light propagation in vacuum are determined by the conformal structure of the underlying riemannian geometry. This implies that the wave covectors satisfy a condition (\ref{lc}), with
${\cal G}^{ij}\sim g^{ij}$. Here $g^{ij}$ denotes the spacetime metric.

We are specifically interested in the following questions: a) What is the
most general constitutive tensor leading to a light cone structure? b) Is it
possible to obtain a light cone structure as a consequence of some
metric-independent conditions? In other words, can a light cone structure be
induced/deduced {\em without postulating the existence of a metric from the 
very beginning}? 
c) If this is the case, how can a conformal metric be explicitly constructed
from the underlying constitutive tensor?

\subsection{Looking for metric-independent conditions}
\label{seclooking}

It is clear that in order to induce a light cone structure additional 
information or assumptions must the added to 
the general pre-metric framework, in the context of which the constitutive tensor is arbitrary. We look for these additional conditions. First, we consider the particular case of vacuum in a riemannian space in order to see if it is possible to find some properties which can be then used as a guide in the pre-metric framework.

We know that the constitutive tensor corresponding to the vacuum in a
riemannian space, i.e.
\begin{equation}\label{chig2}
\chi_{\{g\}}^{ijkl}= 2\sqrt{\frac{\varepsilon_0}{\mu_0}}\sqrt{-g}\,g^{i[k}g^{l]j} ,
\end{equation}
is such that the Fresnel equation reduces to the light cone condition. 
Note that $\chi_{\{g\}}$ is invariant under conformal
transformations $g_{ij}\rightarrow e^{\Psi(x)}g_{ij}\,$. As a consequence, only 9 of the 10 components of the metric can enter $\chi$.

However, since the Fresnel equation is homogeneous, we see that
\begin{equation}\label{chifaxion}
 {\chi}_{\{g,f\}}^{ijkl}=2f(x)\,\sqrt{-g}\,g^{i[k}g^{l]j} ,
\end{equation}
will also induce the same light cone structure, for any scalar function $f(x)$.
This field can be identified with a dilaton field. 
One uses the definition (\ref{G4}) of the Fresnel tensor for this particular
case and, after some algebra, we obtain
\begin{equation}
{\cal G}_{\{g,f\}}^{ijkl}={\rm sgn}(g)\,f^3 \sqrt{|g|}\,g^{(ij}g^{kl)}.
\end{equation}

Let us summarize the properties of the constitutive tensor (\ref{chifaxion}).
Obviously ${}^{(2)}\chi_{\{g,f\}}=0$, because $\chi_{\{g,f\}}$ is
symmetric under the exchange of the first and last pair of indices. Furthermore, 
${}^{(3)}\chi_{\{g,f\}}=0$, i.e., no axion-like term is present. 
Additionally, one verifies that
\begin{equation}\label{closurechig}
 \frac{1}{8}\, \hat\epsilon_{ijkl}\,\chi_{\{g,f\}}^{\, klmn}\,
\hat\epsilon_{mnpq}\,\chi_{\{g,f\}}^{\, pqrs}
=-f^2\,2\delta^{[r}_i\delta^{s]}_j .
\end{equation}
In this expression the metric tensor does not explicitly appear,  
but the negative sign on the right hand side is due to the its lorentzian 
signature.

These two properties, can be rewritten in terms of the corresponding 
linear operator $\kappa$ related to $\chi_{\{g,f\}}$, see (\ref{cl}). 
The symmetry property ${}^{(2)}\chi_{\{g,f\}}=0$ is equivalent to 
\begin{equation}\label{cond1}
A\wedge\left(\kappa B\right) =\left( \kappa A\right) \wedge B  ,
\end{equation}
for any 2-form $A$ and $B$, while the condition (\ref{closurechig}) can be 
written as 
\begin{equation}\label{cond2}
\kappa^2=-f^2 {\bf 1} .
\end{equation} 

From this example, which is realized in GR, we learn that the riemannian vacuum constitutive tensor satisfies some properties with can be written in a 
metric-independent way.  
As we will see in the next sections, an operator satisfying the two conditions 
above defines a {\em duality operator}.  

Hence the question arises whether 
(\ref{cond1}) and (\ref{cond2}) can be taken as additional 
conditions in order to induce a light cone structure 
in the general pre-metric electrodynamic framework and whether these conditions are 
necessary and/or sufficient to induce a light cone structure.
 In fact, it turns out that there exists a relationship 
between dual operators and conformal metrics.

\subsection{Reciprocity and closure}

The two conditions (\ref{cond1}) and (\ref{cond2}) can be motivated 
already in the pre-metric framework. From the discussion of section \ref{seccons} it is clear that the condition 
of symmetry ensures that the medium/spacetime will not possess intrinsic 
dissipative properties. This can then be taken as an additional 
metric-independent condition, see, however, the discussion in sections 
\ref{secnievespal} and \ref{secpropfresnel} about the constitutive tensor of 
Nieves and Pal, for which ${}^{(2)}\chi\neq 0$. 
The condition of closure can be motivated as follows. 
An interesting feature of electrodynamics is its electric/magnetic
{\em reciprocity} property. From the general expression of the kinetic
energy-momentum current
(\ref{defsigma}), one can see that the energy-momentum content of the
electromagnetic field remains the same if one exchanges excitation $H$ and field
strength $F$ in an appropriate way. Consider the transformation
\begin{equation}\label{transreci}
H\rightarrow \zeta F, \qquad F\rightarrow -\frac{1}{\zeta}\, H .
\end{equation}
Here $\zeta$ is a pseudoscalar carrying the same physical dimension as the
constitutive tensor, i.e. $[\zeta]=[H/F]=q^2/h$.
Direct inspection of (\ref{defsigma}) shows that the energy-momentum current
remains invariant under this transformation, i.e.
\begin{equation}
{}^{\rm k}\Sigma_\alpha\rightarrow {}^{\rm k}\Sigma_\alpha.
\end{equation}

In terms of the 3-dimensional forms defined in section \ref{3mas1}, the 
transformation $H\rightarrow \zeta F$ implies ${\cal H}\rightarrow -\zeta E$ and 
${\cal D}\rightarrow \zeta B$ and $F\rightarrow -\frac{1}{\zeta}\, H $ implies 
$E\rightarrow \frac{1}{\zeta} {\cal H}$ and 
$B\rightarrow -\frac{1}{\zeta} {\cal D}$. This clearly shows that 
(\ref{transreci}) exchange electric and magnetic quantities.

So far, this invariance, which we call electric/magnetic 
reciprocity \footnote{In order to distinguish it
from the conventional metric-dependent {\em duality} transformation
$F_{ij}\rightarrow \widetilde{F}_{ij}:=\frac{1}{2}\sqrt{-g}
\,\hat\epsilon_{ijkl}\,g^{km}g^{ln}F_{mn}$, under which the {\em vacuum} ($J=0$) Maxwell equations in a riemannian space are invariant.}, is a property only of the energy-momentum current. 
If one requires, as an additional postulate, the spacetime/constitutive relation also to be electric/magnetic reciprocal, a condition on the constitutive tensor can be found.

Take the spacetime/constitutive relation (\ref{cl}) and perform the transformation
(\ref{transreci}) while leaving the constitutive tensor unchanged. One finds
\begin{equation}
\zeta F_{\alpha\beta}=\frac{1}{4}\,{\hat\epsilon}_{\alpha\beta\gamma\delta}\,
\chi^{\gamma\delta\epsilon\theta}\,(-1)\frac{1}{\zeta}\, H_{\epsilon\theta} .
\end{equation}
Use again (\ref{cl}) to replace the components of the excitation in terms of
those of the field strength, and obtain
\begin{equation}
-\zeta^2 F_{\alpha\beta}=\frac{1}{16}\,{\hat\epsilon}_{\alpha\beta\gamma\delta}\,
\chi^{\gamma\delta\epsilon\theta}{\hat\epsilon}_{\epsilon\theta\mu\nu}\,
\chi^{\mu\nu\lambda\rho}F_{\lambda\rho} .
\end{equation}
This expression leads to a condition for the constitutive tensor, namely to
\begin{equation}\label{chireci1}
-\zeta^2 \delta^{[\lambda}_\alpha\delta^{\rho]}_\beta=
\frac{1}{16}\,{\hat\epsilon}_{\alpha\beta\gamma\delta}\,
\chi^{\gamma\delta\epsilon\theta}{\hat\epsilon}_{\epsilon\theta\mu\nu}\,
\chi^{\mu\nu\lambda\rho} .
\end{equation}
Define now the dimensionless tensor 
\begin{equation}\label{defchi0}
\stackrel{\rm o}{\chi}\!{}^{\alpha\beta\gamma\delta}:=
\frac{1}{\zeta}\,\chi^{\alpha\beta\gamma\delta} ,
\end{equation}
with
\begin{equation}\label{duality4}
  \zeta^2:= -\frac{1}{96}\,({\hat \epsilon}_{ijmn}\,\chi^{mnpq})\,(
  {\hat \epsilon}_{pqr\!s}\,\chi^{r\!sij}).
\end{equation}
Then (\ref{chireci1}) can be rewritten as
\begin{equation}\label{chireci2}
\frac{1}{8}\,{\hat\epsilon}_{\alpha\beta\gamma\delta}
\stackrel{\rm o}{\chi}{}^{\gamma\delta\epsilon\theta}
{\hat\epsilon}_{\epsilon\theta\mu\nu}
\stackrel{\rm o}{\chi}{}^{\mu\nu\lambda\rho}=
- 2\delta^{[\lambda}_\alpha\delta^{\rho]}_\beta .
\end{equation}
It is convenient to define
\begin{equation}\label{circle}
  J_{\alpha\beta}{}^{\gamma\delta}
  :=\frac{1}{2}\,{\hat \epsilon}_{\alpha\beta\eta\sigma}\stackrel{\rm
    o}{\chi}{\!}^{\eta\sigma\gamma\delta}.
\end{equation}
Then the {\em closure relation} (\ref{chireci2}) reads
\begin{equation}\label{chireci2-2}
  \frac{1}{2}\,J_{\alpha\beta}{}^{\eta\sigma} 
  J_{\eta\sigma}{}^{\lambda\rho}=- 
  2\delta^{[\lambda}_\alpha\delta^{\rho]}_\beta
\end{equation}
or, in the corresponding notation in terms of $6\times 6$ matrices,
\begin{equation}\label{close1a}
J^2=-\,1_6.
\end{equation}
We call this condition on the constitutive tensor {\em closure relation}.
The negative sign in (\ref{transreci}b) and therefore the one on
the right hand side of (\ref{chireci2-2}) is a consequence of the negative 
relative sign of the two terms entering in the energy-momentum current, see 
(\ref{defsigma}). 
Below we will see that this negative sign will be responsible for the lorentzian 
signature of the induced metric.

Mathematically, this means that the operator
$J$ represents an almost complex structure on the space of
2-forms.

\subsection{Dual operators and metrics}

A linear operator $J:\Lambda^2(X)\rightarrow 
\Lambda^2(X)$ acting on 2-forms is said to be a {\em dual operator 
defining a complex structure} if it satisfies the property of symmetry, 
\begin{equation}\label{mysym}
A\wedge\left(J B\right) =\left( JA\right) \wedge B  ,
\end{equation}
for any 2-form $A$ and $B$, and if it is additionally {\em closed} such that 
\begin{equation}\label{myclosure}
J^2=-1.
\end{equation} 

Many people have studied this relationship, both as a useful tool 
in GR \cite{Brans71,Brans74} and as a method which could allow to consider 
the metric of spacetime as a secondary quantity constructed from some other 
fields, see for instance \cite{CJD89}. Clearly, the 
dimensionless part of the riemannian constitutive tensor  
$f^{-1}\,\chi_{\{g,f\}}^{ijkl}$ defines a dual operator which is just the 
Hodge dual of the metric $g$.

In the context of electrodynamics, 
it seems that Peres \cite{Peres62} was the first to try to reconstruct the 
conformally invariant part of the spacetime metric from the excitation 
$H$ and the field strength $F$. In his approach, however, the metric is {\em assumed to exist such that} the 
relation between $H$ and $F$ is just the same as the vacuum relation 
in a riemannian space. In our notation, this condition is equivalent to the  postulate that the operator $\kappa$ is a dual operator and that it also equals the Hodge dual of some metric that is to be determined. 
In this sense, the approach of Peres can be considered as a sort of `inverse 
problem'. Contrary to defining the operator $\kappa$ as the dual of the metric 
tensor, as is done in GR, Peres tried to determine a metric such that its Hodge dual operator coincides with $\kappa$. To that goal, it is assumed that the operator $\kappa$ satisfies the conditions of symmetry and closure, as a Hodge dual operator does. However, as we will discuss, Peres did not succeed in deriving the conformally metric since his result is not unique.

Toupin \cite{Toupin65} and Sch\"onberg \cite{Schoenberg71} also studied how a 
conformal metric structure is induced and in particular how the conformally 
invariant part of the metric can be deduced from the spacetime/constitutive 
relation under the assumption of symmetry and closure. They seem to be the 
first who were able to show that a conformal metric structure is actually implied as a consequence of symmetry and closure.
Brans \cite{Brans71,Brans74} also recognized that, within general relativity, 
it is possible to recover the metric from its Hodge dual operator. 
These structures were subsequently discussed by numerous people, by  't
Hooft \cite{tHooft91}, Harnett \cite{Harnett91}, and Obukhov \&
Tertychniy \cite{OT96}, amongst others, see also the references given
there.

From these studies, it seems clear by now that there is a one-to-one 
correspondence between linear operators (of the kind of $\kappa$) satisfying 
symmetry and closure, and conformal metrics. If a metric is available, one 
can immediately construct the corresponding Hodge dual operator which, for a 
metric with lorentzian signature, will satisfy (\ref{mysym}) and 
(\ref{myclosure}). On the other hand, 
the contrary is also valid. If an operator $J$ satisfying (\ref{mysym}) and 
(\ref{myclosure}) is available, 
then a conformal metric structure is {\em induced}, such that the  
operator $J$ can be written as a Hodge dual operator. Furthermore, the metric 
components can be (re-)constructed by using the so-called Sch\"onberg-Urbantke 
formula. 
Sch\"onberg seems to be the first to derive this formula in the context of 
electrodynamics. This formula was also found by 
Urbantke \cite{Urbantke78,Urbantke84}, but in a different context, 
namely in the framework of $SU(2)$ Yang-Mills theories. 
We will discuss some derivations of this important formula below. 

Finally, by using the Sch\"onberg-Urbantke formula in the context of linear 
pre-metric electrodynamics, Obukhov and Hehl \cite{OH99} presented an {\em explicit} construction of the conformally invariant part of the metric tensor in terms of quantities parametrizing a constitutive tensor satisfying symmetry and closure.

After discussing these developments, an alternative procedure to 
deduce the conformal metric and its light cone structure will be presented. It 
relies on the general results about the Fresnel equation governing the local 
properties of electromagnetic waves. 
This alternative approach is therefore much more suitable  
to understand the physics underlying the emergence of the conformal 
structure of spacetime, since it directly involves the properties of 
the propagation of waves. Finally, this approach also allows to study how 
the dropping or modifying of the assumptions of symmetry and 
closure would affect the light cone structure. 
In particular, we will explore the consequences a possible asymmetric 
spacetime/constitutive relation (i.e., with skewon part) would have on the light 
cone, but maintaining the assumption of closure.

\subsubsection{Sch\"onberg-Urbantke formula}
\label{securbantke}

Here we discuss the successful derivation of of the conformal metric by 
Sch\"onberg and the related work of Harnett.


In his work \cite{Schoenberg71} Sch\"onberg studied a different derivation of
the metric from electromagnetism. Additionally, the analysis in \cite{Schoenberg71}
helps to better understand the group theoretical aspects of the emergence
of a lorentzian metric structure and thus of the Lorentz group.

As we already saw before, the electromagnetic
field strength $F$, with its six independent components, can be described as a
vector in a 6-dimensional real vector space. We call this space $S_6$.
One can map a basis of the
space of 2-form, $dx^{i_1}\wedge dx^{i_2}$ to a basis $b^I$ of $S_6$ by means of
the rule $[i_1 i_2]\rightarrow I=01,02,03,23,31,12$. 
Then one can write a vector $A\in S_6$ as $A=A_I\,b^I$.

An important role is played by an inner product $\epsilon$ on $S_6$ induced
by the 4-dimensional Levi-Civita symbol $\epsilon^{ijkl}$.
Given two vectors $A$ and $B$ of $S_6$, one can define their product as
\begin{equation}\label{epsab}
\epsilon(A,B):=\epsilon^{IJ}A_IB_J.
\end{equation}
In 4-dimensional notation, this is equivalent to
\begin{equation}
\epsilon(A,B)\,\hat\epsilon:=A\wedge B 
\end{equation}
or, in components,
\begin{equation}
\epsilon(A,B)=\frac{1}{4}\,\epsilon^{ijkl}A_{ij} B_{kl} .
\end{equation}
Thus, the Levi-Civita symbol acts as a metric in the 6-dimensional space $S_6$.
Notice, however, that $\epsilon(A,B)$ is a 4-dimensional density of weight $+1$. 
Since the eigenvalues of $\epsilon^{IJ}$ are $+1$ and $-1$, each with
multiplicity 3 (see (\ref{chiepsIJ}b)), the signature of $\epsilon^{IJ}$ is
$(+1,+1,+1,-1,-1,-1)$. This
immediately shows that the 6-dimensional space $S_6$ naturally contains a
$SO(3,3)$ group structure. Transformation of $S_6$-vectors under the action of
the $SO(3,3)$ group leaves the product (\ref{epsab}) invariant. It is important
to emphasize that this group structure is {\em always present}, independent of 
any metric, affine, or whatever additional structure on the 4-dimensional 
manifold.

Using this product, one can express the assumption of symmetry of the
constitutive tensor, i.e., ${}^{(2)}\chi=0$ as
\begin{equation}\label{ksim}
\epsilon\left(A,J B\right) = \epsilon\left(J A,B\right)
\end{equation}
for all $A,B\in S_6$, because 
\begin{equation}
\epsilon\left(A,J B\right) = \epsilon^{IJ}A_I\,J_J^{\ K}B_K
= \chi^{IK} A_I B_K 
\end{equation} 
since $\chi^{IK}=\epsilon^{IJ}J_J^{\ K}$, and 
\begin{equation}
\epsilon\left(J A, B\right) = \epsilon^{IJ}J_I^{\ K}A_K B_J
= \chi^{IK} A_K B_J=\chi^{KI} A_I B_K ,
\end{equation} 
so that (\ref{ksim}) requires symmetry of $\chi$, i.e., $\chi^{IJ}=\chi^{JI}$.

Furthermore, the additional condition of closure of the
operator $J$, i.e.,
\begin{equation}\label{kclosure}
J^2=-1, 
\end{equation}
is assumed to hold.

In the 6-dimensional real space $S_6$, the introduction of the linear operator
$J$ satisfying (\ref{ksim}) and (\ref{kclosure}) induces a {\em almost complex
structure}.
In order to be able to discuss this structure and in particular the eigenvectors 
of the operator $J$,
consider the complex extension of $S_6$, i.e. the vector space of the
complex 2-forms $S_6(C)$. The negative sign on
the right hand side of (\ref{kclosure}) implies that the eigenvalues of $J$
are $\pm i$, and since $J$ is a real operator, each eigenvalue must have
multiplicity 3. This means that the corresponding {\em self-dual} and
{\em antiself-dual}
subspaces spanned by vectors with eigenvalues $+i$ and $-i$ are both
3-dimensional. We denote these subspaces as $S_3^+(C)$ and $S_3^-(C)$ 
respectively.
The vectors of $S_3^\pm(C)$ will be denoted as $A^\pm$ and satisfy
\begin{equation}\label{kspm}
J\,A^\pm=\pm i A^\pm . 
\end{equation}
Take any $A^\pm\in S_3^\pm(C)$ and write it as $A^\pm=A+iA^\prime$ with 
$A$ and $A^\prime$ real 6-dimensional vectors (i.e. $A,A^\prime\in S_6$). 
Then from (\ref{kspm}) one finds that
\begin{equation}
J\left(A+iA^\prime\right) =\pm i\left(A+iA^\prime\right),
\end{equation}
with implies
\begin{equation}\label{kap}
J\,A=\mp A^\prime, \qquad J\,A^\prime=\pm A .
\end{equation}
Using (\ref{kap}) one can rewrite every element $S^\pm$ of $S_3^\pm$, in terms 
of only the real $A\in S_6$ and the operator $J$, as
\begin{equation}\label{sakappaa}
A^\pm=A\mp i J A.
\end{equation}

The spaces $S_6^+(C)$ and $S_6^-(C)$ are orthogonal with respect to the product 
$\epsilon$. Take any $A^+\in S_6^+(C)$ and  $B^-\in S_6^-(C)$. Write them in the 
form (\ref{sakappaa}), i.e. $A^+=A-i J A$ and $B^-=B+i J B$, and 
compute their $\epsilon$-product. Using the symmetry and closure properties 
(\ref{ksim}) and (\ref{kclosure}) one finds 
\begin{eqnarray}
\epsilon( A^+,B^-)&=&\epsilon(A- i J\, A,B+ i J\, B) \nonumber \\
&=&\epsilon(A,B)+i\epsilon(A,J B)-i\epsilon(J\, A,B)
+\epsilon(J\, A,J\, B) \nonumber \\
&=& \epsilon(A,B)+i\epsilon(A,J B)-i\epsilon( A,J\,B)
+\epsilon(A,J^2\, B) \nonumber \\
&=& \epsilon(A,B)+\epsilon(A,- B) \nonumber \\
&=& 0.  \label{ortho}
\end{eqnarray} 

Consider a real operator $M$ acting on the real space $S_6$. If $A\in S_6$ then 
also $MA\in S_6$. However, the self-dual element $(MA)^+$ of $S_3^+$ will in 
general not be the result of the application of the linear operator $M$ on 
$A^+$, since 
\begin{equation}
(MA)^+=(MA)-iJ (MA)\neq M(A-iJ A)=M(A^+) ,
\end{equation} 
{\em unless the operators $M$ and $J$ commute}, i.e. $\left[ M,J\right] =0$. 
In other 
words, not every linear operator on $S_6$ is a linear operator in $S_3^+$.  
In particular, not every element of the group $O(3,3)$ which leaves $\epsilon$ 
invariant defines a linear operator in the self-dual space $S_3^+$. 

The corresponding subgroup of $O(3,3)$ which commutes with $J$, i.e. the 
subgroup form by those operators $N$ such that 
\begin{equation}\label{subgroup}
\epsilon\left( NA,NB\right) = \epsilon\left( A,B\right), 
\qquad \left[ N,J\right] =0,
\end{equation} 
can be shown to be isomorphic to the Lorentz group. 
To prove this, consider the product 
$\epsilon$ restricted to the self-dual subspace $S_3^+$, which we denote as 
$\epsilon^+$, and is defined by 
\begin{equation}
\epsilon^+\left( A^+,B^+\right) :=\epsilon\left( A^+,B^+\right) .
\end{equation} 
Consider operators $N$ satisfying (\ref{subgroup}) which are 
therefore also linear operators of $S_3^+$. 
It is clear that they will also leave 
invariant the 3-dimensional product $\epsilon^+$, since in this case 
\begin{eqnarray}
\epsilon^+\left((NA)^+,(NB)^+ \right) 
&=& \epsilon^+\left(NA^+,NB^+ \right) = \epsilon\left(NA^+,NB^+ \right) \nonumber \\
&=&\epsilon\left(A^+,B^+ \right) = \epsilon^+\left(A^+,B^+ \right).
\end{eqnarray} 
This shows that the subgroup of $O(3,3)$ commuting with $J$ forms the 
invariance group of the 3-dimensional metric induced by $\epsilon^+$ in $S_3^+$. 
But since $S_3^+$ is a complex linear space, the corresponding invariance group 
of $\epsilon^+$ is clearly $SO(3,C)$, which is isomorphic to the Lorentz group 
$SO(3,1)$. In other words, an operator $J$ satisfying symmetry 
(\ref{ksim}) and closure (\ref{kclosure}) induces a $SO(3,C)\approx SO(3,1)$ 
group structure, which manifest itself as the symmetry group of the natural 
metric structure on the self-dual space of $J$. Similarly, the 
same arguments can be repeated considering the anti-self-dual space $S_3^+$. 

We consider now how to reconstruct the metric components of the corresponding 
induced lorentzian metric.
Consider a basis for each subspace. We denote the basis of $S^+_3$ by 
$S^{(a)}_+$, $a,b,\ldots=1,2,3$. Then the complex conjugate 
$S^{(a)}_-:=\left( S^{(a)}_+\right)^*$ is a basis of $S^-_3$, see  (\ref{kspm}). 
Since $S_6(C)= S^+_3\oplus S^-_3$, the six $S_6$-vectors 
$\left\{ S^{(a)}_+,S^{(a)}_-\right\} $ form a basis of $S_6(C)$. 
Then the orthogonality property (\ref{ortho}) implies that the components of the 
metric $\epsilon$ of $S_6(C)$ in the basis $\left\{ S^{(a)}_+,S^{(a)}_-\right\} $ 
form a block-diagonal matrix, since $\epsilon\left( S^{(a)}_+,S^{(b)}_-\right)=0$.

One can use the {\em exterior product of vectors in} $S_6(C)$, which we denote 
by $\triangle$, and determine a volume element of $S^+_3$ as 
\begin{equation}\label{w+}
W_+:=S_+^{(1)}\triangle\, S_+^{(2)}\triangle\, S_+^{(3)}
=\frac{1}{3!}\,{\hat\epsilon}_{abc}\,S_+^{(a)}\triangle\, S_+^{(b)}
\triangle\, S_+^{(c)}.
\end{equation} 
Then $W_+$ is a 3-form on $S_6$, i.e. $W_+\in \Lambda^3(S_6)$. 
Since $S^+_3$ is 3-dimensional, $W_+$ is, up to a scalar factor,  
independent of the choice of the basis $S^{(a)}_+$. 
Furthermore, any product of basis vectors $S^{(a)}_+$  of an order higher 
than 3 vanishes identically (e.g., 
$S_+^{(a)}\triangle\, S_+^{(b)}\triangle\, S_+^{(c)}\triangle\, S_+^{(d)}=0$). 
A volume element $W_-$ for $S_3^-$ can be similarly 
defined by using the basis $S^{(a)}_-$ of $S_3^-$, which then satisfies 
$W_-=\left( W_+\right)^*$.

A 3-form $W\in \Lambda^3(S_6)$ has 20 independent components $W_{IJK}$, which 
correspond to a 4-dimensional tensor $W_{i_1i_2j_1j_2k_1k_2}$ of order six, with 
the symmetries $W_{i_1i_2j_1j_2k_1k_2}=-W_{j_1j_2i_1i_2k_1k_2}
=-W_{k_1k_2j_1j_2i_1i_2}=-W_{i_1i_2k_1k_2j_1j_2}$ and, of course, 
$W_{i_1i_2j_1j_2k_1k_2}=$\\ $-W_{i_2i_1j_1j_2k_1k_2}$, etc.
From this tensor, a symmetric second order 4-dimensional tensor density of 
weight $+1$ can be defined as 
\begin{equation}\label{wdd}
{\cal W}_{ij}:=\frac{1}{2}\, \epsilon^{klmn}\,W_{iklmnj},
\end{equation} 
which has thus 10 independent components. The remaining 10 components of 
$W$ can be mapped to a second order contravariant tensor density of weight 
$+2$, defined by 
 \begin{equation}\label{wuu}
{\widetilde {\cal W}}^{\,ij}
:=\frac{1}{4}\, \epsilon^{iklm}\,W_{klmnpq}\,\epsilon^{npqj}.
\end{equation} 

Following (\ref{wdd}) we define the tensor density ${\cal W}_{ij}^+$ associated 
to $W_+$ as 
\begin{equation}\label{wdd+}
{\cal W}^+_{ij}:=\frac{1}{2}\, \epsilon^{klmn}\,W^+_{iklmnj},
\end{equation} 
or, equivalently,  
\begin{equation}\label{defcalw}
{\cal W}^+_{ij}:=\frac{1}{2\cdot 3!}\, \epsilon^{klmn}\,{\hat\epsilon}_{abc}
\,S^{+(a)}_{ik}\,S^{+(b)}_{lm}\,S^{+(c)}_{nj}. 
\end{equation} 

Furthermore, we saw that a change of the basis $S_+^{(a)}$ of $S_3^+$ 
to a new one $S_+^{\prime(a)}$ leads to a new volume $W^\prime_+$ 
which is necessarily proportional to $W_+$, i.e. $W^\prime_+=a_3\,W_+$ 
for some $a_3$. One can therefore define a new tensor density by 
\begin{equation}\label{calw+ij}
\hat{\cal W}^+_{ij}:=\left( \det{{\cal W}_{kl}^+}\right) ^{-1/4}\,{\cal W}_{ij}^+.
\end{equation} 
Then $\hat{\cal W}_+$ is a tensor density of weight $-1/2$ which is {\em 
independent of the choice of the basis $S_+^{(a)}$ of $S_3^+$}. Because of this 
important property, $\hat{\cal W}^+_{ij}$ must describe an intrinsic property 
of the self-dual space $S_3^+$ and therefore of the operator $J$. 
It is then tempting to identify the symmetric tensor density 
(\ref{calw+ij}) with the conformally invariant part of the metric 
tensor $|g|^{-1/4}g_{ij}$, which is also a  tensor density of weight $-1/2$, 
i.e., 
\begin{equation}\label{wisg}
\hat{\cal W}^+_{ij}=a |g|^{-1/4}g_{ij}, 
\end{equation} 
with some (in general complex) factor $a$.

That this identification is consistent with our expectations can be checked as 
follows. 
Consider the linear operator $L$ acting on vectors of $S_6(C)$, defined by 
\begin{equation}
\left( {L} A\right)_{ij}:=\frac{1}{2}\,{L}_{ij}^{\ \ kl} A_{kl},
\end{equation} 
with 
\begin{equation}\label{calLijkl}
{L}_{ij}^{\ \ kl}:=\hat{\cal W}^+_{im}\hat{\cal W}^+_{jn}\,
\epsilon^{mnkl}.
\end{equation} 
This operator is expected to be related to the original operator $J$, 
because the identification (\ref{wisg}) would imply 
\begin{equation}
{L}_{ij}^{\ \ kl}=a^2\,|g|^{-1/2}\,g_{im}\,g_{jn}\,\epsilon^{mnkl}  
\end{equation}  
which is proportional to the Hodge dual operator of the metric $g$, 
when applied to 2-forms.
 
Using (\ref{calw+ij}) one re-writes the definition (\ref{calLijkl}) of the 
operator $L$ as 
\begin{eqnarray}
{L}_{ij}^{\ \ kl}&=&
\left( \det{{\cal W}_{pq}^+}\right)^{-1/2}\,{\cal W}_{im}^+
\,{\cal W}_{jn}^+ \,\epsilon^{mnkl} \label{callww} .
\end{eqnarray} 
Furthermore, it can be proved that 
$L$ is a complex operator. From (\ref{callww}) and the definition 
(\ref{defcalw}) one can verify by direct but rather lengthy calculations 
that the elements of $S_3^+$ and $S_3^-$ are 
eigenvectors of $L$ with eigenvalues $+1$ and $-1$, respectively, i.e. 
\begin{equation}\label{calls+-s}
{L}S^+=S^+, \qquad  {L}S^-=-S^-.
\end{equation} 
Since $S_3^-=\left( S_3^+\right)^*$, the operator $L$ cannot be real. If it 
were, (\ref{calls+-s}a) would imply ${L}S^-=S^-$, contradicting  
(\ref{calls+-s}b). 
It is therefore clear that the real operator $J$ is 
just given by $J=iL$, because 
\begin{equation}\label{kappail}
(i{L}) S^+=iS^+, \qquad  (i{L}) S^-=-iS^-,
\end{equation} 
which coincides with (\ref{kspm}). The property (\ref{kappail}) is valid for 
any vector of $S^\pm_3$ and since $S_6(C)=S^+_3\oplus S^-_3$, it 
implies that the action of $iL$ and $J$ on any vector of $S_6(C)$ is 
exactly the same, hence they are the same operator. 

Thus, we have proved that 
\begin{equation}\label{kappaisil}
J=i{L},
\end{equation} 
so that, from (\ref{callww}), we have 
\begin{equation}
J_{ij}^{\ \ kl}= 
\left( -\det{{\cal W}_{pq}^+}\right)^{-1/2}\,{\cal W}_{im}^+
\,{\cal W}_{jn}^+ \,\epsilon^{mnkl} .
\end{equation} 
With the identification (\ref{wisg}) we write the operator $J$ as 
\begin{equation}
J_{ij}^{\ \ kl}=\sqrt{-g}\,g_{im}g_{jn}
\,\epsilon^{mnkl} .
\end{equation} 
From this expression, one sees that the metric $g_{ij}$ can be taken to be 
real, with {\em lorentzian signature}, since $J$ is a real operator.
In other words, the results and (\ref{defcalw}) allow us to reconstruct 
the metric components as 
\begin{equation}\label{formula}
g_{ij}\sim\epsilon^{klmn}\,{\hat\epsilon}_{abc}
\,S^{+(a)}_{ik}\,S^{+(b)}_{lm}\,S^{+(c)}_{nj}, 
\end{equation} 
with $g_{ij}$ real. The conformal factor remains of course undetermined. 

\subsubsection{Necessary and sufficient conditions for the constitutive tensor}

The results discussed above can be summarized as follows. If the operator $J$,  
defined in (\ref{circle}) by the dimensionless part of the constitutive tensor,  
is symmetric and defines an almost complex structure, i.e., (\ref{mysym}) and 
(\ref{myclosure}) are satisfied, then a lorentzian metric $g$ is determined, 
up to a conformal factor, such that 
\begin{equation}\label{ochig}
\stackrel{\rm o}{\chi}{}^{ijkl}=2\sqrt{-g}\,g^{i[k} g^{l]j} ,
\end{equation} 
with $g:=\det g_{ij}<0$. In other words, the conditions (\ref{mysym}) and 
(\ref{myclosure}) are sufficient to be able to write $\stackrel{\rm o}{\chi}$ in 
the form (\ref{ochig})

But it is clear, see the discussion in section \ref{seclooking}, that 
(\ref{ochig}) is sufficient to ensure that the corresponding operator $J$ 
satisfies (\ref{mysym}) and (\ref{myclosure}).  We thus conclude:

{\em 
The necessary and sufficient conditions to be able to write the dimensionless 
part $\stackrel{\rm o}{\chi}$ of the constitutive tensor as 
\begin{equation}
\stackrel{\rm o}{\chi}{}^{ijkl}=2\sqrt{-g}\,g^{i[k} g^{l]j} ,
\end{equation} 
with a lorentzian metric $g$, are symmetry 
\begin{equation}
\stackrel{\rm o}{\chi}{}^{ijkl}=\stackrel{\rm o}{\chi}{}^{klij} 
\end{equation} 
and closure 
\begin{equation}\label{newcl}
\frac{1}{8}\,{\hat\epsilon}_{\alpha\beta\gamma\delta}
\stackrel{\rm o}{\chi}{}^{\gamma\delta\epsilon\theta}
{\hat\epsilon}_{\epsilon\theta\mu\nu}
\stackrel{\rm o}{\chi}{}^{\mu\nu\lambda\rho}=
- 2\delta^{[\lambda}_\alpha\delta^{\rho]}_\beta .
\end{equation}  
}

\subsection{General solution of the Closure Relation}

In order to be able to find an explicit expression for the metric in terms of 
quantities describing the components of the constitutive tensor, we have to solve 
the closure relation (\ref{chireci1}) or, equivalently, (\ref{chireci2-2}). We 
will solve this equations for the general case in which the constitutive tensor 
is asymmetric, so that we can later investigate the effect of relaxing the 
symmetry condition.

Let us now make the closure relation explicit. We turn back to the
constitutive $6\times 6$ matrix (\ref{close1a}). We define
dimensionless $3\times 3$ matrices $\stackrel{\;{\rm o}}{\cal
  A}\;:={\cal A}/\zeta$, etc. In terms of these dimensionless matrices
(we immediately drop the small circle for convenience), the closure
relation reads,
\begin{eqnarray}  
{\cal A}^{ac}{\cal B}_{cb} +  {\cal C}^a{}_c{\cal C}^c{}_b &=&   
- \delta^a_b,\\  
{\cal C}^a{}_c {\cal A}^{cb} + {\cal A}^{ac}{\cal D}_c^{\ b}   
&=& 0,\\  
{\cal B}_{ac}{\cal C}^c{}_b + {\cal D}_a^{\ c}{\cal B}_{cb}    
&=& 0,\\  
{\cal B}_{ac}{\cal A}^{cb} +  {\cal D}_a^{\ c}{\cal D}_c^{\ b}   
&=& -\delta^b_a.  
\end{eqnarray}  

In $3\times 3$ matrix notation, we then have 
\begin{eqnarray}
{\cal A}{\cal B} + {\cal C}^2 &=&- 1_3,\label{almostclose1} \\
{\cal C}{\cal A} + {\cal A}{\cal D}&=& 0,\label{almostclose2}\\
{\cal B}{\cal C} + {\cal D}{\cal B}&=& 0, \label{almostclose3}\\
{\cal B}{\cal A} + {\cal D}^2&=& -1_3.\label{almostclose4}
\end{eqnarray}
Assume $\det {\cal B}\neq 0$. Then we can find the general 
non-degenerate solution as follows: Define the matrix $K_{ab}$ by 
\begin{equation}  
  K:={\cal BC},\qquad{\rm i.e.}\qquad {\cal C}= {\cal 
    B}^{-1}K,\label{CDKK} 
\end{equation}  
and substitute it into (\ref{almostclose3}), then 
\begin{equation}  
{\cal D} = - K{\cal B}^{-1}.\label{KK}  
\end{equation}  
Next, solve (\ref{almostclose1}) with respect to $\cal A$: 
\begin{equation}  
{\cal A}=-{\cal B}^{-1}-{\cal B}^{-1}K{\cal B}^{-1}K{\cal  
B}^{-1}.\label{ABK}  
\end{equation}  
Multiply (\ref{ABK}) by $\cal C$ from the left and by $\cal D$ from 
the right, respectively, and find with (\ref{CDKK}) and (\ref{KK}), 
\begin{eqnarray}  
{\cal CA} &=& - {\cal B}^{-1}K{\cal B}^{-1}   
- {\cal B}^{-1}K{\cal B}^{-1}K{\cal B}^{-1}K{\cal B}^{-1},\\  
{\cal AD} &=& + {\cal B}^{-1}K{\cal B}^{-1}   
+ {\cal B}^{-1}K{\cal B}^{-1}K{\cal B}^{-1}K{\cal B}^{-1}.  
\end{eqnarray}  
Thus, (\ref{almostclose2}) is automatically satisfied.  
Accordingly, only (\ref{almostclose4}) has still to be checked. 
We compute its first and second term of its left side, 
\begin{eqnarray}  
{\cal BA} &=& -1_3 - K{\cal B}^{-1}K{\cal B}^{-1},\\  
{\cal D}^2 &=& K {\cal B}^{-1} K {\cal B}^{-1}, 
\end{eqnarray} and find that it is fulfilled, indeed. 
  
Summing up, we have derived the general solution of the closure 
relation (\ref{close1a}) in terms of two arbitrary matrices ${\cal B}$ 
and $K$ as 
\begin{eqnarray}  
{\cal A} &=& - {\cal B}^{-1} - {\cal B}^{-1}K{\cal B}^{-1}K{\cal  
B}^{-1},\\  
{\cal C} &=& {\cal B}^{-1}K,\\  
{\cal D} &=& - K{\cal B}^{-1}, 
\end{eqnarray}  
or, in components, 
\begin{eqnarray}  
{\cal A}^{ab} &=& - {\cal B}^{ab} - {\cal B}^{ac}K_{cd}{\cal B}^{de}K_{ef}
{\cal B}^{fb}, \label{abk}\\  
{\cal B}^{ab}&=& {\cal B}^{ab}, \\
{\cal C}^a_{\ b} &=& {\cal B}^{ac}K_{cb}, \label{cbk}\\  
{\cal D}_a^{\ b} &=& - K_{ac}{\cal B}^{cb} \label{dkb}. 
\end{eqnarray}  
Here ${\cal B}^{ab}$ are the components of the inverse ${\cal B}^{-1}$, i.e.
${\cal B}^{ab}{\cal B}_{bc}=\delta^a_b$.
The solution thus has $2\times 9 = 18$ independent components. 
Alternatively, one can write the solution in a more compact notation as 
\begin{eqnarray}  
{\cal A} &=&
- (1_3+{\cal C}^2){\cal B}^{-1}\, ,\\ {\cal D} &=& - {\cal BCB}^{-1} , 
\end{eqnarray}  
which is parametrized by the arbitrary matrices ${\cal B}$ and ${\cal C}$  
with altogether 18 independent components. In components, this means 
\begin{eqnarray}
{\cal A}^{ab} &=&- \left(\delta^a_d+{\cal D}_c^{\ a}{\cal D}_d^{\ c}\right)
{\cal B}^{db} ,\\
{\cal B}_{ab} &=& {\cal B}_{ab}  , \\
{\cal C}^b_{\ a} &=& - {\cal B}_{ac}{\cal D}_d^{\ c}{\cal B}^{db} , \\
{\cal D}_b^{\ a} &=& {\cal D}_b^{\ a} . \\
\end{eqnarray}

\subsubsection{Explicit derivation of the metric components from the 
constitutive tensor}

Here, the Sch\"onberg-Urbantke formula will be used to find an explicit expression 
for the conformal metric in terms of quantities parametrizing a constitutive 
tensor satisfying the assumptions of symmetry and closure. From the symmetry 
condition one finds that the constitutive tensor has the form (\ref{chiepsIJ}), 
but with 
\begin{equation}\label{ABCDsym}
{\cal A}^{ab}={\cal A}^{ba}, \qquad  {\cal B}_{ab}={\cal B}_{ba}, 
\qquad {\cal D}_b^{\ a}={\cal C}^a_{\ b}  ,
\end{equation} 
so that ${}^{(2)}\chi=0$, see (\ref{chi2epsIJ}). These symmetry conditions 
restrict the general solution of the closure relation found in the last section.
From (\ref{cbk}) and (\ref{dkb}) we see that (\ref{ABCDsym}c) implies 
$-K_{bc}{\cal B}^{ca}={\cal B}^{ac}K_{cb}$. Multiplying this equation by 
${\cal B}_{ad}$ and using (\ref{ABCDsym}c) we obtain $-K_{bd}=K_{db}$ which 
tells us that the matrix $K$ must be antisymmetric. Then (\ref{abk}) implies 
that $\cal A$ is automatically symmetric, so no further conditions on $K$  arise 
from (\ref{ABCDsym}a). 
In summary, a symmetric constitutive tensor satisfying the closure relation is 
given by (\ref{abk})--(\ref{dkb}) with
\begin{equation}\label{ABCDsym2}
{\cal B}_{ab}={\cal B}_{ba}, \qquad  K_{ab}=-K_{ba} .
\end{equation} 

As we saw in section \ref{securbantke}, to construct the metric using the 
Sch\"onberg-Urbantke formula, we need a basis of the 3-dimensional space 
$S_3^+$ of self-dual 2-forms of the duality operator $J$.

 To that end, we decompose the basis $b^I$ of $S_6$ into two 3-dimensional
column vectors, according to 
\begin{equation}\label{bIdec}
 b^I = \left(\begin{array}{c} \beta^a \\ \gamma_b 
 \end{array}\right),\quad a,b, \dots = 1,2,3. 
\end{equation} 
Then we can find their self-dual parts, 
\begin{equation}
 b_+^I = \frac{1}{2}(b^I - iJb^I),
\end{equation} 
and decompose them similarly as in (\ref{bIdec}), into
\begin{equation} 
b_+^I = \left(\begin{array}{c}  
 \beta_+^a \\  
\gamma_b^+ \end{array}\right) . 
\end{equation}

Assuming, as we did before to find a solution of the closure relation, that 
the matrix $\cal B$ is nonsingular, we can show that one can take $\gamma^+$ as a 
basis of the self-dual space $S_3^+$, since the remaining self-dual 2-forms 
$\beta_+$ can be written as linear combinations of the former. This is expected 
since $S_3$ is 3-dimensional. Explicitly, we have 
\begin{equation}
\beta_+^a=(i\delta^a_b + {\cal B}^{ac}K_{cb}){\cal B}^{bd}\gamma^+_d.  
\end{equation}  
This allows us to choose the 2-forms $\gamma^+_a$ or,
equivalently, the triplet
\begin{equation}  
 S_+^a:= -{\cal B}^{ab}\gamma^+_b ,
\end{equation}  
as basis of $S_3^+$. 
 
The information of the constitutive tensor $\chi$ is now encoded into the 
triplet of 2-forms $S_+^{(a)}$.

The $\left( S_+^a\right) _{ij}$ are the components of the 2-form triplet 
$S_+^a =\left( S_+^a\right) _{ij}  dx^i\wedge dx^i/2$.  
If we substitute the self-dual 2-forms $S_+^a$ into (\ref{formula}), we can 
display the metric explicitly in terms of the constitutive coefficients:
\begin{equation}\label{metric1} 
g_{ij} =\phi \left(\begin{array}{c|c} \det {\cal B} & 
-\,k_a \\ \hline -\,k_b & -\,{\cal B}_{ab} + (\det {\cal B})^{-1}\,k_a\,k_b 
\end{array}\right) \, .
\end{equation}  
Here $k^a:=\epsilon^{abc}K_{bc}/2$, $k_a:={\cal B}_{ab}k^b$ and $\phi$ is an 
arbitrary factor. The determinant of this metric is found to be 
$g=-\phi^4\left( \det {\cal B}\right)^2$ so that we verify that the metric in 
(\ref{metric1}) has lorentzian signature.

\subsubsection{Properties of the metric}

The inverse of (\ref{metric1}) can be found to be 
\begin{equation}\label{metricinv}
 g^{ij}= \frac{1}{\phi\det {\cal B}}\left(\begin{array}{c|c}  
  1- (\det {\cal B})^{-1} k_c k^c & -k^b\\ \hline  
  -k^a & -(\det {\cal B}){\cal B}^{ab}\end{array} \right)\, . 
\end{equation} 
With the help of (\ref{metric1}) and (\ref{metricinv}), we can compute 
the Hodge duality operator ${}^*$ attached to this metric. In
terms of the components of the 2-form $F$, we have
\begin{equation} 
^* F_{ij}:=\frac{\sqrt{-g}}{2}\,\hat\epsilon_{ijkl}\,
g^{km}g^{ln}F_{mn}.
\end{equation} 
This equation can be rewritten, in analogy to (\ref{cl}), by
using the constitutive tensor $\chi_{\{g\}}$, defined in (\ref{chig2}), so that 
\begin{equation}
 {}^* F_{ij}=\frac{1}{4}\hat\epsilon_{ijkl}\, \chi_{\{g\}}^{klmn} F_{mn} , 
\end{equation}

In order to compare $\chi_{\{g\}}$ with the original constitutive tensor 
$\chi$, we compute the corresponding 3-dimensional constitutive matrices 
of  $\chi_{\{g\}}$ according to (\ref{cd01}) and (\ref{cd02}).

Then straightforward calculations yield
\begin{eqnarray}
 {\cal A}_{\{g\}}^{ab}&=&\sqrt{-g}\left( g^{00}g^{ab}-g^{0a}g^{0b}\right)
 ={\cal A}^{ab} , \\ 
 ({\cal B}_{\{g\}})_{ab}&=&\frac{1}{4}\sqrt{-g}\left(g^{ce}g^{df}
 -g^{de}g^{ef}\right) \hat\epsilon_{acd}\, \hat\epsilon_{efb}={\cal B}_{ab} , \\
 ({\cal C}_{\{g\}})^a_{\ b}&=&\frac{1}{2}\sqrt{-g}\left(g^{0c}g^{ad}
 -g^{ac}g^{0d} \right) \hat\epsilon_{bcd}={\cal B}^{ad} K_{db}={\cal }C^a_{\ b}.
\end{eqnarray} 
Thus, $\chi_{\{g\}}=\chi$, i.e., the metric extracted allows us to write 
the original duality operator $J$ as Hodge duality operator, $J={}^*$, when 
applied on 2-forms.  

\subsubsection{Alternative derivation}

Here we provide an alternative derivation of the conformal metric, this time 
based on a direct computation of the Fresnel equation for the symmetric solution 
of the closure relation given by (\ref{abk}), (\ref{dkb}) and (\ref{ABCDsym2}).
Using this equation and our general expressions (\ref{ma1})-(\ref{ma4}) for 
the Fresnel equation in terms of the 3-dimensional constitutive matrices, we 
find after a straightforward calculation that
\begin{eqnarray} 
M &=&-\,{\frac 1 {\det{\cal B}}}\left(1 - 
{\frac {k_ak^a}{\det{\cal B}}}\right)^2,\label{r1}\\ 
M^a &=& {\frac 1 {\det{\cal B}}}\,4k^a\left(1 - {\frac {k_bk^b}{\det{\cal
B}}}
\right), \label{r2}\\ 
M^{ab} &=& -\,{\frac 1 {\det{\cal B}}}\,4k^ak^b + 2{\cal B}^{ab} 
\left(1 - {\frac {k_ck^c}{\det{\cal B}}}\right), \label{r3}\\ 
M^{abc} &=& -\,4\,{\cal B}^{b(a}\,k^{c)}, \label{r4}\\ 
M^{(abcd)} &=& -\,(\det{\cal B})\,{\cal B}^{(ab}{\cal B}^{cd)}.\label{r5} 
\end{eqnarray}
Substituting all this into the general Fresnel equation (\ref{fresnel}),
we find 
\begin{eqnarray} 
{\cal W}&=& -\frac{\theta^2}{\det{\cal B}}
\left[q_0^2 \left(1 - \frac{k_ak^a}{\det{\cal B}}\right) 
- 2q_0(q_ak^a) 
 \right. \nonumber \\ && \qquad \qquad \qquad \left. 
- \left( \det{\cal B}\right) 
\,\left( q_aq_b{\cal B}^{ab}\right) \right]^2 .
\label{reduc}
\end{eqnarray} 
Therefore we find that the Fresnel equation, ${\cal W}=0$, can be written as 
\begin{equation}
\left(g^{ij}q_iq_j\right)^2=0,
\end{equation} 
where $g^{ij}$ is the (inverse) 4-dimensional conformal metric which arises from 
the duality operator and the closure relation. Direct comparison with 
(\ref{reduc}) shows that 
\begin{eqnarray} 
g^{00}&=& \psi\,
\left(1 - {\frac {k_ak^a}{\det{\cal B}}}\right),\\ 
g^{0a}&=& -\psi\,k^a,\\ 
g^{ab}&=& -\psi\,\left( \det{\cal B}\right) \,{\cal B}^{ab}. 
\end{eqnarray} 
Here $\psi$ in the undetermined conformal factor. 

Thus we indeed recover the null cone structure for the propagation of 
electromagnetic waves from our general analysis. 
The quartic surface degenerates to the null cone for the conformal metric $g$. 

\subsection{Relaxing the symmetry condition}
\label{metric}

Having verified and explicitly constructed the metric tensor from the constitutive 
tensor in the case in which it is symmetric and satisfies the closure relation, 
we want to study what consequences a nonvanishing skewon piece in the constitutive tensor could have on the emergence of a conformal structure. In particular, 
can the spacetime metric still be constructed? We have seen that the skewon 
piece does influence the light cone structure, and we also saw that an 
asymmetric constitutive tensor satisfying closure can accommodate 18 independent 
functions. Could this case, for instance, correspond to the emergence of 
two light cones, each with its 9 independent components?

Unfortunately, a computation of the Fresnel tensor and the Fresnel equation 
using the general asymmetric solution (\ref{abk}) to (\ref{dkb}) of the closure relation does 
not help much to recognize a possible double light cone structure. 
The Fresnel equation is still a quartic equation, in general.

In what follows, we will study a particular case in order to try to get an idea of 
the qualitative properties that an additional skewon piece can induce. More details 
can be found in \cite{ROH02}.  
We will consider the case in which $K=0$, which implies ${\cal D} = 0$, 
${\cal C} =0$ \footnote{i.e., a medium with no `magneto-electrical' properties.} 
and ${\cal A} = - {\cal B}^{-1}$. Furthermore, we 
decompose the arbitrary matrix ${\cal B}$ into its symmetric and antisymmetric 
parts,
\begin{equation}
  {\cal B}_{ab} = b_{ab} + \hat{\epsilon}_{abc}n^c,\quad{\rm
    with}\quad b_{ab}:= {\cal B}_{(ab)},\quad n^c:= \epsilon^{cab}
  \,B_{[ab]}.\label{Bn}
\end{equation}
Note that $ b_{ab}$ contributes only to $^{(1)}\chi$ and $n^c$ only to
$^{(2)}\chi$.  
 In this case, by substituting (\ref{Bn}) into (\ref{fresnel}), we obtain:
\begin{equation}
{\cal W} = -{\frac {q_0^2}{(\det{b}+n^2)}}\left[q_0^4
- 2q_0^2\left({\bar q}^2 - (qn)^2\right) + \left({\bar q}^2 +
(qn)^2\right)^2 \right]. \label{freC0}
\end{equation}
Here we used the abbreviations $n^2:=b_{ab}n^an^b$, ${\bar q}^2:={\bar b}^{ab}\,q_aq_b$, 
$qn:=q_a\,n^a$, and ${\bar b}^{ab}$ is the (symmetric) matrix of 
the minors of $b_{ab}$.

In general, the right hand side of (\ref{freC0}) is neither a square of a quadratic
polynomial nor a product of two quadratic polynomials. In other words,
neither a light cone nor a birefringence (double light cone) structure
arises generically. 

Formally, one can rewrite (\ref{freC0}) as

\begin{equation}\label{freC02} 
{\cal W} = -\,\frac{q_0^2}{(\det{b}+n^2)} 
\left[ q_0^2+\left((qn) +\sqrt{-\overline{q}^2}\right)^2\right] 
\left[ q_0^2+\left((qn) - \sqrt{-\overline{q}^2}\right)^2\right] 
\end{equation}

Thus, the question of the reducibility of the Fresnel equation
translates into the algebraic problem of whether the square root
$\sqrt{-\overline{q}^2}$ is a real linear polynomial in $q_a$. There
are three cases, depending on the rank of the $3\times 3$ matrix
$b_{ab}$, see \cite{ROH02}.

(i) When $b_{ab}$ has rank 3, {\it no factorization into light cones} is
possible (the roots $\alpha$ are complex), unless $n^a=0$. This latter
condition implies  ${}^{(2)}\chi=0$, and the previous results are recovered.

(ii) When $b_{ab}$ has rank 2, i.e., $\det b =0$, but at least one of the
minors is nontrivial. Then, without loss of generality, one can assume:
\begin{equation}
b_{ab} = \left(\begin{array}{ccc}b_{11}&b_{12}&0\\ b_{12}&b_{22}& 0\\
0 & 0 & 0\end{array}\right).\label{b-deg}
\end{equation}
In order to avoid complex solutions, we have to assume that the minor $\overline{b}^{33}
= -\,\mu^2 < 0$, so that $\sqrt{-{\bar q}^2}=\mu q_3$. In this case we find
\begin{eqnarray}
{\cal W} &=& -\,{\frac {q_0^2}{b_{11}(n^1)^2+2b_{12}n^1n^2+b_{22}(n^2)^2}}
\nonumber \\
&& \times\, \left[q_0^2+\left( q_1n^1 + q_2n^2 + q_3(n^3+\mu)\right)^2\right]
\nonumber \\
&& \times\, \left[q_0^2+\left( q_1n^1 + q_2n^2 + q_3(n^3-\mu)\right)^2)\right] ,
\end{eqnarray}
i.e. two different light cones. We can read off, up to conformal factors, 
the components of the two corresponding `metric' tensors defining the light cones:
\begin{equation}
g_1^{ij}=\left(   \begin{array}{cccc}
                1 & 0 & 0 & 0 \\
                0 & (n^1)^2 & n^1 n^2 & n^1(n^3+\mu) \\
                0 & n^1n^2 & (n^2)^2 & n^2n^3 \\
                0 & n^1(n^3+\mu) & n^2n^3 & (n^3+\mu)^2
         \end{array}\right) \, ,
\end{equation}
\begin{equation}
g_2^{ij}=\left(   \begin{array}{cccc}
                1 & 0 & 0 & 0 \\
                0 & (n^1)^2 & n^1 n^2 & n^1(n^3-\mu) \\
                0 & n^1n^2 & (n^2)^2 & n^2n^3 \\
                0 & n^1(n^3-\mu) & n^2n^3 & (n^3-\mu)^2
         \end{array}\right) \, .
\end{equation}
One verifies that $\det (g_1^{ij})=\det (g_2^{ij})=
(n^1)^2(n^2)^2{\bar{b}}^{33}=-(n^1)^2(n^2)^2\mu^2<0$, so that both metrics
have the correct Lorentzian signature.

(iii) When the $3\times 3$ matrix $b_{ab}$ has rank 1. In this case
corresponds to the case 2 with $\mu=0$. The Fresnel equation reduces to a single 
light cone, but the resulting metric is degenerated, since $\det (g^{ij})=0$.

Thus, we have demonstrated with a special case that some of the asymmetric 
solutions of the closure relation can yield birefringence. 

We saw that the conditions of closure and symmetry of
$\chi$ are sufficient for the existence of a well-defined light
cone structure. If any of these conditions is violated, the light cone
structure seems to be lost. The necessary conditions have still to be
found.

\section{Conclusions and Prospects}

In this work, we have developed a general framework for describing classical 
electrodynamics in a general 4-dimensional medium. We have paid special attention to identify the structures which are metric-independent, and which can therefore be applied to a great variety of particular cases, from generalized models describing the electromagnetic properties of spacetime to  classical optics in material media. 

One of the central results of this work is summarized in the Fresnel equation 
(\ref{Fresnel}) and in particular in the Fresnel tensor (\ref{G4}). This equation 
determines the local properties of the propagation of waves, in particular it 
describes the geometry of the wave covectors. It is a remarkable result that 
this important equation could be derived for {\em any linear medium}. 
Furthermore, the result can also be used to describe the effective properties of 
electromagnetic perturbations in nonlinear media. Its structure, i.e., its 
dependence on the constitutive tensor of the corresponding medium is highly 
nontrivial. Cubic structures are not very common in physics. 
To the best of my knowledge, no previous derivation of the Fresnel 
equation has been given which is as general as that in section \ref{secfresnel}, see 
also the early work of Tamm \cite{Tamm25}.

It is also important to emphasize that the Fresnel equation and the whole 
formalism is generally covariant. This means that no artificial or 
particular coordinate choices are necessary and that only quantities describing 
intrinsic properties, in our 
case the constitutive tensor, of the physical system under consideration enter 
in the formalism. Of course, inside a given medium, for example, one can 
choose specific coordinates which are useful for concrete calculations. This 
is usually the case if the medium possesses some symmetry, as defined in section 
\ref{secsymlin}. Then an adapted coordinate system can be used which exploits 
this symmetry, so that calculations become simpler. This is, however, only a 
convenient choice, determined by the properties of the system (the constitutive 
tensor), but not a necessary a priori ingredient. 
In a material medium which is not 
isotropic, for instance, there is no a priori reason to use cartesian 
coordinates, the later are rather coordinates adapted to an isotropic medium, 
as for instance the minkowskian vacuum.
Therefore, it is highly satisfactory to be able to describe the electromagnetic 
properties of a material medium in a generally covariant way.

We also studied the properties of the three different irreducible pieces 
which a general constitutive tensor can contain. We have proved that the first, 
the symmetric 
piece ${}^{(1)}\chi$ is essential for the medium to admit well behaved wave properties. 
This piece is therefore the most important object determining the geometry of wave covectors 
and the light cone structure under some particular conditions. The second piece 
${}^{(2)}\chi$ was shown to describe dissipative properties of the medium. A nonvanishing 
${}^{(2)}\chi$ piece can lead to dissipation of electromagnetic energy, even 
is the material's properties are time independent. 
Furthermore, the skewon piece does influence the wave propagation and therefore the light 
cone structure. We have shown, studying some particular cases, that the skewon field can 
even give rise to a double light cone structure. 
The third possible irreducible piece ${}^{(3)}\chi$ of 
a constitutive tensor, the abelian axion, is more elusive. 
It does not contribute to the energy-momentum density of the electromagnetic 
field and does not affect the local properties of the 
electromagnetic waves and thus not the light cone structure either. 
It is possible, however, that it could influence the properties of wave propagation over 
finite (long) distances. Examples of all the three irreducible pieces of a constitutive can 
be found in the literature. The skewon field has been discussed so far only for material 
media violating $CP$ symmetry, but up to now not as a model for the classical electromagnetic 
properties of spacetime. 
If one follows the common assumption that a fundamental physical system should be describable 
in terms of an action principle, then the detection of a skewon piece could be interpreted 
as indicating an underlying substructure of the system under study. Of course, we do not 
claim here that the skewon field is actually nonvanishing in spacetime. We do believe, 
however, that it is a possible piece which can be used to quantify well defined, in general 
dissipative, properties of spacetime.
Additionally, the skewon piece can be used to model possible effects violating 
local Lorentz invariance. Within the framework of GR, where a metric is present, one can show 
that the skewon piece necessarily violates local Lorentz invariance, see appendix \ref{lli}.

Furthermore, we have studied the conditions under which a light cone structure is induced. 
We have seen that, if the constitutive tensor is symmetric and satisfies a closure relation, 
a light cone is, in fact, induced. 
We also saw that the necessary and sufficient conditions for a constitutive tensor to be 
written as proportional 
to the Hodge dual operator of some metric, are symmetry and closure. This 
one-to-one relationship is valid when we formulate it in terms of the 
{\em constitutive tensor}. However, it is still an open question whether 
symmetry and closure are necessary conditions for defining a light cone, in the sense that the 
quartic Fresnel equation reduces to a quadratic equation for the wave covectors, as 
discussed at the beginning of section \ref{sectcl}. We proved that symmetry and closure are 
sufficient conditions for such a reduction. Strictly speaking, closure of the 
whole constitutive tensor is not a necessary condition since an additional axion piece does 
not disturb the Fresnel equation. 
However, it may be that closure of only the principal irreducible piece ${}^{(1)}\chi$ and symmetry, i.e. ${}^{(2)}\chi=0$, could be the necessary conditions. A proof of 
this conjecture is, however, still missing. Non-trivial examples supporting this 
conjecture can be found in \cite{OR02}.

Other interesting open issues which could be investigated are: We have seen that 
the constitutive tensor defines a certain generalization of the conformal 
properties of spacetime. Does it also defines a (generalized) affine structure? To answer this question, it would be interesting to study some additional properties of the {\em propagation} of waves, for instance, how the polarization vector propagates along a light ray. It should be possible to find an answer to this question within our general formalism and it is expected to involve some affine properties (under which parallel displacement is the polarization vector constant along the light ray?).

\appendix

\section{The situation so far}
\begin{figure}[h]
 \centering\epsfig{file=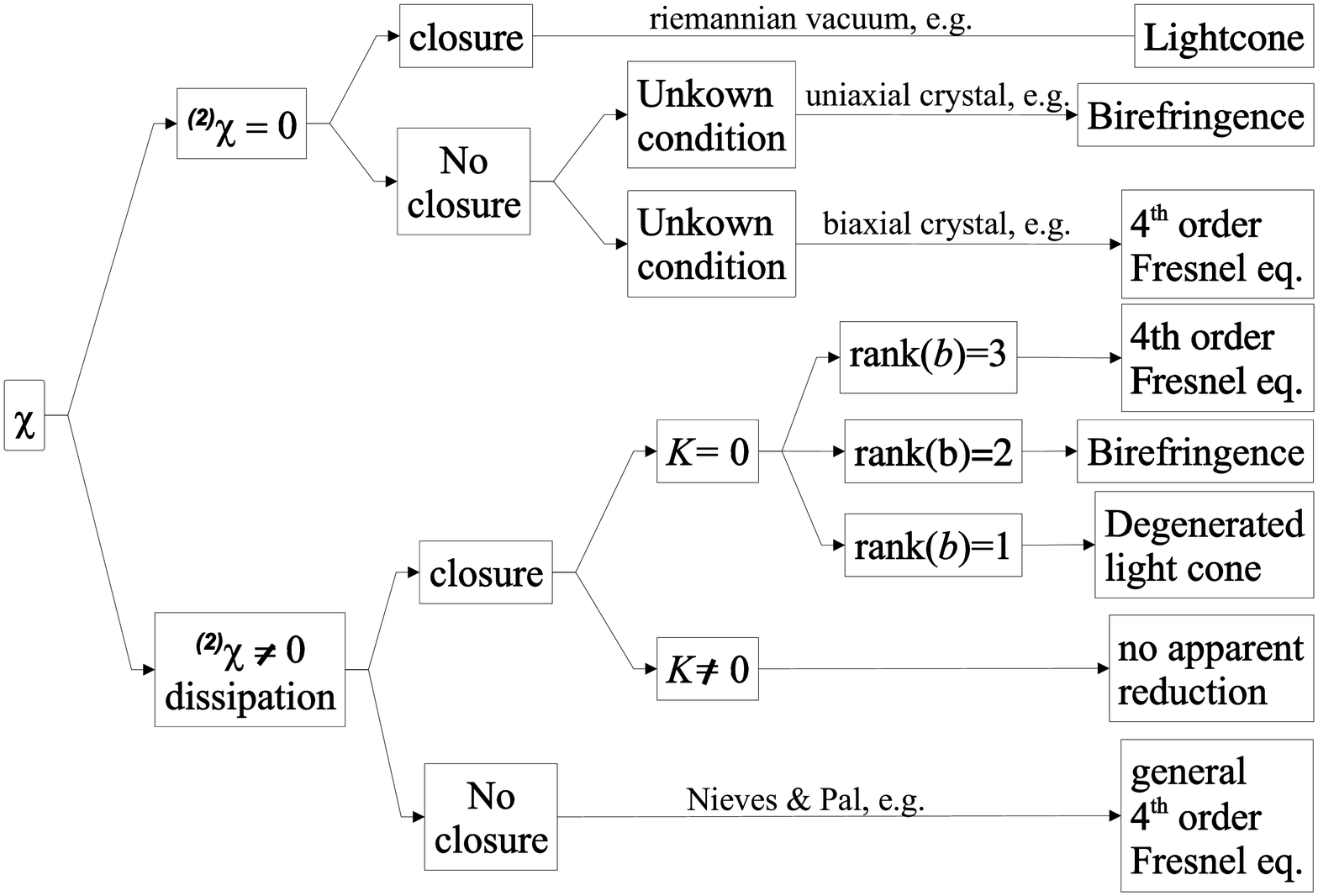, height=7cm, width=12cm}
 \caption{Different particular cases for the constitutive tensor 
and the corresponding behavior of the Fresnel equation and, therefore, of the light cone structure . 
The quantities $K$ and $b$ refer to the 3-dimensional matrices in section \ref{metric}.}
 \label{chart}
\end{figure}

\section{Electrodynamics in a material medium}
\label{appmat}

In a {\em material medium}, treated as a continuum, one can describe
the properties of electromagnetic fields by means of {\em macroscopic} Maxwell's
equations. A material medium can be defined in general terms as a 
region with a given total charge at macroscopic scales, but with a microscopic charge substructure. This substructure corresponds to the individual electrons, ions, etc. which form the medium. The particular distribution and dynamics of the charges forming the material are in general unknown or difficult to model. Therefore, it is useful to describe the electrodynamics of materials only in term of the so called `external' charges, which are those that can, in principle, be manipulated in experiments.

The procedure to define macroscopic Maxwell equations for the material medium
starting from the microscopic ones consist then in the separation of the
{\em total charge} in the sum of the two
contributions originating `from the inside' (bound or polarizational charge)
and `from the outside' (free or external charge):
\begin{equation}\label{sc}
J=J^{\rm mat}+J^{\rm ext} .
\end{equation}
Bound charges and bound currents are inherent characteristics of
matter determined by the medium itself. In contrast, external charges and
external currents appear in general outside and inside matter and can
be prepared for a specific purpose.

The external charge is assumed to be conserved, and consequently the bound
charge is conserved, too:
\begin{equation}
dJ^{\rm mat}=0, \qquad dJ^{\rm ext}=0 .
\end{equation}
As in the case of vacuum, see Refs. \cite{OH99,HOR00,HO02}, this allows us to
introduce the corresponding excitation $H^{\rm mat}$ as a `potential' for
the bound current:
\begin{equation}\label{dhmat}
dH^{\rm mat}=J^{\rm mat} .
\end{equation}
In a (1+3) decomposition, the 6 components of $H^{\rm mat}$ are identified
with the {\em polarization} $P$ and {\em magnetization} $M$.

Defining now the external excitation $H^{\rm ext}$ as
\begin{equation}
H^{\rm ext}:=H-H^{\rm mat},
\end{equation}
we find the inhomogeneous Maxwell equation
\begin{equation}
dH^{\rm ext}=J^{\rm ext}\, .
\end{equation}
Here, only external quantities are involved.

It remains to specify the constitutive law. In addition to the spacetime
relation, $H=H(F)$, the knowledge of the internal structure of a medium
yields the macroscopic excitation $H^{\rm mat}$ (i.e., the polarization
and magnetization) as a function of the electromagnetic field strength
$F$ (and possibly of some thermodynamical variables describing the material
continuum). Then the constitutive law of the material is given by
\begin{equation}
H^{\rm ext}= H^{\rm ext}(F)=H(F)-H^{\rm mat}(F) .
\end{equation}

\section{Local Lorentz invariance}
\label{lli}

Given a lorentzian metric, we can define the notion of local Lorentz invariance. 
Let $T^{i_1\dots i_p}$ be the contravariant coordinate components of a
tensor field and $T^{\alpha_1\dots\alpha_p} :=e_{i_1}{}^{\alpha_1}
\cdots e_{i_p}{}^{\alpha_p}\, T^{i_1\dots i_p}$ its frame
components with respect to an {\em orthonormal} frame $e_\alpha =
e^i{}_\alpha\,\partial_i$.  A tensor is said to be local Lorentz invariant 
at a given point, if its frame components are invariant under a
local Lorentz rotation of the orthonormal frame. Similar considerations
extend to tensor densities.

There are only two geometrical objects which are numerically invariant
under local Lorentz transformations: the Minkowski metric
$\eta_{\alpha\beta}$ and the Levi-Civita tensor
density $\hat\epsilon_{\alpha\beta\gamma\delta}$. Thus
\begin{equation}\label{iso}
{\cal T}^{\alpha\beta\gamma\delta}=\phi(x)\left( \eta^{\alpha\gamma}
\eta^{\beta\delta}-\eta^{\beta\gamma}\eta^{\alpha\delta}\right) + \alpha(x)\,
\epsilon^{\alpha\beta\gamma\delta}  
\end{equation}
is the most general form of the frame components of a locally Lorentz invariant 
contravariant fourth rank tensor with the symmetries ${\cal T}^{ijkl}=-{\cal
T}^{jikl} =-{\cal T}^{ijlk}$. Here $\phi$ and
$\alpha$ are scalar and pseudo-scalar fields, respectively.  Therefore, back in
coordinate components, we find that
\begin{equation}\label{isocomp}
\chi^{ijkl}=\phi(x)\, \sqrt{-g}\left( g^{ik}g^{jl} - g^{jk} g^{il}\right)
+ \alpha(x)\, \epsilon^{ijkl}
\end{equation}
is the most general form of a locally Lorentz invariant constitutive tensor 
allowed in a space with a lorentzian metric. 
Notice that this constitutive tensor is necessarily symmetric.
In other words, any additional asymmetric piece $\sim {}^{(2)}\chi$ would
violate local Lorentz invariance. This was found before in the particular 
examples of constitutive tensors studied by Nieves and Pal, see section 
\ref{secnievespal}.

\vspace*{0.25cm} \baselineskip=10pt{\small \noindent  
The author would like to thank Prof. F.W. Hehl 
for his support and for the discussions and joint work on the subject of this 
work. I also thank Dr. Y. Obukhov (Moscow) for the joint work, and 
the members of the gravity group at Cologne for their help and kindness. 
Finally, I would like to thank the German Academic 
Exchange Service (DAAD) for financial support.

\end{document}